\documentclass[aps,prl,reprint,balancelastpage,nofootinbib,preprintnumbers,superscriptaddress]{revtex4-1}

\usepackage{slashed}
\usepackage{bm}
\usepackage{latexsym,amssymb,amsmath,float,url,mathrsfs}
\usepackage{latexsym}
\usepackage{graphicx}
\usepackage{epstopdf}
\usepackage{amsmath}
\usepackage{subfigure}
\usepackage{natbib}
\usepackage{hyperref}
\usepackage{pifont}
\usepackage{blindtext}
\usepackage{multirow}
\usepackage{tabularx}
\usepackage[table]{xcolor}

\usepackage{graphics, appendix,afterpage,makecell}

\newcolumntype{P}[1]{>{\centering\arraybackslash}p{#1}}

\newcommand\Tstrut{\rule{0pt}{2.6ex}}         
\newcommand\Bstrut{\rule[-0.9ex]{0pt}{0pt}}   

\hypersetup{
     colorlinks   = true,
     citecolor    = violet,
     urlcolor     = violet,
     linkcolor    = violet
}

\begin{document}

\title{Dark Matter Strikes Back at the Galactic Center}
\preprint{MIT-CTP/5104}

\author{Rebecca K. Leane}
\thanks{{\scriptsize Email}: \href{mailto:rleane@mit.edu}{rleane@mit.edu}; {\scriptsize ORCID}: \href{http://orcid.org/0000-0002-1287-8780}{0000-0002-1287-8780}}
\affiliation{Center for Theoretical Physics, Massachusetts Institute of Technology, Cambridge, MA 02139, USA}

\author{Tracy R. Slatyer}
\thanks{{\scriptsize Email}: \href{mailto:tslatyer@mit.edu}{tslatyer@mit.edu}; {\scriptsize ORCID}:
\href{http://orcid.org/0000-0001-9699-9047}{0000-0001-9699-9047}}
\affiliation{Center for Theoretical Physics, Massachusetts Institute of Technology, Cambridge, MA 02139, USA}
\affiliation{School of Natural Sciences, Institute for Advanced Study, Einstein Drive, Princeton, NJ
08540, USA}

\date{\today}

\begin{abstract}
Statistical evidence has previously suggested that the Galactic Center GeV Excess (GCE) originates largely from point sources, and not from annihilating dark matter. We examine the impact of unmodeled source populations on identifying the true origin of the GCE using non-Poissonian template fitting (NPTF) methods. In a proof-of-principle example with simulated data, we discover that unmodeled sources in the \textit{Fermi} Bubbles can lead to a dark matter signal being misattributed to point sources by the NPTF. We discover striking behavior consistent with a mismodeling effect in the real \textit{Fermi} data, finding that large artificial injected dark matter signals are completely misattributed to point sources. Consequently, we conclude that dark matter may provide a dominant contribution to the GCE after all.
\end{abstract}

\maketitle

\noindent
\textbf{\textit{Introduction.}}
There has been an extensive debate in the literature over the origins of the Galactic Center Excess (GCE), an extended and roughly spherically symmetric gamma-ray source filling the region within $\sim1.5$ kpc of the Galactic Center (GC), with energy spectrum peaking at $1-3$ GeV \cite{Goodenough:2009gk, Hooper:2010mq, Hooper:2011ti, Hooper:2013rwa, Daylan:2014rsa,Calore:2014xka,TheFermi-LAT:2015kwa}. The leading hypotheses are a new population of unresolved gamma-ray pulsars, individually too faint to be detected but in aggregate yielding the excess \cite{Abazajian:2014fta, Abazajian:2012pn, Hooper:2013nhl, Mirabal:2013rba, Calore:2014oga, Cholis:2014lta, Yuan:2014yda, OLeary:2015qpx,Ploeg:2017vai,Hooper:2018fih, Bartels:2018xom,Bartels:2018eyb}, or alternatively a signal from annihilating dark matter (DM) (e.g. \cite{Goodenough:2009gk, Daylan:2014rsa, Karwin:2016tsw}). The latter explanation, if confirmed, would be of extraordinary importance for our understanding of the Universe, as the first non-gravitational probe of the properties of DM and its interactions with visible particles.

The hypothesis that unresolved point sources (PSs) generate much or all of the GCE can potentially be distinguished from the DM hypothesis via photon statistics~\cite{Malyshev:2011zi,Lee:2014mza}. The DM signal is expected to be dominated by the smooth Galactic halo (although see \cite{Agrawal:2017pnb}), and the probability of seeing a certain number of photons from a spatial pixel is obtained from the Poisson distribution based on the expected number of photons. For a population of unresolved sources, the positions of the individual sources are not known, and the scatter of the number of \emph{sources} (of a given brightness) in a pixel must be taken into account. Qualitatively, a population of unresolved PSs will generally have a greater probability to generate pixels with a large number of counts (due to the presence of one or more sources) or a very small number of counts (due to an absence of sources) compared to an extended diffuse source with the same overall expected number of photons.

Standard ``template fitting'' methods, where the sky is modeled as a linear combination of components with distinct spatial morphologies, can be adapted to incorporate these differences in statistical behavior. Templates for diffuse components consist of an expected number of photons in each pixel, possibly with a free overall normalization factor; templates for populations of unresolved PSs are characterized additionally by the ``source count function'' (SCF), which describes the probability that a given source has a certain brightness (i.e. produces a certain expected number of photons). It is then possible to calculate the probability to observe a certain number of photons in each pixel, as a function of the coefficients of the various templates and the source-count function parameters, and to study the resulting overall likelihood as a function of these parameters. This approach is called non-Poissonian template fitting (NPTF)~\cite{Malyshev:2011zi,Lee:2014mza,Lee:2015fea}. 

The NPTF method has previously been used to study the inner Galaxy in gamma rays, modeling the sky as a linear combination of Galactic diffuse emission, the large structures known as the $\emph{Fermi}$ Bubbles, isotropic extragalactic diffuse emission, isotropically distributed extragalactic PSs, Galactic PSs tracing the disk of the Milky Way, and the GCE. Modeling the GCE as a linear combination of a DM signal and an unresolved point-source population, with identical spatial morphologies for the signal but differing statistics, Ref.~\cite{Lee:2015fea} found that there was a strong statistical preference for the presence of the point-source population, and that the DM contribution was consistent with zero.

In this \textit{Letter}, we explore the robustness of the NPTF to the presence of additional physical contributions to the gamma-ray data, which are not captured by the standard choice of templates. Some such modifications -- such as changing the exact morphology of the disk-correlated or GCE-correlated PSs -- have already been explored \cite{Lee:2015fea} and shown not to qualitatively change the preference for PSs in the GCE. Motivated by the results of wavelet studies that find enhanced small-scale power in this region~\cite{Bartels:2015aea,Balaji:2018rwz}, we focus on the possibility that there could be Galactic PSs present that are not physically associated with the Galactic disk or the GCE. If such small-scale power is present but corresponds to a source population unrelated to the GCE, we hypothesize that the template-based techniques employed in Ref.~\cite{Lee:2015fea} could incorrectly attribute these PSs to the GCE, and consequently bias the reconstruction of any DM contribution.

We first explore this possibility by simulating a proof-of-principle scenario, where one set of templates is used to construct the simulated data, and these data are then fitted with a different set of templates, similar to the NPTF pipeline in Ref.~\cite{Lee:2015fea}. This allows us to directly observe and explore a bias reducing the reconstructed DM contribution, in a case where all inputs are known. In particular, we show that this bias can occur, can drive the reconstructed DM contribution to a value consistent with zero and inconsistent with its true simulated value, and that this remains true even when an additional DM signal is injected into the simulated data. We then examine real \textit{Fermi} gamma-ray data, where the distribution of the underlying source populations is not known. We demonstrate that when an additional simulated DM signal is injected into the real data, it is also (incorrectly) attributed to PSs, qualitatively similar to the behavior we expect in the biased case from our proof-of-principle example.\\

\noindent
\textbf{\textit{Methodology and Data Selection.}}
To employ the NPTF method, we use the NPTF package \texttt{NPTFit} \cite{Mishra-Sharma:2016gis}, interfaced with the Bayesian interference tool \texttt{MultiNest}~\cite{Feroz:2008xx}. The total number of live points for all \texttt{MultiNest} runs is \mbox{\texttt{nlive = 500}}.
Simulated contributions from Poissonian templates are generated by taking a random Poisson draw in \texttt{Python} of the product of the template and a predetermined normalization. Mock data for PS populations are generated using \texttt{NPTFit-Sim}~\cite{NPTFSim}. We analyze eight years of \textit{Fermi} \texttt{Pass 8} gamma-ray data, for energies $2-20$ GeV, restricting to the top quartile of events graded by angular reconstruction. Our region of interest (ROI) is within a $30^\circ$ radius of the GC, with a $|b|<2^\circ$ plane mask, where $b$ is Galactic latitude. PSs contained in the 3FGL catalog~\cite{Acero:2015hja} are masked at a $99\%$~C.L. containment radius of 0.778 degrees for 2 GeV photons. Further details are supplied in the supplementary material.

We define the ``standard NPTF pipeline'' to be a fit including Poissonian templates for the Galactic diffuse emission (``Diffuse''), isotropic emission (``Iso''), the $\emph{Fermi}$ Bubbles (``Bub''), and the GCE (``DM''), and non-Poissonian templates for PSs tracing the Galactic disk (``Disk PS''), isotropic emission (``Iso PS''), and the GCE (``NFW PS''); template definitions are included in the supplementary material. ``NFW'' here stands for Navarro-Frenk-White~\cite{Navarro:1995iw}, a commonly-employed prescription for the DM density profile that has previously been used in modeling the GCE. We perform a fit on the real data using this standard pipeline; we calculate the medians of the posterior distributions for the various model parameters, and simulate data based on those parameter values. We also define a ``baseline NPTF pipeline'' which contains the same templates except that it omits the NFW PS template; the Bayes factor between the standard and baseline fits thus describes the strength of evidence in favor of GCE-distributed PSs.

\begin{figure*}[t!]
\leavevmode
\centering
\subfigure{\includegraphics[width=0.85\columnwidth]{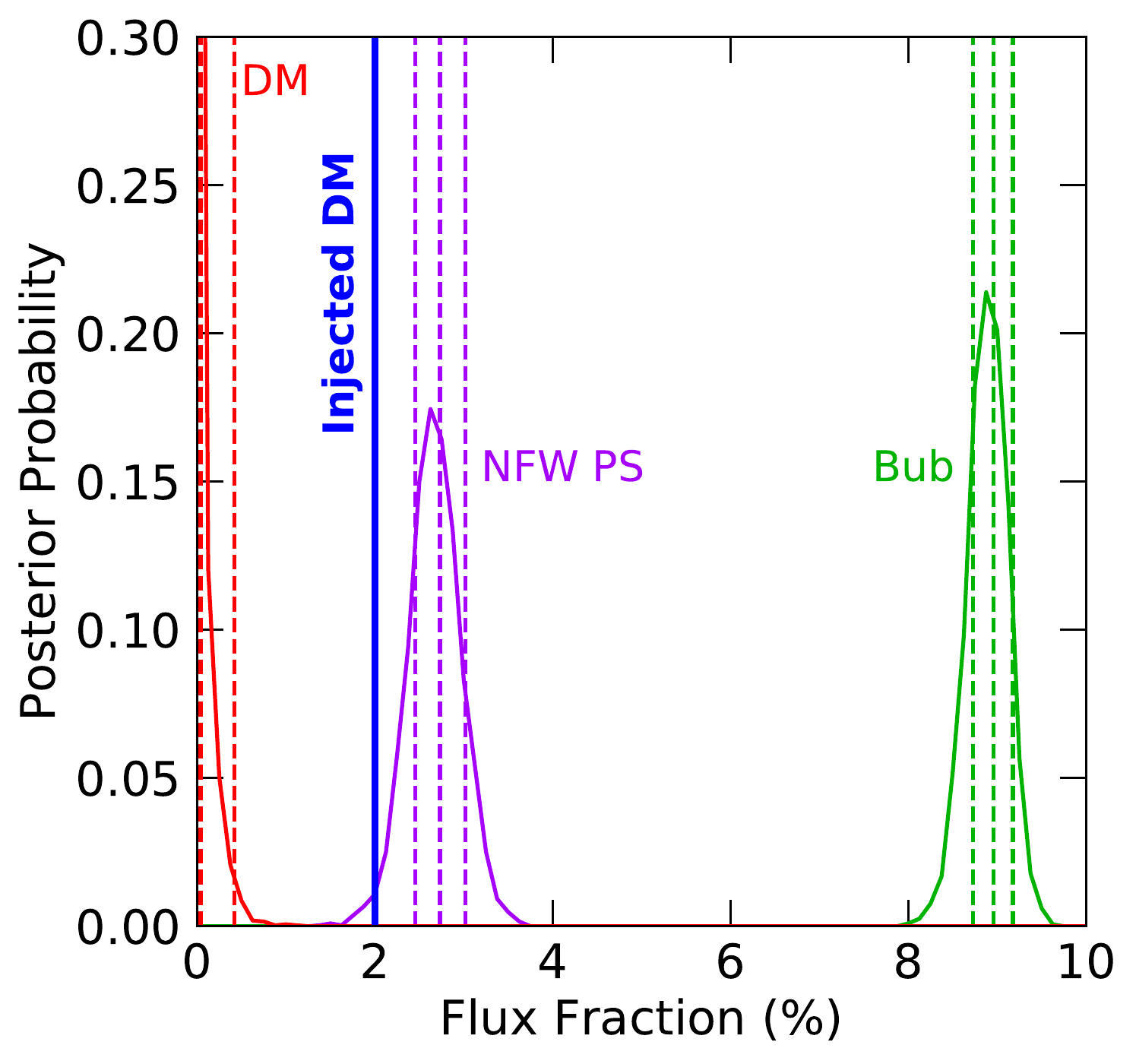}}
\hspace{1mm}
\subfigure{\includegraphics[width=0.85\columnwidth]{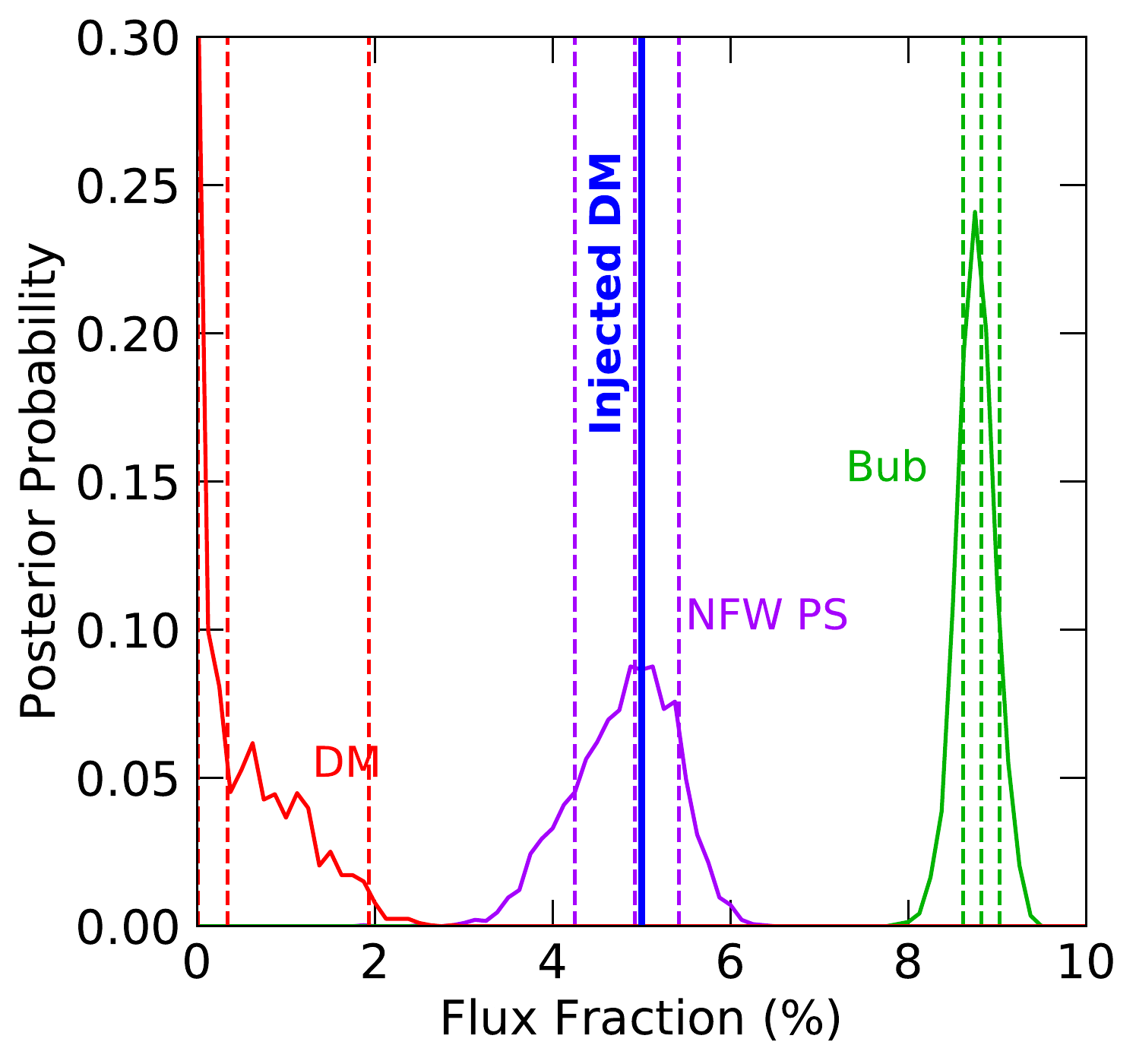}}\\
\subfigure{\includegraphics[width=0.85\columnwidth]{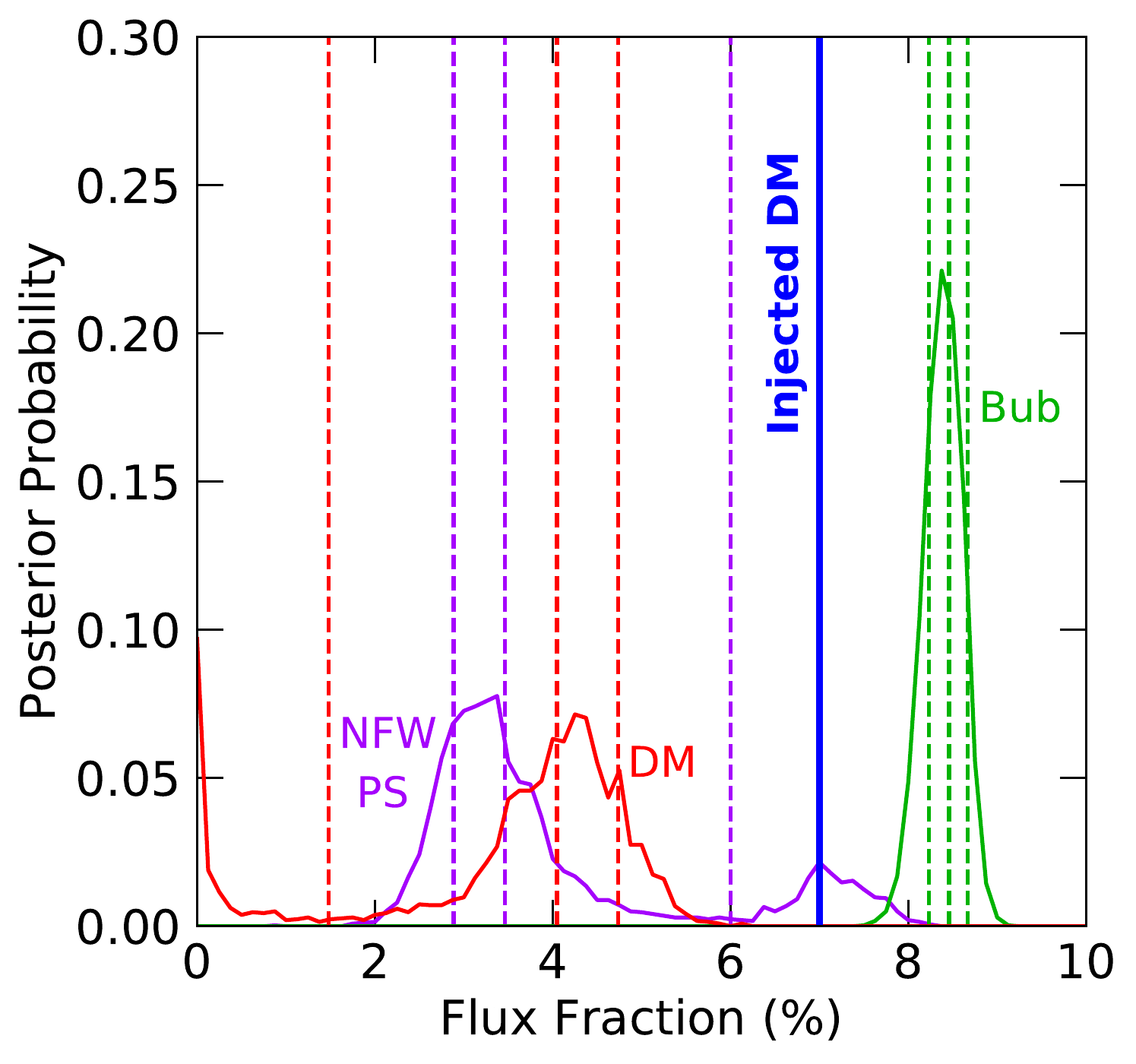}}
\hspace{1mm}
\subfigure{\includegraphics[width=0.85\columnwidth]{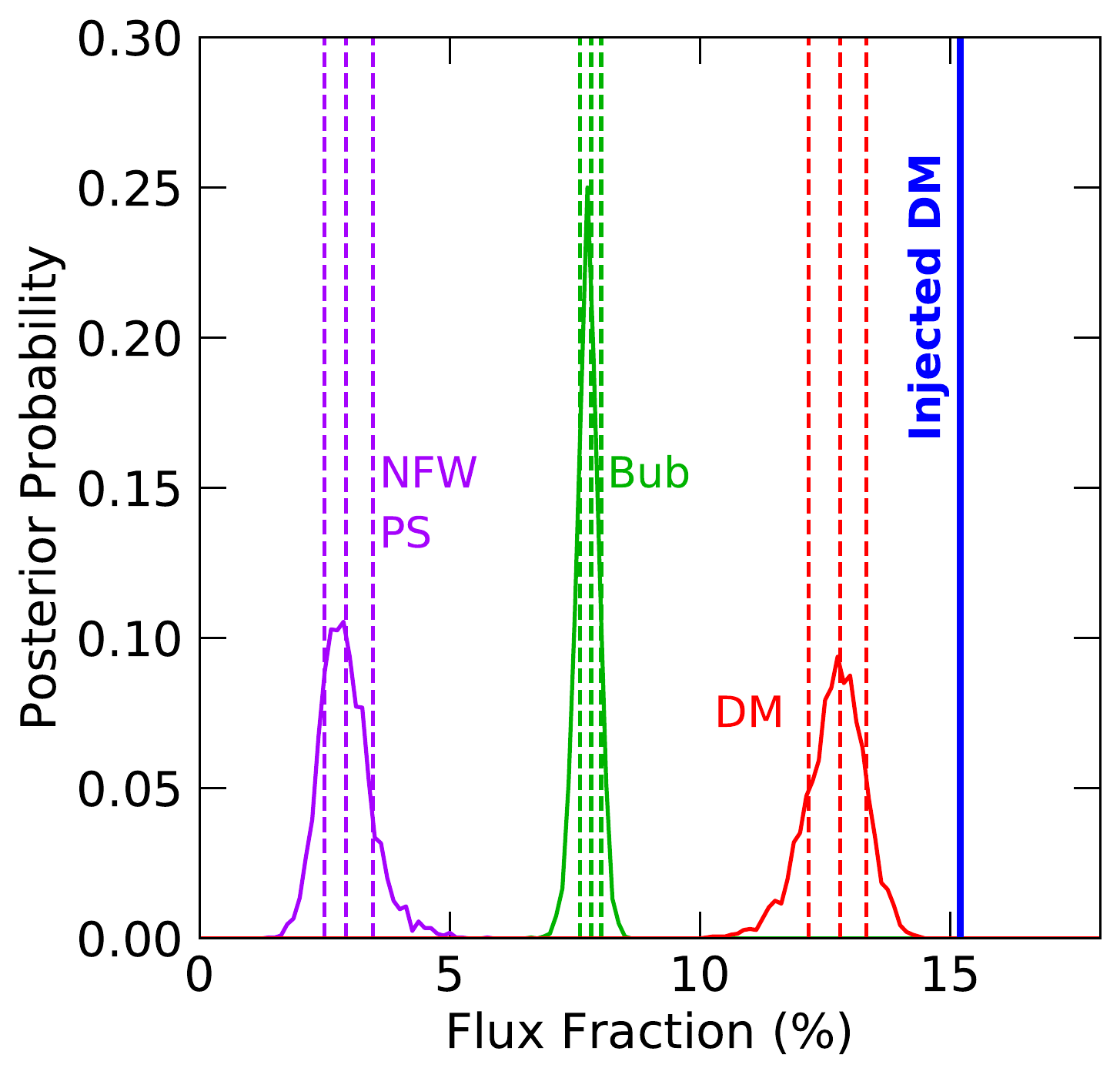}}
\caption{Evidence that DM can be misidentified when templates used in the analysis do not accurately describe the data. We simulate Bubbles PSs and Disk PSs, and smooth DM, Bubbles, Isotropic, and Galactic Diffuse templates. We show the flux posteriors for analyses using the same templates as the simulated data, except the fit includes NFW PS and not Bubbles PS (note the smooth isotropic and disk PS templates yield zero flux so are not shown on the plots for simplicity). Vertical dashed lines indicate posterior medians and 68\% containment bands. Different amounts of DM flux have been injected each time; the correct amount that should have been recovered is shown as the blue line labeled ``Injected DM''. \textbf{Top-Left:} $2\%$ DM injection. \textbf{Top-Right:} $5\%$ DM injection. \textbf{Bottom-Left:} $7\%$ DM injection. \textbf{Bottom-Right:} $15\%$ DM injection.}
\label{fig:GCEbias_sim}
\end{figure*}

\begin{table*}
\renewcommand{\arraystretch}{1.55}
\centering
\begin{tabular}{|P{2.2cm}|P{2cm}|P{5.2cm}|P{2.5cm}|P{1.2cm}|P{1.2cm}|P{1.2cm}|}
\hline
\multicolumn{7}{|c|}{\textsc{Simulated Data, 3FGL Masked}} \\
\hline
 \multirow{2}{*}{\textbf{Simulation}} &  \textbf{Simulated} &  \multirow{2}{*}{\textbf{Analysis Templates}} & \textbf{DM Flux}  &  \multicolumn{3}{c|}{\multirow{2}{*}{\textbf{Bayes Factor}} } \\
                                    &    \textbf{DM Flux}      &                                         & \textbf{(95\%)} &  \multicolumn{3}{c|}{} \\ \hline\hline

Bubbles PS  &               & Same as simulated                         & $[ \, 1.3, 2.5 \,]\  \%$                                     & $7\times10^{20}$  & \cellcolor{black!25}             & $1\times10^{26}$                          \\ \cline{3-4} \cline{6-7}
Disk PS     &  $2 \%$  & Same but Bubbles PS $\rightarrow$ NFW PS  & $ \textcolor{violet}{[ \, \textbf{0.0, 0.4} \,]\  \%} $                   &                                  & $2\times10^{5}$      & \cellcolor{black!25}  \\ \cline{3-5} \cline{7-7}
DM      &               & Same but no Bubbles PS                    & $[ \, 0.8, 2.2 \,]\  \%$                                     &  \cellcolor{black!25}            &                                  &                                       \\ 
\hline \hline
Bubbles PS &    & Same as simulated                         & $[ \, 4.6, 5.8 \,]\  \%$ & $9\times10^{16}$        & \cellcolor{black!25}             & $8\times10^{23}$            \\ \cline{3-4} \cline{6-7}
Disk PS   &   $5 \%$     & Same but Bubbles PS $\rightarrow$ NFW PS    & $\textcolor{violet}{[ \, \textbf{0.0, 1.8} \,]\  \%}$     &                                    & $9\times10^{6}$      & \cellcolor{black!25}  \\ \cline{3-5} \cline{7-7}
DM &    & Same but no Bubbles PS                        & $[ \, 4.3, 5.4 \,]\  \%$ &  \cellcolor{black!25}              &                                  &                       \\ \hline \hline

Bubbles PS  &    & Same as simulated                         & $[ \, 14.6, 15.8 \,]\  \%$ &    $5\times10^{16}$     & \cellcolor{black!25}             & $5\times10^{21}$            \\ \cline{3-4} \cline{6-7}
Disk PS    &  $15 \%$      & Same but Bubbles PS $\rightarrow$ NFW PS  & $\textcolor{violet}{[ \, \textbf{11.4, 13.7} \,]\  \%}$   &                                    & $1\times10^{5}$     & \cellcolor{black!25}  \\ \cline{3-5} \cline{7-7}
DM &  & Same but no Bubbles PS                        & $[ \, 14.3, 15.6\,]\  \%$ &  \cellcolor{black!25}              &                                  &                       \\  \hline 
\end{tabular}
\caption{Bayes factors and 95\% containment intervals on the DM flux posterior, for various analyses on simulated data. Note that in addition to the templates listed, in all cases we also simulate smooth Bubbles, Isotropic, and Galactic Diffuse templates. In the simulation, the DM flux is not correctly recovered in the case where the fit is performed with the standard pipeline (shown in purple). The data analysis that is favored has the relative Bayes factor appear next to the analysis row, where the two analyses that are being compared are not grayed out.}
\label{tab:bayes_sim}
\end{table*}

\noindent
\textbf{\textit{Proof-of-Principle Example of the Impact of Unmodeled Point Sources.}}
We first investigate the performance of the NPTF pipeline on simulated data, in the case where the data contain contributions that are \emph{not} well-modeled by the standard templates (in the supplementary material we explore the case where the mock data matches the templates). We will show that it is possible to induce significant biases to the reconstructed parameters, depending on the additional contributions. As an example, we consider unmodeled populations of PSs tracing the \emph{Fermi} Bubbles. Small gas clumps have recently been identified~\cite{2018ApJ...855...33D} in the Bubbles, and this could generate apparent gamma-ray PSs if the gas density in the clumps is sufficiently high. 

More generally, the Bubbles serve as an example of a component with a spatial morphology that is not degenerate with the disk or with isotropic PSs, and which could conceivably be absorbed by a GCE template in preference to the other PS templates. As we show in
separate work \cite{Leane:2019}, we do not detect PSs associated with this Bubbles template in real data. As such, we do
not expect that this example precisely describes the real gamma-ray sky; it is simply a proof-of-principle example
that (as we will see) the reconstructed DM-associated flux can be significantly biased if the templates used in the fit do not adequately describe the real data.

Studying the \textit{Fermi} data with the standard pipeline yields simulated data with essentially no DM contribution, consistent with the result of Ref.~\cite{Lee:2015fea}. In order to simulate a DM contribution of comparable size to the GCE, we also perform a fit with the NFW PS template removed from the standard pipeline, and take the posterior median of the DM template normalization in this case. This gives a DM flux of about 2 percent of the total photon flux in the masked ROI. To simulate a hypothetical source population tracing the \textit{Fermi} Bubbles, we use the same source count function parameters as simulated for the NFW PS population. 

We simulate PSs correlated with the Bubbles and Disk, as well as a DM component and diffuse Bubbles, Isotropic and Galactic Diffuse templates. We then apply the standard NPTF pipeline to this scenario (minus the omitted isotropic PS template); in particular, the fit includes a template for NFW PSs, but not Bubbles-correlated PSs. We also repeat the same fit with the analogue of the baseline NPTF pipeline for this case (again omitting the isotropic PS template). 

Table~\ref{tab:bayes_sim} summarizes the relative Bayes factors and flux containment intervals in our analyses. Comparing the Bayesian evidences for these two models yields the Bayes factor in favor of NFW PSs; we consistently find a preference in favor of the model containing NFW PSs, despite the fact that no such PSs are present in the simulation. 

Figure~\ref{fig:GCEbias_sim} shows the best-fit fluxes attributed to the various templates in the ``standard pipeline'' case, as a fraction of the total simulated photon flux in the ROI. These are the first of our main results: \textit{The fit misattributes the simulated DM signal to point sources.} The DM-sourced photon flux appears to be divided between NFW PSs and smooth emission in the Bubbles. Increasing the amount of injected DM (but keeping the other simulated components the same), the DM flux continues to be absorbed by the NFW PS template, until it reaches a threshold of absorption, where some DM is eventually identified.

We note that a bias-inducing population of unmodeled PSs could also be consistent with wavelet studies indicating an excess of small-scale power at the GCE, exceeding expectations from the Galactic diffuse emission; we leave a detailed study of consistency to future work. Non-NPTF studies~\cite{Bartels:2015aea} have argued that this small-scale power cannot be easily explained by the unresolved PSs associated with known populations, but have not been able to clearly associate it spatially with the GCE. Consequently, their results also do not place a firm upper limit on any DM contribution to the GCE.

\noindent
\textbf{\textit{Dark Matter Injection Test in Real Gamma-Ray Data.}}
Above we have identified a proof-of-principle example of how a DM signal can be misattributed to PS by the NPTF, when an unmodeled PS population is present. We now seek to test whether similar effects could be occurring in the real gamma-ray data. 

The complicating factor is that in real data, the distributions of any unmodeled PS populations are (by definition) unknown. One possible test is to add additional physically-motivated templates for PS populations, to see if the GCE is then attributed to a DM component, but we have not yet identified an additional template that has this effect (and as the number of templates tested proliferates, caution would be needed in statistical interpretation of the results). 
\begin{figure*}[t!]
\leavevmode
\centering
\subfigure{\includegraphics[width=0.85\columnwidth]{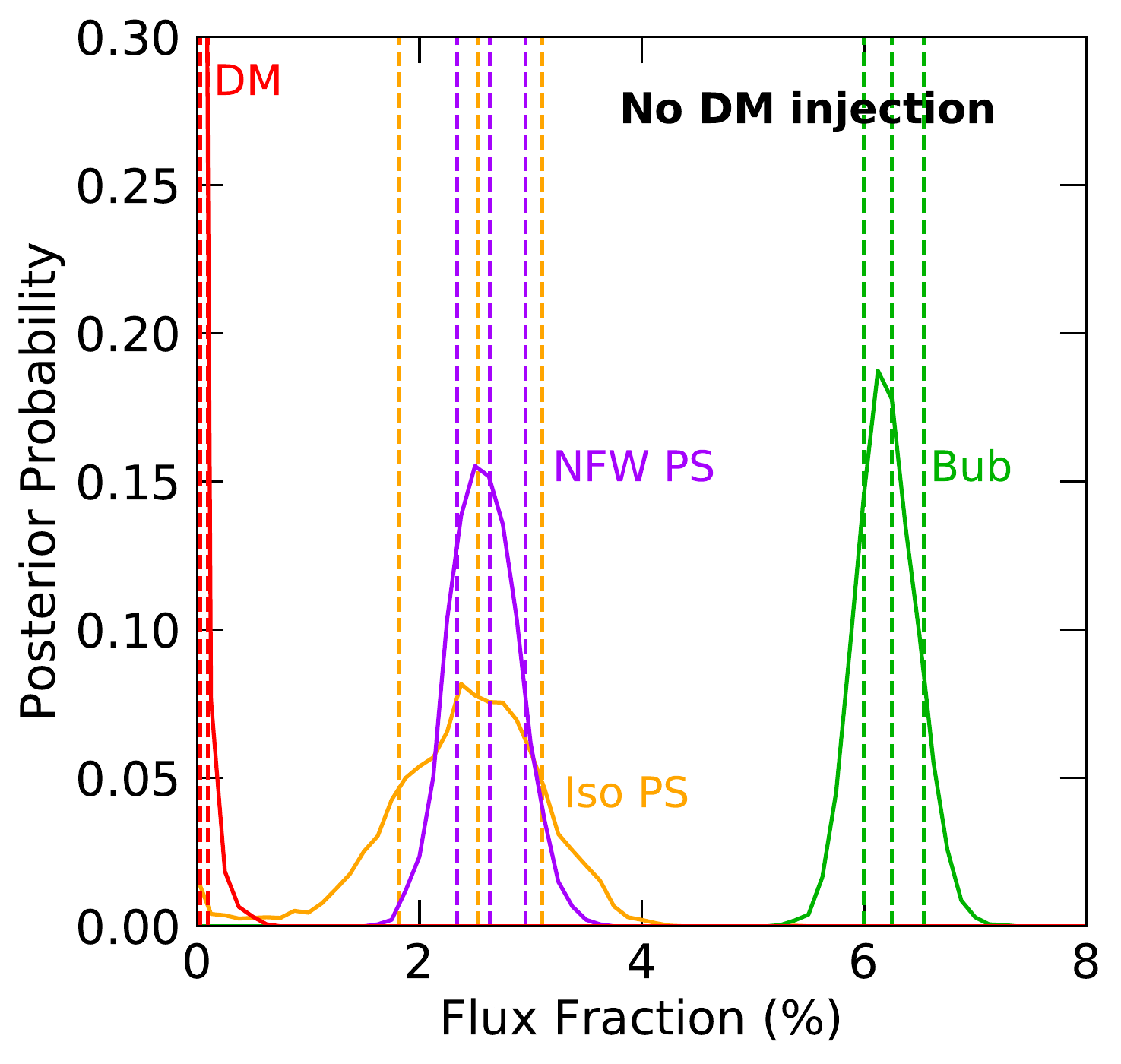}}
\hspace{1mm}
\subfigure{\includegraphics[width=0.85\columnwidth]{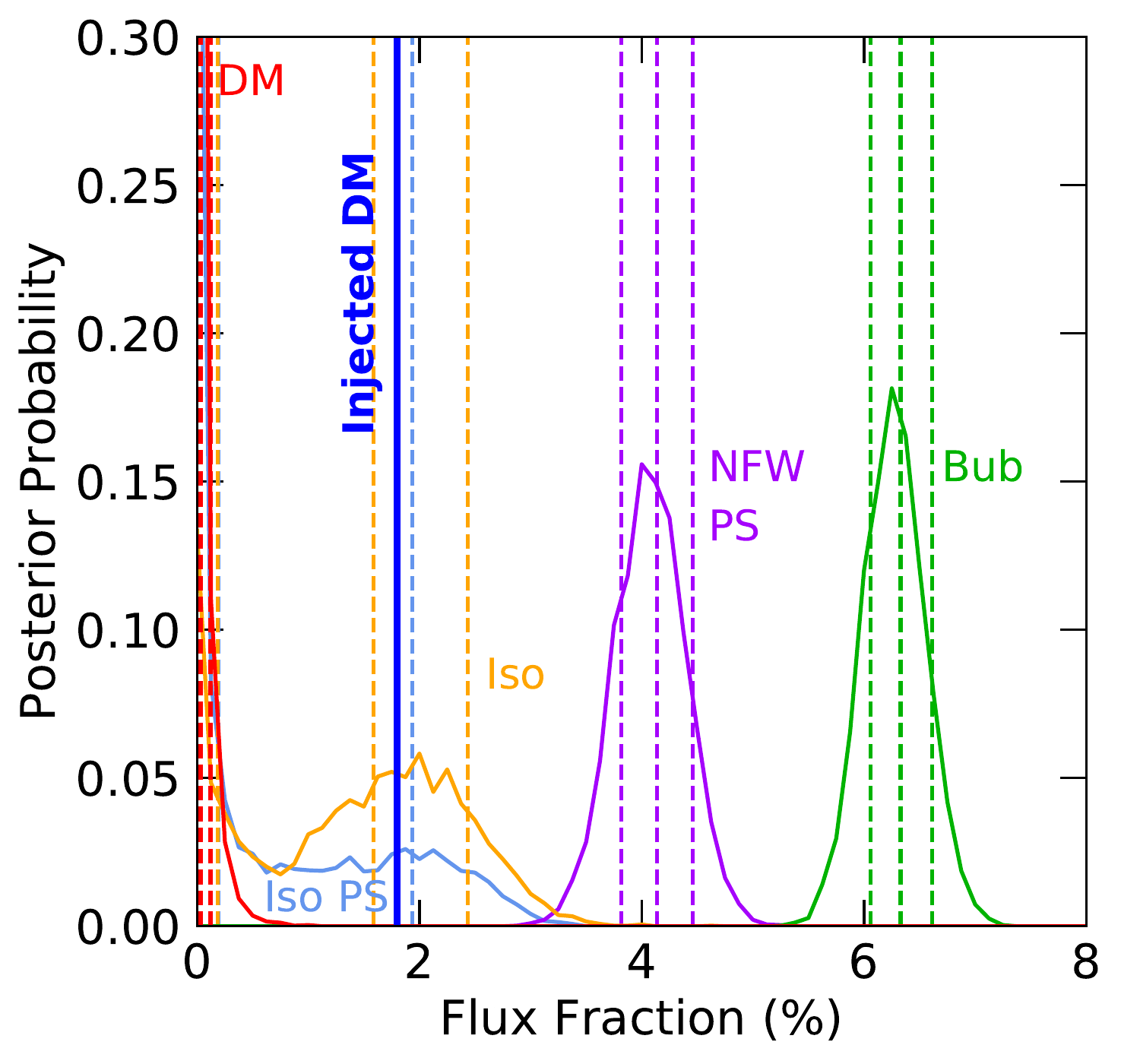}}
\\
\subfigure{\includegraphics[width=0.85\columnwidth]{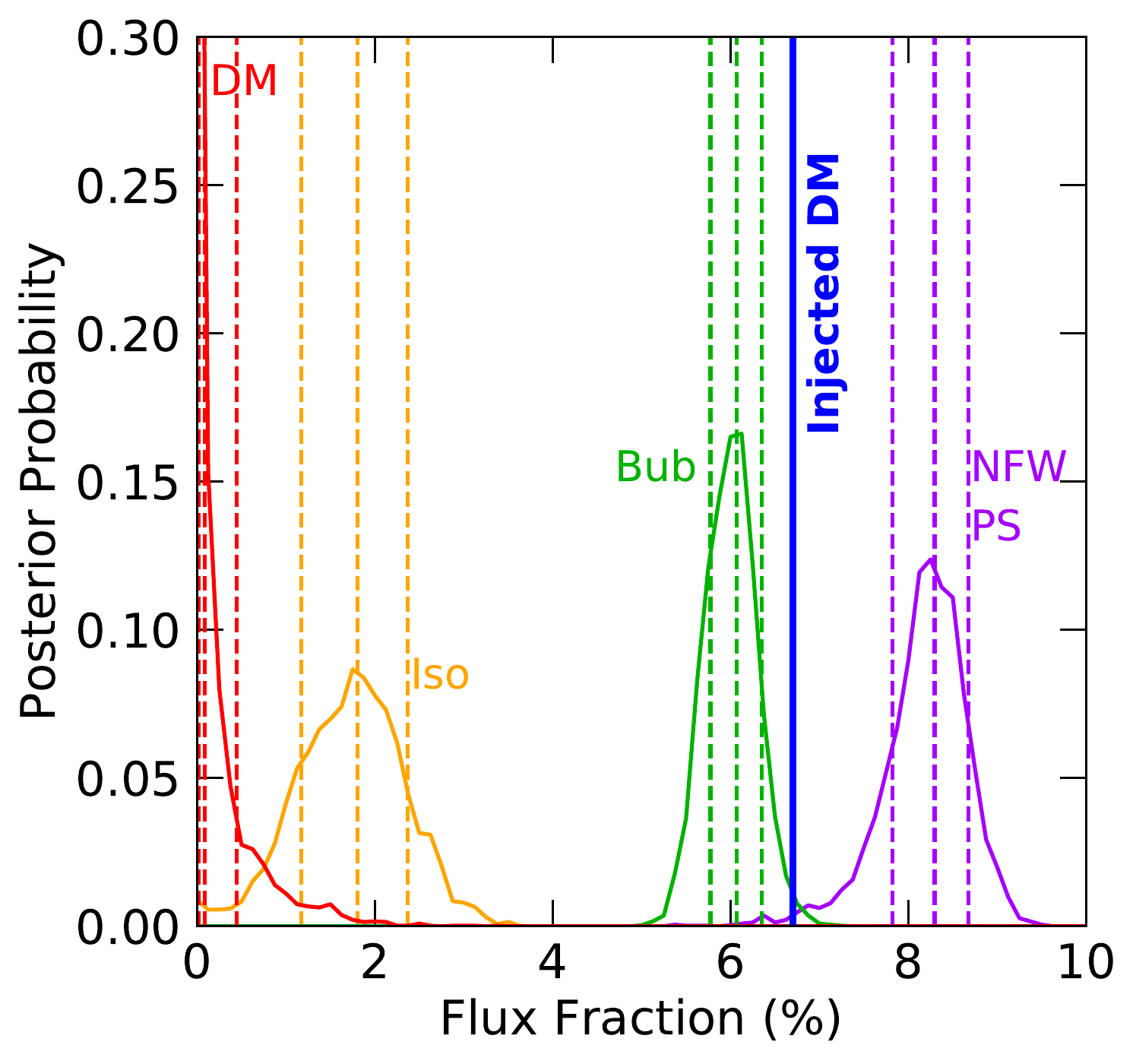}}
\hspace{1mm}
\subfigure{\includegraphics[width=0.85\columnwidth]{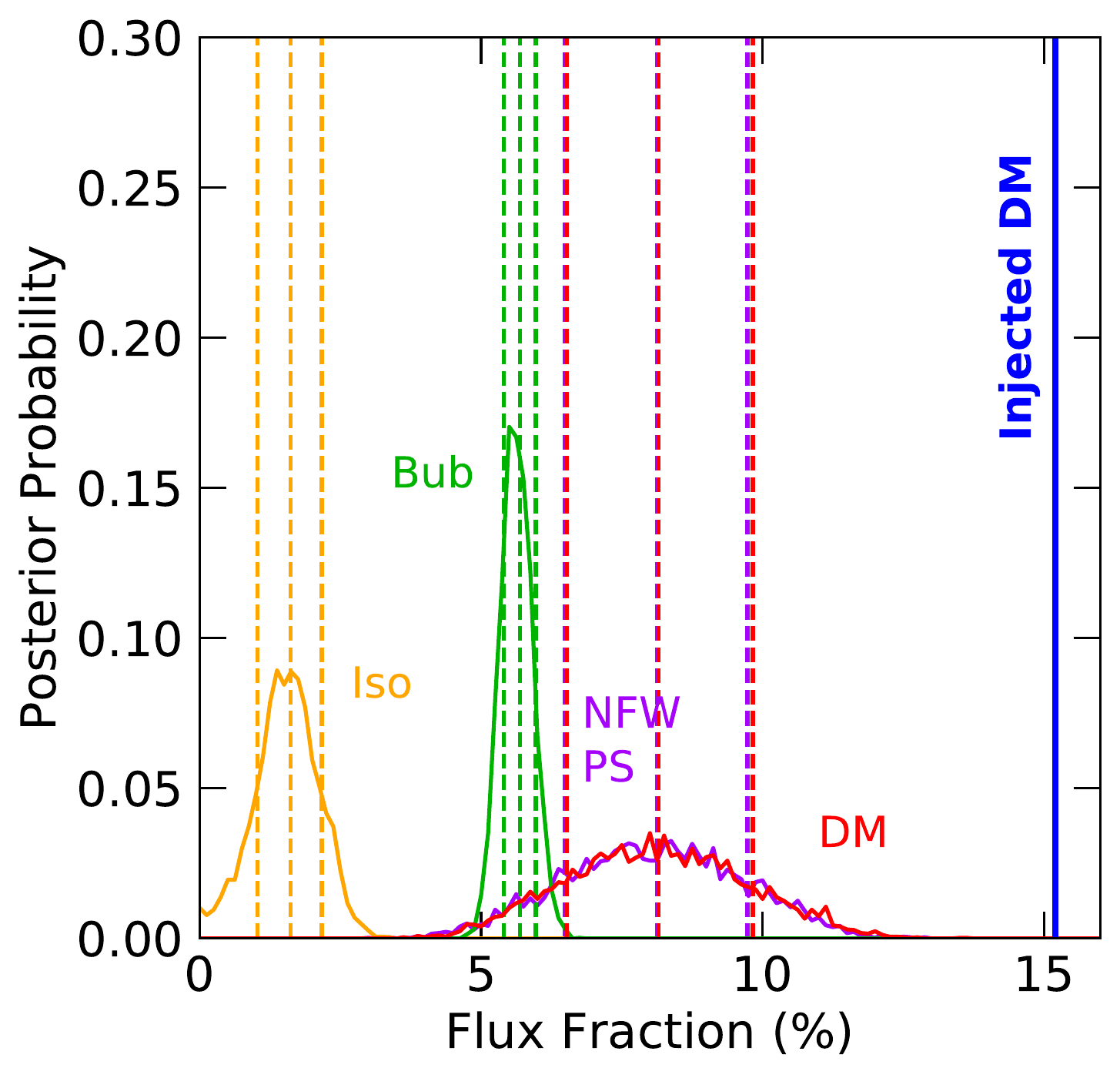}}
\caption{Flux posteriors when an artificial DM signal with increasing normalization is injected into the \textit{Fermi} data, and the data are analyzed with NFW PS, Disk PS, Isotropic PS, DM, Bubbles, Isotropic and Galactic Diffuse templates (note if any template has flux peaked below 0.1\% (other than DM), it is omitted from the plots for simplicity). Vertical dashed lines indicate posterior medians and 68\% containment bands. Different amounts of DM flux have been injected in each plot, the correct amount that should have been recovered is shown as the blue line labeled ``Injected DM''.  \textbf{Top-Left:} Zero DM injection. \textbf{Top-Right:} $1.8\%$ DM flux injection. No DM is recovered, and DM is instead attributed to NFW PS. \textbf{Bottom-Left:} $6.7\%$ DM flux injection. DM is still not recovered, and the NFW PS flux has been pushed up further. \textbf{Bottom-Right:} $15.2\%$ DM flux injection. Some DM flux is finally identified, albeit clearly not all of it.}
\label{fig:bias_data}
\end{figure*}

However, as a possible diagnostic that does not rely on specifying the biasing population, we can artificially inject a DM contribution into the data and see if it is correctly reconstructed. In our proof-of-principle example, increasing the DM contribution well above the GCE level still resulted in the attribution of the DM component to NFW PSs. It would be a coincidence if, in the presence of an unmodeled PS population (or another effect that biased the normalization of the DM component), the bias was sufficient to incorrectly attribute the observed GCE to PSs, but could not absorb any additional DM injections. In other words, if this effect is present, there is no reason to think it should be saturated.

As previously, we simulate this additional DM injection by taking a random Poisson draw in \texttt{Python} of the DM template multiplied by an appropriate normalization factor; the normalization is varied to test a range of injection levels.

Table~\ref{tab:bayes_data} presents our results for injected DM fluxes as a fraction of the total photon flux in the real + simulated data in the ROI (post-injection); the reconstructed fluxes associated with various templates, within the standard and baseline NPTF pipelines; and the Bayes factors between pairs of models. We also perform fits where the DM template is fixed to the injected value, to test the Bayes factor for this scenario compared to the cases where DM is allowed to float freely. 

In general, we find that when the injected DM signal is up to a factor of $\sim 5$ larger than the baseline GCE, the injected signal is completely attributed to the NFW PSs template, with the reconstructed bounds on the DM signal being consistent with zero flux and strongly inconsistent with the injected value. We find that the model where the NFW PS and DM templates are allowed to float, with the DM template being driven toward zero flux, is preferred over both the model where the DM contribution is fixed to its injected value, and (to a larger degree) the model without NFW PSs. This behavior -- where a model that is inconsistent with the injected signal is favored by the fit -- is qualitatively similar to the simulated proof-of-principle case, and {\it consistent with a mismodeling effect capable of driving any actual DM contribution to the GCE to zero flux}.

Once the injected signal is sufficiently large, the flux associated with the DM template becomes inconsistent with zero, albeit still inconsistent with the injected value; however, this requires a very large injection. For example, injecting a DM signal almost an order of magnitude brighter (15.2$\%$ post-injection) than the GCE itself results in around half the injection (only $\sim8\%$) being attributed to DM.

\begin{table*}
\renewcommand{\arraystretch}{1.3}
\centering
\begin{tabular}{|P{2.5cm}|P{6.2cm}|P{3cm}|P{1.2cm}|P{1.2cm}|P{1.2cm}|}
\hline
\multicolumn{6}{|c|}{\textsc{Real Data, 3FGL Masked}} \\
\hline
 \textbf{Injected} &  \multirow{2}{*}{\textbf{Analysis Templates}} &  \textbf{DM Flux}                          &  \multicolumn{3}{c|}{\multirow{2}{*}{\textbf{Bayes Factor}} } \\
\textbf{DM Flux}   &                                               &   \textbf{(95\%)}                     &     \multicolumn{3}{c|}{}                                 \\ \hline\hline

\multirow{4}{*}{None} &  Disk PS + Iso PS               &     \multirow{2}{*}{ $[ \, 1.1, 2.4 \,]\  \%$}      &   \multicolumn{3}{c|}{ }          \\ 
                    &  Diffuse + Iso + Bub + DM    &                                      &      \multicolumn{3}{c|}{}        \\ \cline{2-3}
                    &  Disk PS + Iso PS + NFW PS         &      \multirow{2}{*}{ $\textcolor{violet}{[ \, \textbf{0.0, 0.3} \,]\  \%}$}     &      \multicolumn{3}{c|}{\multirow{2}{*}{$1\times10^{9}$}}          \\ 
                    &  Diffuse + Iso + Bub + DM    &                                      &     \multicolumn{3}{c|}{}         \\ \hline\hline
 
\multirow{6}{*}{$1.8 \%$}    &  Disk PS + Iso PS                  &     \multirow{2}{*}{ $[ \, 2.8, 4.2 \,]\  \%$}    &      &  \cellcolor{black!25}         &          \\ 
                                &  Diffuse + Iso + Bub + DM   &                                    &                                &   \cellcolor{black!25}        &         \\ \cline{2-3} \cline{5-6} 
                    
                                &  Disk PS + Iso PS + NFW PS         &      \multirow{2}{*}{ $\textcolor{violet}{[ \, \textbf{0.0, 0.3} \,]\  \%}$}   &        \multirow{2}{*}{$2\times10^{12}$ }                          &     \multirow{2}{*}{$4\times10^{2}$}     &  \cellcolor{black!25}       \\ 
                                &  Diffuse + Iso + Bub + DM    &                                   &                                  &           &   \cellcolor{black!25}    \\ \cline{2-4} \cline{6-6} 
                                
                                &  Disk PS + Iso PS + NFW PS         &      Fixed at injection         &        \cellcolor{black!25}      &             &    \multirow{2}{*}{$7\times10^{9}$}  \\ 
                                &  Diffuse + Iso + Bub + Fixed DM    &       value ($1.8 \%$)                            &          \cellcolor{black!25}    &           &   \\ \hline\hline

\multirow{4}{*}{$6.7 \%$}  &  Disk PS + Iso PS   &     \multirow{2}{*}{ $[ \, 7.7, 9.0 \,]\  \%$}                   &   \multicolumn{3}{c|}{ }           \\ 
                    &  Diffuse + Iso + Bub + DM   &                         &      \multicolumn{3}{c|}{}        \\ \cline{2-3}
                    &  Disk PS + Iso PS + NFW PS   &      \multirow{2}{*}{ $\textcolor{violet}{[ \, \textbf{0.0, 1.3} \,]\  \%}$}                   &     \multicolumn{3}{c|}{\multirow{2}{*}{$7\times10^{11}$}}         \\ 
                    &  Diffuse + Iso + Bub + DM   &                        &      \multicolumn{3}{c|}{}       \\ \hline\hline

\multirow{4}{*}{$15.2 \%$}  &  Disk PS + Iso PS   &     \multirow{2}{*}{$[ \, 15.7, 16.9 \,]\  \%$}                   &  \multicolumn{3}{c|}{ }         \\ 
                    &  Diffuse + Iso + Bub + DM   &                         &       \multicolumn{3}{c|}{}       \\ \cline{2-3}
                    &  Disk PS + Iso PS + NFW PS   &      \multirow{2}{*}{$\textcolor{violet}{[ \, \textbf{5.0, 11.5} \,]\  \%}$}                   &   \multicolumn{3}{c|}{\multirow{2}{*}{$8\times10^{9}$}}           \\ 
                    &  Diffuse + Iso + Bub + DM   &                      &       \multicolumn{3}{c|}{}      \\ \hline
\end{tabular}
\caption{Bayes factors for analyses of the real \textit{Fermi} data injected with a DM signal. The DM flux is not correctly recovered when the standard PS templates are considered (shown as purple), suggesting a large downward bias to the DM flux that could hide a real signal, similar to the proof-of-principle case detailed in Tab.~\ref{tab:bayes_sim}. The data analysis that is favored has the relative Bayes factor appear next to the analysis row, where the two analyses that are being compared are not grayed out.}
\label{tab:bayes_data}
\end{table*}

Figure~\ref{fig:bias_data} shows results for recovered flux fractions in the ROI for varying levels of DM injection: 0, 1.8, 6.7, and 15.2$\%$ (post-injection). The top-left plot shows results of a fit on the real data, without any injected DM signal, to serve as a baseline for comparison. By comparison with the no-injection case, we see two important effects: firstly that (as noted above) the flux attributed to the DM template is consistent with zero and inconsistent with the injected value, and secondly that the NFW PS flux fraction increases, approximately absorbing the injected DM signal. As the DM injection amount increases, we see that the NFW PS flux fraction continues to increase, until it reaches a saturation point and the DM template begins to absorb some of the flux. In order for the DM to be detected with non-zero flux, the injected DM signal appears to require a total flux a factor $\gtrsim 5$ larger than the GCE itself.

\noindent
\textbf{\textit{Conclusions and Outlook.}}
We have studied examples of how NPTF methods can be biased in both real and simulated gamma-ray data, and how this could impact explanations of the GCE. We have showed a proof-of-principle example in simulated data where a DM signal can incorrectly be attributed to PSs by the NPTF, as a result of PSs with a spatial distribution that is not described by the standard templates.

We have searched for behavior consistent with this effect in the real \textit{Fermi} gamma-ray data. If an unmodeled source population is biasing results of NPTF studies, causing the NFW PS template to absorb the DM flux, it is likely not saturated in its ability to do so. We injected an artificial DM signal into the real data, with various normalizations, to test if similar behavior is observed to the simulated-data proof-of-principle example. Strikingly, the injected DM signal was misattributed to PS, with the recovered NFW PS flux fraction increasing proportionally to the size of the DM signal injected. We increased the normalization of the DM signal, and found this effect persists until the amount of DM injected exceeds the GCE itself by almost an order of magnitude, at which point the fit recovers a DM component, but not with the correct injected normalization. This behavior is qualitatively similar to the simulated proof-of-principle example; in both cases there are strong statistical preferences for models that reconstruct too little dark matter, compared to the injected values. This suggests that the data are mis-modeled in some way, and that DM may be the dominant contribution to the GCE after all - although this does not constitute positive evidence for DM, and the GCE could still be composed mainly of PSs. In the supplementary materials, we briefly explore the degree to which this evidence of mis-modeling persists across different diffuse models and ROIs.

An improved future version of the NPTF pipeline, with greater flexibility to accommodate a diverse range of source populations or a different range of ROIs, could potentially characterize and eliminate biases similar to those demonstrated in our examples. Upcoming observations by radio telescopes such as \textsc{MeerKat} and the \textsc{Square Kilometer Array (SKA}) should allow some currently unresolved source populations to be identified~\cite{Calore:2015bsx}. Complementary studies such as those using wavelet techniques~\cite{Bartels:2015aea,Balaji:2018rwz}, or focusing on improved diffuse emission modeling with new frameworks such as \textsc{SkyFact}~\cite{Storm:2017arh}, could also help disentangle sources of emission from the GC.

\noindent
\textbf{\textit{Acknowledgments.}}
We thank Tim Linden, Mariangela Lisanti, Nick Rodd, and Ben Safdi for helpful discussions and comments. We thank Tim Linden for suggesting the injection test on real data, the authors of Ref.~\cite{2018NatAs...2..387M} for providing their Boxy Bulge template, and the \textit{Fermi} Collaboration for the use of \textit{Fermi} public data. RKL thanks the DESY Theory Group and the Princeton Institute for Advanced Study for their hospitality during this work. RKL and TRS are supported by the Office of High Energy Physics of the U.S. Department of Energy under Grant No. DE-SC00012567 and DE-SC0013999; TRS is partially supported by a John N. Bahcall Fellowship.
\bibliography{DM_strikes_back}

\clearpage
\newpage
\maketitle
\onecolumngrid
\begin{center}
\textbf{\large Dark Matter Strikes Back at the Galactic Center} \\ 
\vspace{0.05in}
{ \it \large Supplementary Material}\\ 
\vspace{0.05in}
{Rebecca K. Leane and Tracy R. Slatyer}
\end{center}
\onecolumngrid
\setcounter{equation}{0}
\setcounter{figure}{0}
\setcounter{section}{0}
\setcounter{table}{0}
\setcounter{page}{1}
\makeatletter
\renewcommand{\theequation}{S\arabic{equation}}
\renewcommand{\thefigure}{S\arabic{figure}}
\renewcommand{\thetable}{S\arabic{table}}

The supplementary material is organized as follows. We first describe our methodology, which includes an overview of the NPTF method, our template choices and data selection. We then present additional analyses on simulated data, which includes studies where the ROI has the 3FGL PSs unmasked, evidence that parameters are recovered (approximately) correctly when analysis templates match those of the simulated data, and additional plots for simulated PS in the Bubbles (including the relevant SCFs).  We then provide supplemental figures (including the relevant SCFs) and supplemental analyses for our DM injection tests in real data. We show results for varying Galactic diffuse background models, and describe an alternative way to see the issue driving the results of the injection test, by allowing the DM template to have a negative coefficient. We perform a survival function test on the data, quantifying the number of pixels at a given level of unlikeliness; this distribution can be compared with predictions from simulated data. Finally, we show the two dimensional posterior values of floated parameters in our analysis.

\section{Template Fitting Methodology}
\label{sec:method}

In this section we review the template fitting methods employed in this work, describe the templates we use in our analysis, and detail the regions of interest and dataset studied.

\subsection{Poissonian Template Fitting}

Template fitting is a widely used approach that fits a linear combination of templates for physical sources of photon emission (in this case, gamma rays) to observational data. In general, templates may vary with both position and energy, but for our purposes we will work within a single broad energy bin and characterize templates purely by their spatial distributions (and photon statistics). 

In the simplest scenario, such spatial templates describe the expected mean photon counts per pixel, photons are assumed to be independent events, and thus the probability of observing a certain number of photons sourced by a given template in a specific pixel is given by the Poisson distribution. This assumption of independence is valid for diffuse emission and for PS populations where the locations of the PSs are known.

Contributions from different templates can be trivially combined in this case. Let us denote the pixel-dependent expected photon counts from the individual spatial templates by $\mu_{p,\ell}({\bf \theta})$, where $p$ is a pixel index, $\ell$ indexes the choice of template, and ${\bf \theta}$ is a set of parameters controlling the properties of the templates. In the Poissonian case, we will generally choose the parameters ${\bf \theta}$ to control the coefficients of the various templates, so that:
\begin{equation}
 \mu_{p,\ell}( {\bm \theta}) = A_\ell ({\bm \theta}) \ T_{p,\ell} \,,
\label{eq:poissnorm}
\end{equation}
where $T_{p,\ell}$ defines a spatial template with some fixed normalization. The mean number of total counts within a specified pixel, $\mu_p( {\bm \theta})$, is then given by:
\begin{equation}
 \mu_p^{n_p}( {\bm \theta}) = \sum_\ell \mu_{p, \ell} ({\bm \theta}) \,,
\label{eq:meancounts}
\end{equation}
 and the probability of drawing $n_p$ counts in a given pixel $p$, as a function of the parameters $\bm{\theta}$, is:
\begin{equation}
 P_{n_p}^{(p)}({\bm \theta}) = {\mu_p^{n_p}( {\bm \theta}) \over n_p !} e^{- \mu_p( {\bm \theta}) } \,.
\label{eq:poisson}
\end{equation}

In this case, for given model parameters $\bm{\theta}$, a model  $\mathcal{M}$ can be fit to a dataset $d$ with $n_p$ counts in each pixel $p$.  The likelihood function is then just the product of Poisson likelihoods,
\begin{equation}
P(d |{\bm \theta}, \mathcal{M}) = \prod_p P_{n_p}^{(p)}({\bm \theta}) \,.
\label{eq:likelihood}
\end{equation}

\subsection{Non-Poissonian Template Fitting}

In the situation where the gamma-ray emission contains spatial correlations between photons that are not captured by the spatial template itself, the photons from a given pixel can no longer be treated as independent events, and Poissonian statistics are not appropriate. This is the case, for example, where there are PSs producing two or more photons in the data, but the positions of these sources are not included in the spatial template; they are unresolved. The presence of the sources manifests itself in an increased probability of seeing multiple photons from the same location, compared to Poissonian expectations; we can search for this effect while remaining agnostic as to the exact positions of the sources.

The detailed statistical treatment of this situation was first discussed in the context of gamma-ray data in Ref.~\cite{Malyshev:2011zi}, and expanded in Ref.~\cite{Lee:2014mza} (other techniques for estimating contributions from unresolved PS populations are discussed in e.g. \cite{Zechlin:2015wdz,Zechlin:2016pme}). Ref.~\cite{Lee:2015fea} realized that these techniques could be combined with template-fitting methods to include information on the spatial distributions of the backgrounds and potential point-source populations, and tested the resulting NPTF pipeline on \emph{Fermi} data; follow-up studies were conducted in Refs.~\cite{Linden:2016rcf,Lisanti:2016jub}, and a public code package (which we employ) was presented in Ref.~\cite{Mishra-Sharma:2016gis}. We summarize the key features of the mathematical framework below.

In the NPTF, the photon count probability distribution $P_{n_p}^{(p)}$ discussed above has an additional dependence on a pixel-dependent PS source-count distribution $dN_p/dF$, where $F$ is the source flux. We model the source count function as a multiply broken power law, and assume it is the same in all pixels up to a pixel-dependent normalization factor:
\begin{equation}
 \frac{dN_p}{dF} (F; {\bm \theta}) = A ( {\bm \theta}) T^{({\rm PS})}_p \left\{ \begin{array}{lc} \left( \frac{F}{F_{b,1}} \right)^{-n_1}, & F \geq F_{b,1} \\ \left(\frac{F}{F_{b,1}}\right)^{-n_2}, & F_{b,1} > F \geq F_{b,2} \\ \left( \frac{F_{b,2}}{F_{b,1}} \right)^{-n_2} \left(\frac{F}{F_{b,2}}\right)^{-n_3}, & F_{b,2} > F \geq F_{b,3} \\ \left( \frac{F_{b,2}}{F_{b,1}} \right)^{-n_2} \left( \frac{F_{b,3}}{F_{b,2}} \right)^{-n_3} \left(\frac{F}{F_{b,3}}\right)^{-n_4}, & F_{b,3} > F \geq F_{b,4} \\ \\
\ldots & \ldots \\ \\
\left[ \prod_{i=1}^{k-1} \left( \frac{F_{b,i+1}}{F_{b,i}} \right)^{-n_{i+1}} \right] \left( \frac{F}{F_{b,k}} \right)^{-n_{k+1}}, & F_{b,k} > F \end{array} \right. .
\label{eq:brokenpower}
\end{equation}
Here, the source-count distribution is parameterized with an arbitrary number of breaks $k$, denoted by $F_{b,i}$ with \mbox{$i \in [1,2, \ldots, k]$}, and $k+1$ indices $n_i$ with \mbox{$i \in [1,2, \ldots , k+1]$}. The spatial distribution of the PSs is described by the template $T_p^{({\rm PS})}$, with a pixel-independent normalization $A ( {\bm \theta})$ which is a function of the model parameters $\bm \theta$. Note that while the number of sources varies between pixels, in our work the locations of the flux breaks and the indices are fixed across pixels.

In practice we need the source count function as a function of photon counts $S$, rather than photon flux $F$; the two are related by the pixel-dependent \textit{Fermi} exposure map $E_p$. The conversion formula is:
\begin{equation}
 {dN_p \over dS} (S; {\bm \theta}) = \frac{1}{E_p} {dN_p \over dF} (F = S / E_p; {\bm \theta}) \, .
 \label{eq:fluxtocounts}
\end{equation}

To calculate the pixel likelihoods for models containing templates with non-Poissonian templates, the method of generating functions is employed; the generating function $\mathcal{P}^{(p)}(t)$ for the complete model (consisting of a sum over templates) is the product of the individual template generating functions, and the probability $P_k^{(p)}$ to observe $k$ photons in pixel $p$ can be derived from this generating function as:
\begin{equation}
 P_k^{(p)} = \frac{1}{k!} \left. \frac{d^k \mathcal{P}^{(p})(t)}{dt^k} \right|_{t=0} \,.
\end{equation}
Here $t$ is an auxiliary variable. The generating function for a Poissonian template is given by $\mathcal{P}_\ell^{(p)}(t) = e^{\mu_{p,\ell}(t-1)}$, while for the non-Poissonian PS templates we have instead:
\begin{equation}
\mathcal{P}_{\rm NP}(t; {\bm \theta}) = \prod_p \exp \left[ \sum_{m=1}^{\infty} x_{p,m}( {\bm \theta}) ( t^m - 1) \right] \,,
\end{equation}
where
\begin{equation}
 x_{p,m}( {\bm \theta}) =\int_0^{\infty} dS \frac{dN_p}{dS}(S;{\bm \theta}) \int_0^1 df \rho(f) \frac{(fS)^m}{m!} e^{-fS} \,.
\end{equation}
The point spread function (PSF) of the {\it Fermi} instrument is taken into account through $\rho(f)$, which describes the dilution of the flux from a PS into neighboring pixels due to smearing by the PSF. For a distribution $ dN_p(S;{\bm \theta})/dS $, $x_{p,m}$ describes the expected number of $m$-count PSs in the pixel $p$. For more details, see Refs.~\cite{Malyshev:2011zi, Lee:2014mza}.

We generate results using the public  \texttt{Python} and \texttt{Cython} NPTF package \texttt{NPTFit} \cite{Mishra-Sharma:2016gis}, interfaced with the Bayesian interference tool \texttt{MultiNest}~\cite{Feroz:2008xx}, which implements the likelihood framework described in this section. The total number of live points for all \texttt{MultiNest} runs is \mbox{\texttt{nlive = 500}}. Typically about $10^6$ posterior samples are generated for each run.

\subsection{Data Selection}

We use the \texttt{Pass 8} \textit{Fermi} data, which were collected in the energy range $2-20$ GeV over 413 weeks, from August 4th 2008 to July 7th 2016. The event class {\sc UltraCleanVeto} (1024) was applied, which has the highest cosmic ray rejection. This is further restricted to the top quartile of events graded by angular reconstruction (PSF3 (32)), with quality cuts \texttt{DATA\_QUAL==1 \&\& LAT\_CONFIG==1}. The maximum zenith angle is $90^\circ$. 

\subsection{Modeling and Masking of Resolved Sources}

The gamma-ray sky contains a number of high-significance PSs. To mask and/or model these sources, we use as a reference the \textit{Fermi}-LAT 3FGL catalog~\cite{Acero:2015hja}. While there are more up-to-date catalogs, in particular the 1FIG catalog \cite{TheFermi-LAT:2015kwa}, the main goal of masking PSs is to eliminate very bright pixels where computing the relevant photon count probabilities is time-consuming, and the 3FGL catalog has proved adequate for this purpose (it has also been demonstrated that masking the 1FIG PSs does not affect the GCE \cite{Bartels:2017xba}).

\subsection{Regions of Interest}

We study two different regions of interest (ROI) in the inner Galaxy, one with known PSs masked, and one with no masking of PSs.

In detail, the regions are:
\begin{itemize}
 \item \textbf{Inner Galaxy (masked):} within $30^\circ$ of the GC, and $|b|>2^\circ$, with all 3FGL PSs masked at a $99\%$ containment at an energy of 2 GeV (corresponding to a radius of 0.778 degrees), and
 \item \textbf{Inner Galaxy (unmasked):} within $30^\circ$ of the GC, and $|b|>2^\circ$,
\end{itemize}
where $\ell$ is longitude, and $b$ is latitude, in the galactic coordinate system. The unmasked case provides more ROI close to the GC, as the 3FGL PSs are clustered around the GC, and masking the 3FGL sources removes a substantial fraction of the near-GC solid angle. We perform all tests in the $30^\circ$-radius ROI with the 3FGL sources masked, and then check the key results with the unmasked ROI.
 
In principle, \textit{Fermi}'s exposure should be accounted for individually in each pixel, when converting between the flux of a source and its expected counts as in Eq.~\ref{eq:fluxtocounts}.  However, in practice this is not necessary, and we instead subdivide the ROI into a number of ``exposure regions''. Within each such subregion, the exposure is approximated as its average value over the exposure region, for the purpose of converting between flux and counts. For the Poissonian templates, where the exposure simply modifies the expected number of counts in each pixel, we do not perform this approximation and instead implement the full exposure map. In our Inner Galaxy analyses, we break the sky into 5 exposure regions. To ensure this is an accurate description of the data, we have checked stability of our main results (discussed in the main text) up to at least 30 exposure regions.

\subsection{Spatial Templates}

By default we use the templates provided publicly with the  \texttt{NPTFit}  package. We review the properties of these templates below, and note when we differ from the \texttt{NPTFit}  baseline.

All templates are implemented in a \texttt{HEALPix}~\cite{Gorski:2004by} pixelization of the data with \texttt{nside = 128}, such that the photon map contains 196,608 equal area pixels over the full sky. All Poissonian templates are exposure-corrected; they are thus maps of expected photon counts/pixel (in contrast to non-Poissonian templates which are maps of sources/pixel). The Galactic diffuse emission model, which accounts for the bulk of the gamma-ray emission, is smoothed by the correct \textit{Fermi}-LAT PSF using the Fermi Science Tools. The DM and \textit{Fermi} Bubbles diffuse templates are smoothed with a Gaussian with $\sigma=0.1812$ degrees. All templates are then rescaled to have a mean of one in the region with $|b|>2^\circ$ within $30^\circ$ of the GC.

Table~\ref{tab:priors} details the priors used for each template parameter. When floating any template, we scan with a log-flat prior distribution for the template normalization. The remaining parameters have linear-flat prior distributions. We describe each spatial template below.

\begin{table*}
\renewcommand{\arraystretch}{1.5}
\setlength{\tabcolsep}{5.2pt}
\begin{center}
\begin{tabular}{ c  P{3.2cm}  P{3.2cm}  }
\hline
\multicolumn{3}{c}{\textsc{Prior Ranges}}\Tstrut\Bstrut		\\  
\hline Parameter 	& Inner Galaxy (masked) & Inner Galaxy (unmasked) 	\Tstrut\Bstrut \\
\hline 
\hline
$\log_{10}A_\text{iso}$  & $[-3,1]$ & $[-3,1]$ \Tstrut\Bstrut \\ 
$\log_{10}A_\text{dif}$  & $[0,2]$ & $[0,2]$   \Tstrut\Bstrut \\
$\log_{10}A_\text{bub}$  & $[-3, 1]$ & $[-3, 1]$  \Tstrut\Bstrut \\ 
$\log_{10} A_\text{NFW}$ & $[-3, 1]$ & $[-3, 1]$   \Tstrut\Bstrut	\\
$\log_{10}A_\text{PS}^\text{NFW}$ & $[-6, 1]$ & $[-6, 1]$  \Tstrut\\
$S_b^\text{NFW}$  & $[0.05 ,60]$      & $[0.05 ,60]$     \\
$n_1^\text{NFW}$   & $[2.05, 60]$     & $[5, 45]$    \\
$n_2^\text{NFW}$   & $[ -3 ,1.95]$    & $[ -3 ,1.95]$ \\
$\log_{10}A_\text{PS}^\text{disk}$ & $[-6, 1]$ & $[-6, 1]$  \Tstrut\\
$S_b^\text{disk}$ & $[0.05 ,60]$      & $[0.05 ,60]$     \\
$n_1^\text{disk}$  & $[2.05, 60]$     & $[2.05, 5]$    \\
$n_2^\text{disk}$   & $[ -3 ,1.95]$    & $[ -3 ,1.95]$ \\
$\log_{10}A_\text{PS}^\text{iso}$  & $[-6, 1]$ & $[-6, 1]$ \Tstrut\\
$S_{b_1}^\text{iso}$  & $[5,40]$ & $[1,40]$    \\
$S_{b_2}^\text{iso}$  & $[0.05,30]$ & $[0.05,30]$    \\
$n_1^\text{iso}$ & $[2.05, 5]$& $[2.05, 5]$  \\
$n_2^\text{iso}$ & $[1.5, 4.5]$  & $[0.5, 4.5]$  \\
$n_3^\text{iso}$  & $[-1.95, 1.95]$    & $[-1.95, 1.95]$\Bstrut\\
\hline
\hline
\end{tabular}
\end{center}
\caption{Parameters and associated prior ranges for the Inner Galaxy analyses, when using the \textit{Fermi} \texttt{p6v11} diffuse model, as in the main text.}
\label{tab:priors}
\end{table*}

\begin{itemize}
\item {\bf Galactic Diffuse Emission}

Diffuse gamma-ray emission from the Milky Way dominates the gamma-ray sky. Such emission originates from three main sources: (1) cosmic-ray (CR) protons colliding with the gas and producing photons via pion production, \mbox{$pp\rightarrow X+\pi^0\rightarrow X+ \gamma\gamma$}, (2) CR electrons upscattering photons of the interstellar radiation field to gamma-ray energies via inverse Compton scattering, and (3) CR electrons scattering on the gas to produce photons via bremsstrahlung (this third process is generally subdominant to the other two).

To model this diffuse background emission, we use the \mbox{\texttt{Pass 6}} \textit{Fermi} diffuse model \texttt{p6v11}; this is the most recent diffuse model released by the \textit{Fermi} Collaboration that does not incorporate a built-in template for the \textit{Fermi} Bubbles or add a component for otherwise-unmodeled diffuse emission, allowing us to study these contributions separately. This was also the principal diffuse background model used in earlier NPTF studies of the GCE, facilitating comparisons \cite{Lee:2015fea}.

However, this diffuse model was fitted to much earlier data from \textit{Fermi}, with different instrument response functions to the current dataset; to translate the flux model into counts we use the appropriate instrument response functions for our current dataset, but errors in the instrument response modeling (either now or in the construction of the \texttt{p6v11} model)  could in principle introduce difficult-to-model systematic uncertainties. 

As a cross-check of our results, we also employ two models for the Galactic diffuse emission generated using \texttt{GALPROP}~\cite{Strong:1998pw}, labeled models A and F in Ref.~\cite{Calore:2014xka}. These models were found to provide a better description of the data than the \texttt{p6v11} model at GeV+ energies. They contain two component templates which are floated independently, corresponding on one hand to gas-correlated emission from pion production and bremsstrahlung, and on the other to emission from inverse Compton scattering.

The Galactic Diffuse emission template is specified solely by its normalization value $A_{\rm dif}$, or by two normalization values $A_{\pi^0\text{brem}}$ and $A_{\rm ics}$ in the case of the \texttt{GALPROP}-based diffuse models.

\item {\bf Fermi Bubbles}

The \textit{Fermi} Bubbles are a large gamma-ray structure extending nearly $\sim10$~kpc on either side of the galactic disk~\cite{2010ApJ...724.1044S}, and approximately centered at the Galactic Center.

We use the \textit{Fermi} Bubbles template from Ref.~\cite{2010ApJ...724.1044S}. This is taken to have uniform emission intensity before exposure correction. The smooth Bubbles template is characterized by only the normalization $A_{\rm bub}$, with log-flat prior distribution. We leave an analysis of possible alternate models of the Bubbles \cite{Fermi-LAT:2014sfa} to later work.

Hypothetical PS emission from the \textit{Fermi} Bubbles region is specified by extra model parameters, ${\bm\theta}=\{A_{\rm bub}^{\rm PS},n_1^{\rm bub},n_2^{\rm bub},S_b^{\rm bub}\}$. The PS source-count function in a given pixel $p$ is taken to be a singly broken power law:
\begin{equation}
 {d N_p(S) \over dS} = A_{p} \left\{ 
\begin{array}{cc}
\left( {S \over S_b} \right)^{-n_1} & S \geq S_b \\
\left( {S \over S_b} \right)^{-n_2} & S < S_b  \, ,
\end{array}
\right.
\label{eq:singlepower}
\end{equation}
where $S_b=S_b^{\rm bub}$ is the break, $n_1=n_1^{\rm bub}$ and $n_2=n_2^{\rm bub}$ are the slopes above and below the break, and $A_p=A_p^{\rm bub}$ is the pixel-dependent normalization. In order for the total number of photons from PS to be finite, it is required that  $n_1$ and $n_2$ are greater than and less than 2 respectively. The source-count function in counts in Eq.~(\ref{eq:singlepower}) is related to the function in flux via Eq.~(\ref{eq:fluxtocounts}).

\item {\bf Isotropic Background}

Gamma-ray emission from extragalactic sources is expected to be approximately isotropic on the sky; this component can also absorb residual cosmic-ray contamination (although in the {\sc UltraCleanVeto} dataset we use, such contamination should be very small). We include both a Poissonian isotropic component, specified by its normalization $A_{\rm iso}$ (with log-flat prior), and also a component for isotropically distributed PSs. In high-latitude studies of the isotropic PS population, it has been shown that the choice of priors can determine the source-count function at low flux~\cite{Lisanti:2016jub}, and that more than one break is required to accurately describe the source-count function at high flux. Consequently, we model the isotropic PS source count function with two breaks, so it is specified by the model parameters ${\bm\theta}=\{A_{\rm iso}^{\rm PS},n_1^{\rm iso},n_2^{\rm iso},n_3^{\rm iso},S_{b_1}^{\rm iso},S_{b_2}^{\rm iso}\}$.

Explicitly, the PS source-count function in a given pixel $p$ is a doubly broken power law:
\begin{equation}
\frac{dN_p}{dS} = A_p \left\{ \begin{array}{lc} \left( \frac{S}{S_{b_1}} \right)^{-n_1}, & S \geq S_{b_1} \\ \left(\frac{S}{S_{b_1}}\right)^{-n_2}, & S_{b_1} > S \geq S_{b_2} \\ \left( \frac{S_{b_2}}{S_{b_1}} \right)^{-n_2} \left(\frac{S}{S_{b_2}}\right)^{-n_3}, & S_{b_2} > S \end{array}  \right.
\label{eq:doublebreak}
\end{equation}
where $S_{b_1}=S_{b_1}^{\rm iso}$ is the first break, $S_{b_2}=S_{b_2}^{\rm iso}$ is the second break, $n_1$ and $n_2$ are the slopes above and below the first break, $n_2$ and $n_3$ are the slopes above and below the second break,  and $A_p=A_p^{\rm iso}$ is the pixel-dependent normalization. The NP model priors other than normalization are taken to be linear flat. The source-count function in counts in Eq.~(\ref{eq:doublebreak}) is related to the function in flux via Eq.~(\ref{eq:fluxtocounts}).

\item {\bf Generalized NFW Profile (Galactic Center Excess)}

The flux profile of the GCE has been previously found to be well described by the square of a generalized NFW profile~\cite{Daylan:2014rsa,Calore:2014xka}, consistent with DM annihilation if the DM density profile follows the (generalized) NFW profile. As such, we include a template for a signal with spatial distribution matching that of a squared generalized NFW profile~\cite{Navarro:1995iw} integrated along the line of sight (l.o.s.). 

The functional form of the density profile is:
\begin{equation}
 \rho(r)=\rho_0 \frac{(r/r_s)^{-\gamma}}{(1+r/r_s)^{3-\gamma}},
\end{equation}
where $r_s=20$ kpc, $\gamma=1.25$, and the normalization is allowed to float. (We also check the impact of including a template of a standard NFW profile with $\gamma=1.0$, and find no significant difference in our results.) The resulting flux is determined by:
\begin{equation}
 J(\phi)=\int_{\rm l.o.s.} \rho^2(r) ds
\end{equation}
where $s$ is the l.o.s. distance and $\phi$ is the angle from the GC.

The Poissonian generalized-NFW template, as would be appropriate for a DM signal, is characterized by its normalization $A_{\rm NFW}$. PSs distributed following this profile are characterized by the model parameters ${\bm\theta}=\{A_{\rm NFW}^{\rm PS},n_1^{\rm NFW},n_2^{\rm NFW},S_b^{\rm NFW}\}$, which determine the source count function via Eq.~(\ref{eq:singlepower}). We will refer to ``DM'' to mean the generalized NFW profile ($\gamma=1.25$) template, rather than the traditional NFW profile ($\gamma=1.0$).

\item {\bf Galactic Disk}

We model PS emission from the galactic disk through a doubly exponential thin-disk source distribution,
\begin{equation}
n(z, R) \propto \exp\left[ \frac{-R}{5~\text{kpc}} \right] \, \exp \left[\frac{-|z|}{1~\text{kpc}}\right] \, ,
\end{equation}
with radius $R=5$~kpc and a scale height $z=1~$kpc. (This disk model, and a thinner disk with scale height $z=0.3~$kpc, were both tested in Ref.~\cite{Lee:2015fea}; the resulting differences in the fit were found to be small.) We integrate along the line of sight to obtain the expected number of sources per pixel.

PSs distributed along the disk are characterized by the model parameters ${\bm\theta}= \{A_{\rm disk}^{\rm PS},n_1^{\rm disk},n_2^{\rm disk},S_b^{\rm disk}\}$, which determine the source count function via Eq.~(\ref{eq:singlepower}). 
 
\item {\bf Boxy Bulge}

The Boxy Bulge is an X-shaped structure at the center of the Milky Way, which appears to be correlated with gamma-ray emission from the inner Galaxy~\cite{Bartels:2017vsx,2018NatAs...2..387M,Macias:2019omb,Song:2019nrx}. We do not use the corresponding template in the standard or baseline pipeline, but since it has been suggested that stars associated with the Boxy Bulge could contribute significantly to the GCE, we will later perform the test of replacing the NFW PS template with a PS population tracing this structure.

We use the template for the Boxy Bulge from Ref.~\cite{2018NatAs...2..387M}, and pixelize the template with $\texttt{nside=128}$.
Smooth emission from the Boxy Bulge template is characterized by the normalization parameter $A_{\rm bul}$. PS emission associated with the Boxy Bulge is described by the model parameters ${\bm\theta}=\{A_{\rm bul}^{\rm PS},n_1^{\rm bul},n_2^{\rm bul},S_b^{\rm bul}\}$, which determine the source count function via Eq.~(\ref{eq:singlepower}). 
\end{itemize}

\begin{table*}
\renewcommand{\arraystretch}{1.5}
\setlength{\tabcolsep}{5.2pt}
\begin{center}
\begin{tabular}{ccc}
\hline
\multicolumn{3}{c}{\textsc{Simulation Parameters}} 		\Tstrut\Bstrut \\
\hline
Parameter 	& Inner Galaxy (masked) & Inner Galaxy (unmasked) 	\Tstrut\Bstrut \\
\hline 
\hline
$\log_{10}A_\text{iso}$  & ${-1.94}$ & ${-2.20}$  \Tstrut\Bstrut \\ 
$\log_{10}A_\text{dif}$  & ${1.11}$ & ${1.11}$  \Tstrut\Bstrut \\
$\log_{10}A_\text{bub}$  & ${-0.08}$ & ${-0.06}$ \Tstrut\Bstrut \\ 
$\log_{10} A_\text{NFW}$ & ${-0.56}$ & ${-0.42}$    \Tstrut\Bstrut	\\
$\log_{10}A_\text{PS}^\text{NFW}$, $\log_{10}A_\text{PS}^\text{bub}$ & ${-2.08}$ & ${-2.37}$  \Tstrut\\
$S_b^\text{NFW}$, $S_b^\text{bub}$  & 13.47      & $20.53$    \\
$n_1^\text{NFW}$, $n_1^\text{bub}$ & 35.13     & $32.98$    \\\
$n_2^\text{NFW}$, $n_2^\text{bub}$   & $-1.68$    & $-1.41$  \\
$\log_{10}A_\text{PS}^\text{disk}$ & ${-4.15}$ & $-3.71$  \Tstrut\\
 $S_b^\text{disk}$ & 37.69      & $42.70$    \\
 $n_1^\text{disk}$  &  31.59    & $2.40$    \\
 $n_2^\text{disk}$   & $-0.88$    & $-1.50$  \\
 $\log_{10}A_\text{PS}^\text{iso}$ & $-4.78$ & $-4.18$  \Tstrut\\
 $S_{b,1}^\text{iso}$ & $22.59$     & $23.85$    \\
  $S_{b,2}^\text{iso}$ & $3.22$     & $5.22$    \\
 $n_1^\text{iso}$  &  $3.89$   & $3.41$    \\
 $n_2^\text{iso}$   & $3.84$    & $3.48$  \\
  $n_3^\text{iso}$   & $-0.27$    & $-0.18$  \\

\hline
\hline
\end{tabular}
\end{center}
\caption{Parameter values used to generate the simulated data from Poissonian and non-Poissonian templates for the proof-of-principle scenario, for cases where the 3FGL PSs are masked or unmasked. Note that for simplicity, in the proof-of-principle scenario, the Isotropic PS are not simulated when the 3FGL are masked, however they are included in the unmasked case to ensure the bright 3FGL are correctly captured.}
\label{tab:sim_values}
\end{table*}

\section{Proof-of-Principle with Simulated Data: Supplementary Figures and Analyses}

In this section, we supplement the main text by checking that we correctly recover the simulated templates, when the same templates are used in the construction of simulated data and at the analysis stage.

We also study simulated data biases in a ROI without a PS mask (in contrast to the masked ROI in the main text) for the example of a PS population tracing the \textit{Fermi} Bubbles. In this case, we also include isotropic PSs in the fit and the resulting simulated data. In the previous (masked) analyses, the masking of the 3FGL PSs ensured that the only isotropic PSs in the data were faint and unresolved, so the effect of leaving them out was expected to be relatively minor; in contrast, in the unmasked case, the isotropic PS population is expected to include some number of very bright sources. 

Table~\ref{tab:sim_values} details all the model parameters used to create the simulated data, for the two inner Galaxy ROIs (with 3FGL PSs masked and unmasked, respectively). Note that not all templates are simulated in every analysis.

\subsection{Recovery of Parameters from Simulated Data}

We simulate mock data containing disk PSs, bubbles PSs, DM, and all other Poissonian templates contained in the standard pipeline (including isotropic PS in the unmasked ROI). We test that analyzing this simulated data with the same pipeline correctly recovers the simulated parameters within the uncertainties (for this realization), and in particular correctly identifies the smooth DM component as well as the two PS templates. 

\begin{figure*}[t!]
\leavevmode
\centering
\subfigure{\includegraphics[width=0.48\columnwidth]{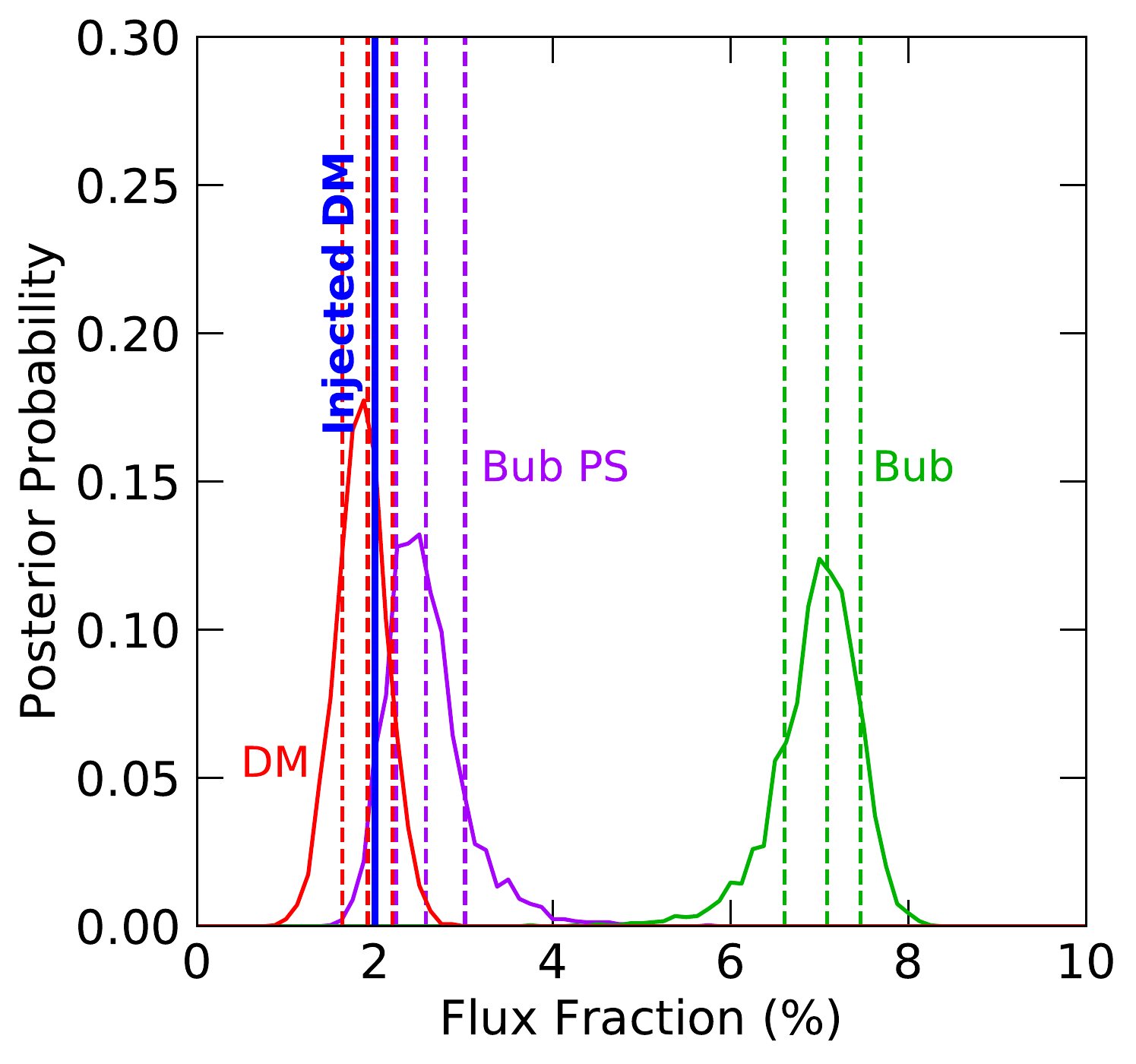}}
\hspace{1mm}
\subfigure{\includegraphics[width=0.48\columnwidth]{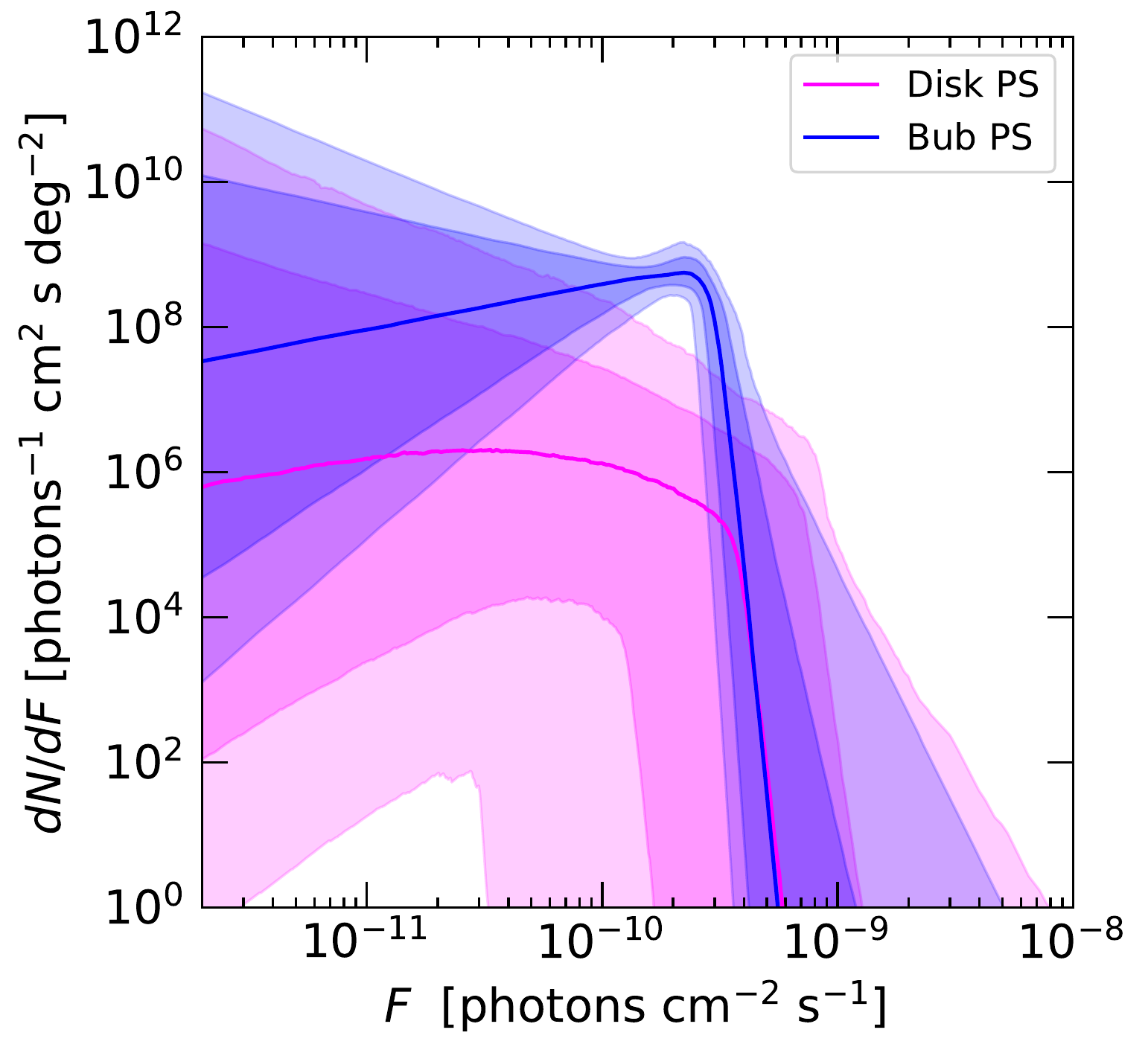}}
\caption{Inner Galaxy (masked) results for simulated Bubbles PS and Disk PS, and smooth DM, Bubbles, Isotropic and Galactic Diffuse templates. \textbf{Left:} Flux posteriors for analysis using the same
templates as the simulated data. \textbf{Right:} Luminosity functions for this scenario for Bubbles PS, Disk
PS, and Isotropic PS.}
\label{fig:GCEbiasrecover}
\end{figure*}

Figure~\ref{fig:GCEbiasrecover} shows the source count functions and flux fractions attributed to the various templates when the templates used in the simulation match those used in the analysis. We also check if increasing the normalization of the simulated DM component impacts the analyzed results. We find it correctly recovers all simulated parameter values and fluxes in all cases.

\begin{figure*}[t!]
\leavevmode
\centering
\subfigure{\includegraphics[width=0.48\columnwidth]{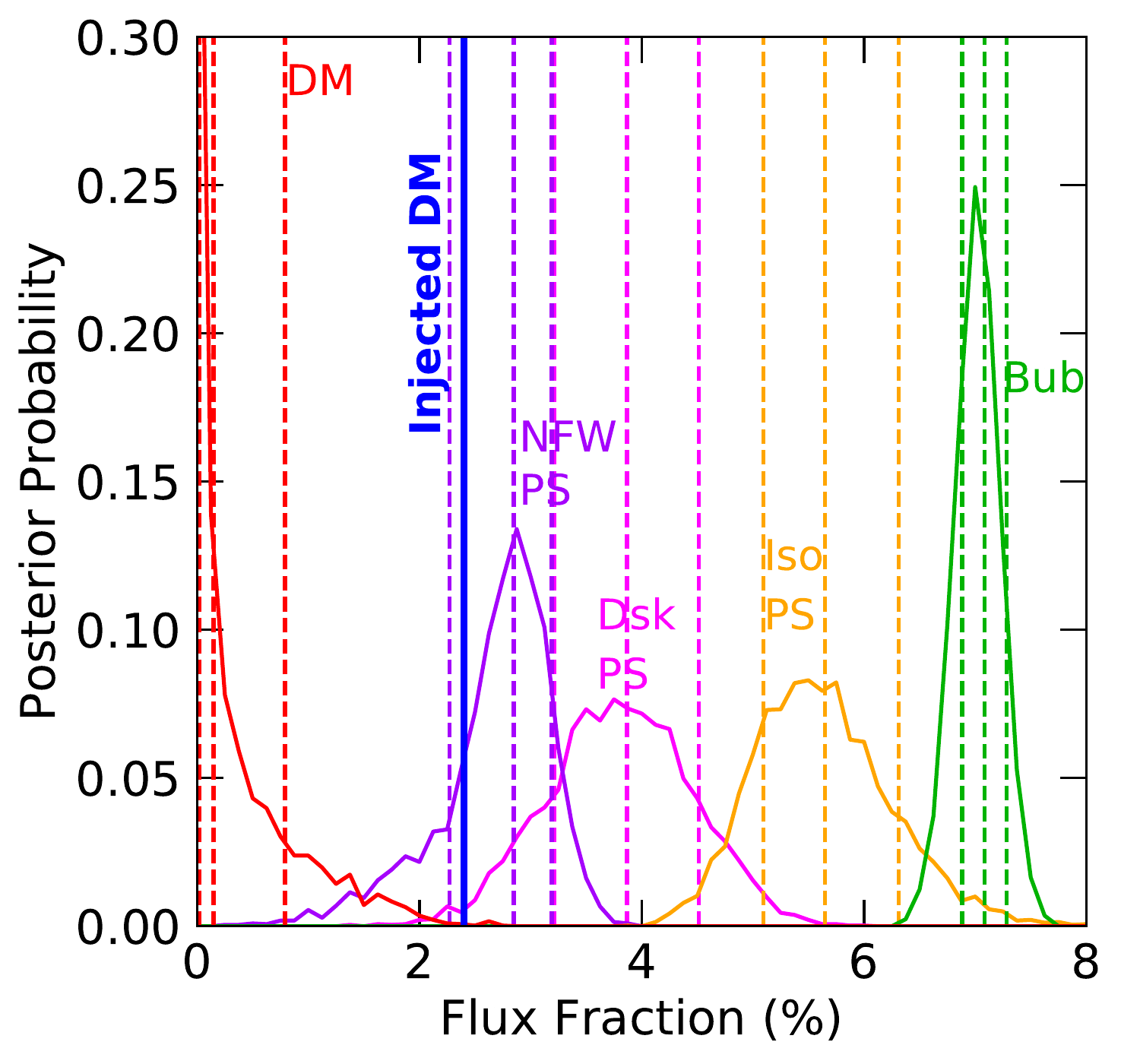}}
\hspace{1mm}
\subfigure{\includegraphics[width=0.49\columnwidth]{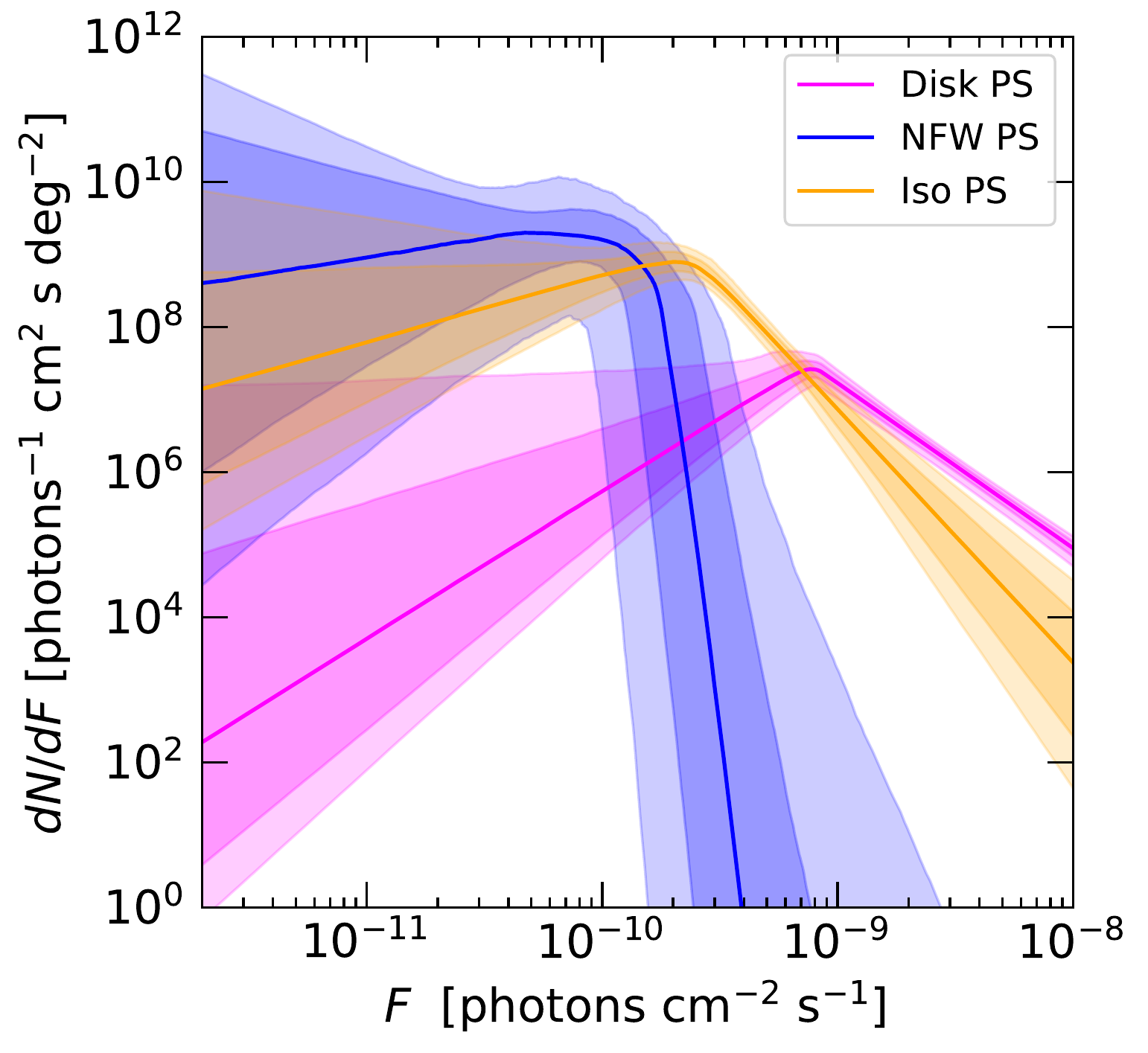}}
\caption{Bias injection test on the simulated data, but with the 3FGL unmasked (and isotropic PS included in simulation and analysis). $2.4\%$ DM was injected. All templates are present, but those with fluxes peaked below 0.1\% (except for DM) are not shown for clarity.}
\label{fig:GCEbias_unmasked}
\end{figure*}

\subsection{Supplementary Results for Simulated Point Sources in the \emph{Fermi} Bubbles}

\begin{table*}
\renewcommand{\arraystretch}{1.5}
\centering
\begin{tabular}{|c|P{1.8cm}|P{5.2cm}|P{2.5cm}|P{1.6cm}|P{1.6cm}|P{1.6cm}|}
\hline
\multicolumn{7}{|c|}{\textsc{Simulated Data, 3FGL Unmasked}} \\
\hline
 \multirow{2}{*}{\textbf{Simulation}} &  \textbf{Injected} &  \multirow{2}{*}{\textbf{Analysis Templates}} & \textbf{DM Flux}  &  \multicolumn{3}{c|}{\multirow{2}{*}{\textbf{Bayes Factor}} } \\
                                    &    \textbf{DM Flux}      &                                         & \textbf{(95\%)} &  \multicolumn{3}{c|}{} \\ \hline\hline

Bubbles PS  &               & Same as simulated                         & $[ \, 2.3, 3.1 \,]\  \%$                                     & $1\times10^{34}$  & \cellcolor{black!25}             & $4\times10^{36}$                         \\ \cline{3-4} \cline{6-7}
Disk PS     &  $2.4 \%$  & Same but Bubbles PS $\rightarrow$ NFW PS  & $\textcolor{violet}{[ \, \textbf{0.0, 1.7} \,]\  \%}$                   &                                  & $3\times10^2$      & \cellcolor{black!25}  \\ \cline{3-5} \cline{7-7}
DM      &               & Same but no Bubbles PS                    & $[ \, 2.4, 3.4 \,]\  \%$                                     &  \cellcolor{black!25}            &                                  &               \\ \hline 

\end{tabular}
\caption{Same as the simulated data proof-of-principle scenario in Tab.~\ref{tab:bayes_sim}, but an unmasked analysis. Purple signals the standard pipeline analysis.}
\label{tab:bayes_sim_unmasked}
\end{table*}


\begin{figure*}[t!]
\leavevmode
\centering
\subfigure{\includegraphics[width=0.48\columnwidth]{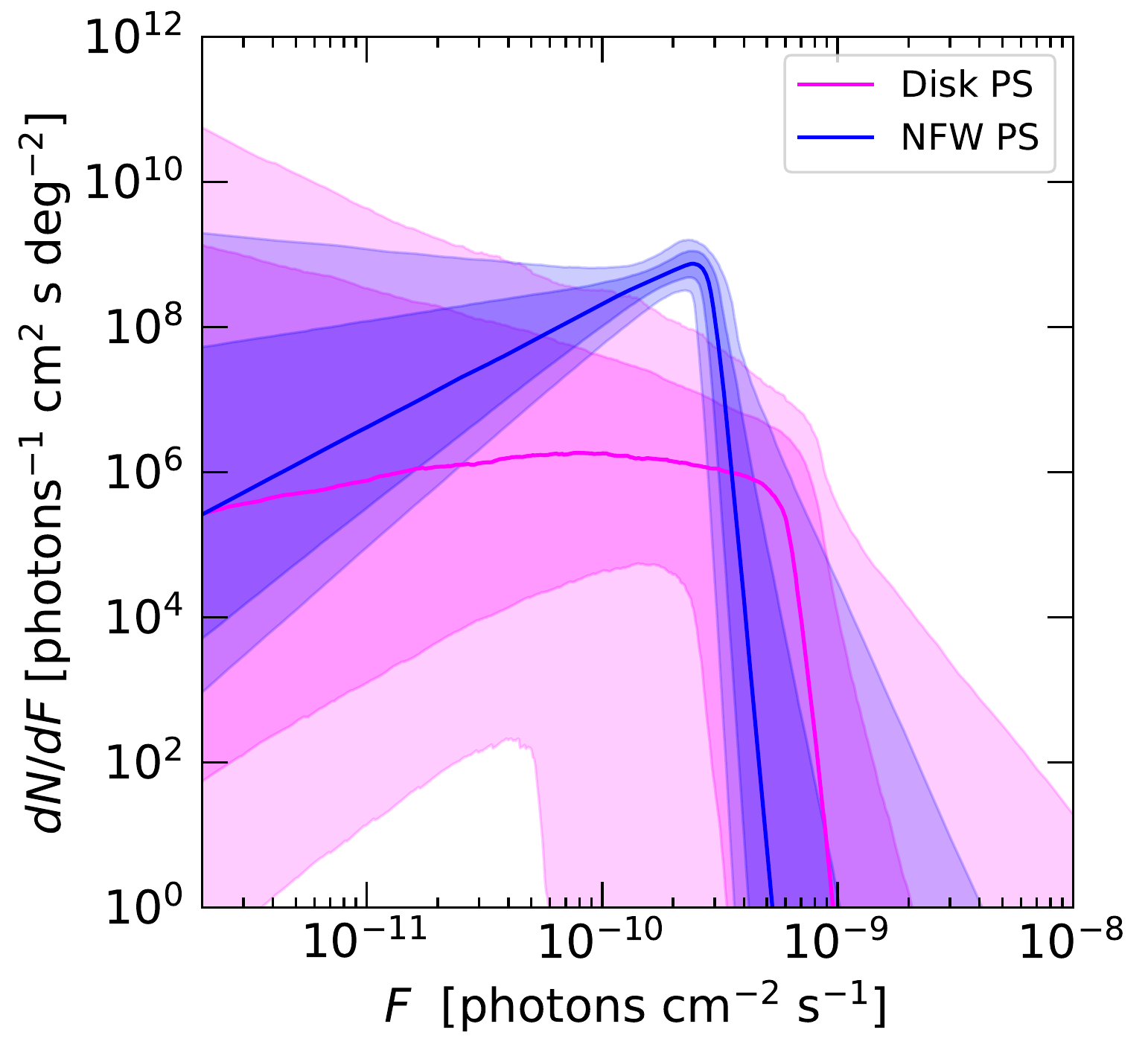}}
\hspace{1mm}
\subfigure{\includegraphics[width=0.48\columnwidth]{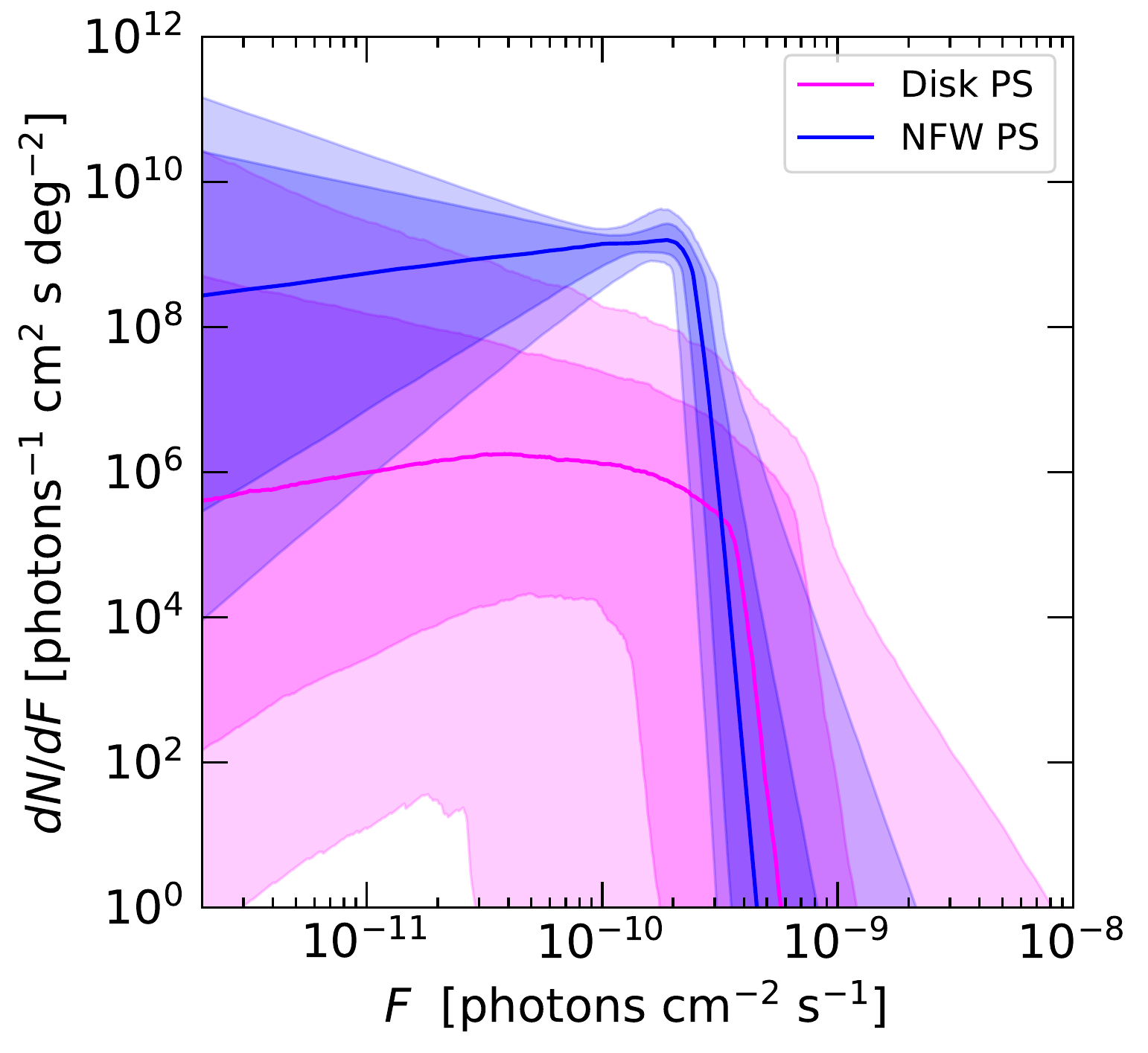}}\\
\subfigure{\includegraphics[width=0.48\columnwidth]{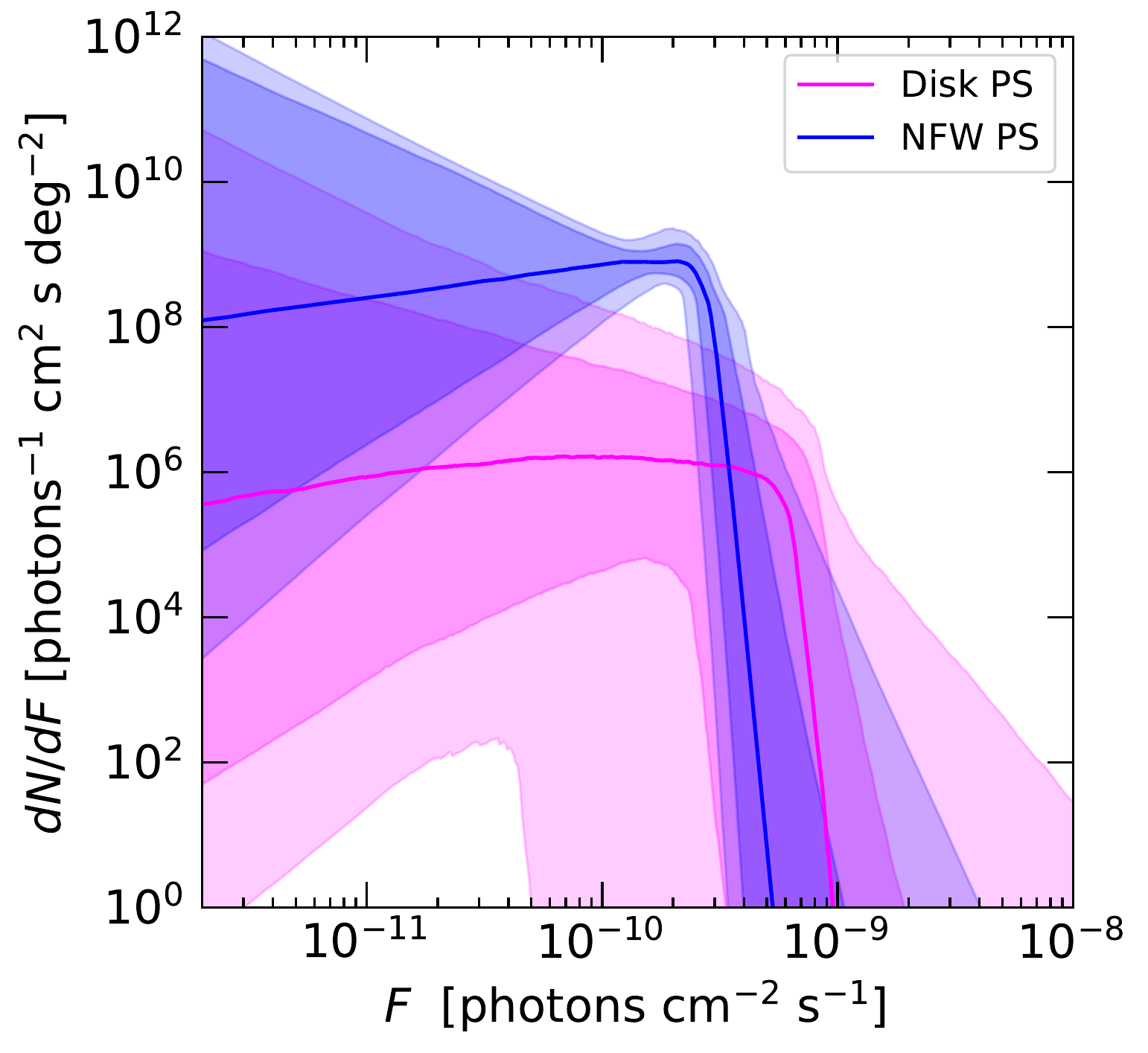}}
\hspace{1mm}
\subfigure{\includegraphics[width=0.48\columnwidth]{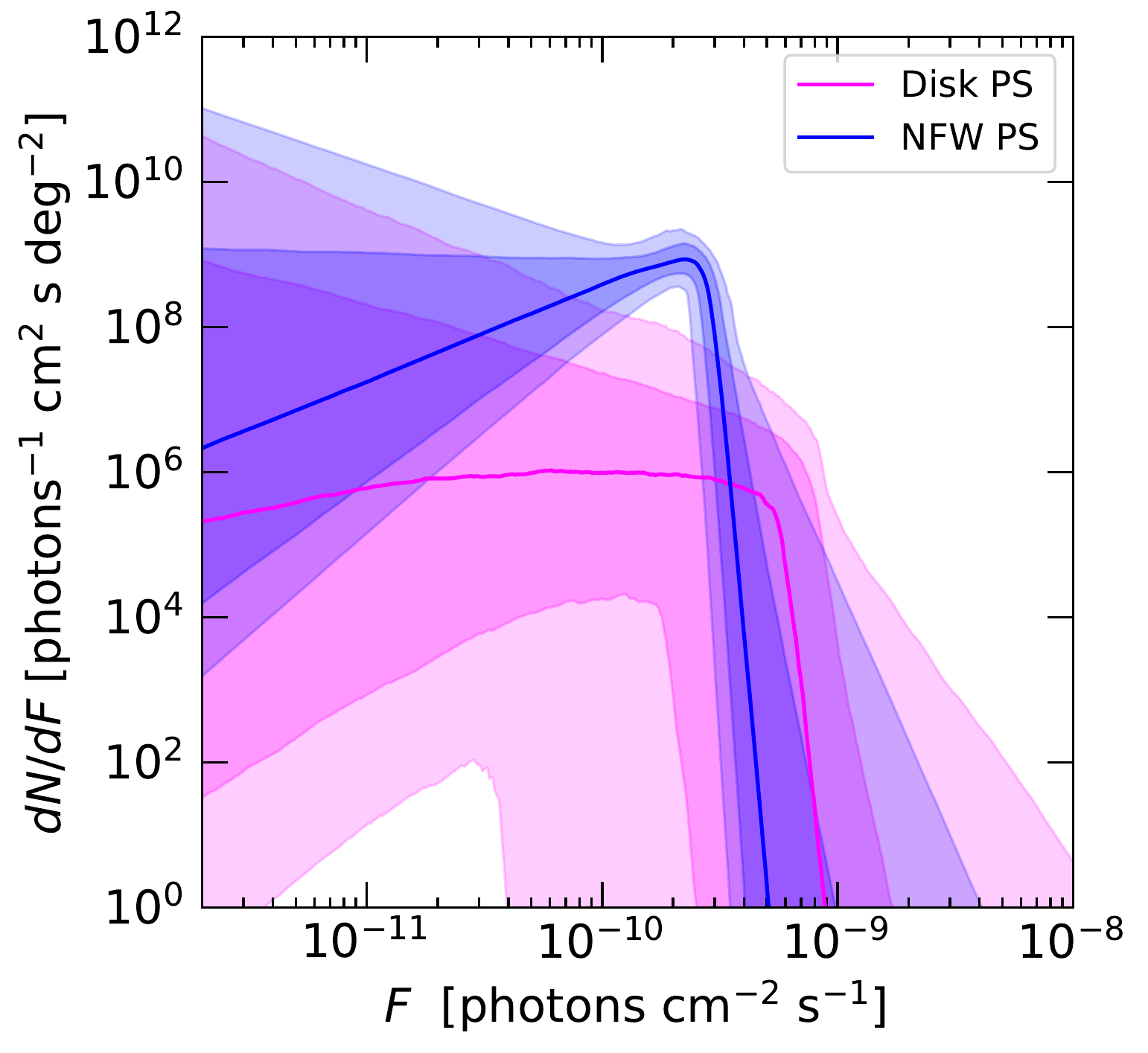}}
\caption{Inner Galaxy (masked) results for the recovered SCF when we simulate Bubbles PS and Disk PS, and smooth DM, Bubbles, Isotropic and Galactic Diffuse templates (the corresponding flux fractions are given in Fig.~\ref{fig:GCEbias_sim}). \textbf{Top-Left:} $2\%$ DM injection. \textbf{Top-Right:} $5\%$ DM injection. \textbf{Bottom-Left:} $7\%$ DM injection. \textbf{Bottom-Right:} $15\%$ DM injection.}
\label{fig:GCEbias}
\end{figure*}

In addition to the results in the main text, corresponding to the ($30^\circ$ radius, $|b|<2^\circ$ plane mask) masked ROI, we test the same proof-of-principle example in the ($30^\circ$ radius, $|b|<2^\circ$ plane mask) unmasked ROI. 

Table~\ref{tab:bayes_sim_unmasked} shows the result of applying the standard and baseline NPTF pipelines to simulated data containing DM and Bubbles PSs in this case, when an additional DM component is injected. As in the masked ROI, we find that the fit incorrectly reconstructs the simulated DM component as NFW PSs when given the option.

Figure~\ref{fig:GCEbias_unmasked} shows the posterior probability distributions for the relevant template flux fractions, and the reconstructed source count functions for the PSs.

Figure~\ref{fig:GCEbias} shows, as a supplement to the main text, the evolution in the reconstructed source count functions for disk and NFW PSs, for 4 different injected DM fluxes (2, 5, 7, and $15\%$ post-injection), in the masked ROI. These results correspond to the flux posteriors shown in Fig.~\ref{fig:GCEbias_sim}, where the mock data contains Bubbles PSs and a DM component, but is fitted with the standard NPTF pipeline (i.e. with NFW PSs and without Bubbles PSs). We observe that as the DM normalization is increased, the source count function for the reconstructed NFW PSs remains fairly stable; there is some trend toward higher $dN/dF$ at low $F$, but the difference remains within the uncertainty bands.

Note that in general for our Bubbles analyses, as the \textit{Fermi} Bubbles extend to much higher latitudes than our ROIs, it is also possible to fix the diffuse emission within the Bubbles to its high-latitude normalization, instead of allowing it to float. In this case, we find that even more flux is attributed to the NFW PSs. Similarly, fixing the diffuse model to its high-latitude value provides qualitatively comparable results.

We generate 20 sets of simulated data, and find this result persists, with comparable posteriors and Bayes factors.

It is possible that biases of this type could be detected and modeled by an improved version of the NPTF
pipeline. For example, giving extra spatial freedom to the NFW PS template, or allowing the source count
function to vary between different spatial subregions of the template, could identify a mismatch between the
NFW template and the true PS distribution. Varying the ROI and comparing the results to those expected in
simulated data could also help distinguish scenarios with different spatial PS distributions. For example, the Bayes factor in favor of NFW PSs may change between scenarios when the resolved 3FGL sources are unmasked or masked, and may not change in a manner matching simulations. Reducing the ROI to the region where the GCE is brightest can similarly help eliminate biases from populations outside that
region. Both strategies, however -- increasing the number of degrees of freedom, and reducing the size of
the ROI -- are also expected to inflate the statistical uncertainties, reducing constraining power. We have studied
some possible improvements to the NPTF pipeline in this section and the following sections, but leave a comprehensive study to future work.

\begin{table*}
\renewcommand{\arraystretch}{1.3}
\centering
\begin{tabular}{|P{2.8cm}|P{6.2cm}|P{3cm}|P{1.6cm}|P{1.6cm}|P{1.6cm}|}
\hline
\multicolumn{6}{|c|}{\textsc{Real Data, 3FGL Unmasked}} \\
\hline
 \textbf{Injected} &  \multirow{2}{*}{\textbf{Analysis Templates}} &  \textbf{DM Flux}                          &  \multicolumn{3}{c|}{\multirow{2}{*}{\textbf{Bayes Factor}} } \\
\textbf{DM Flux}   &                                               &   \textbf{(95\%)}                     &     \multicolumn{3}{c|}{}                                 \\ \hline\hline

 \multirow{4}{*}{None} &  Disk PS + Iso PS               &     \multirow{2}{*}{ $[ \, 1.9, 2.9\,]\  \%$}      &   \multicolumn{3}{c|}{ }          \\ 
                    &  Diffuse + Iso + Bub + DM    &                                      &      \multicolumn{3}{c|}{}        \\ \cline{2-3}
                    &  Disk PS + Iso PS + NFW PS         &      \multirow{2}{*}{ $\textcolor{violet}{[ \, \textbf{0.0, 0.2}\,]\  \%}$}     &      \multicolumn{3}{c|}{\multirow{2}{*}{$4\times10^{18}$}}          \\ 
                    &  Diffuse + Iso + Bub + DM    &                                      &     \multicolumn{3}{c|}{}         \\ \hline\hline
                    
\multirow{6}{*}{$2.4 \%$}    &  Disk PS + Iso PS                  &     \multirow{2}{*}{$[ \, 3.7, 4.7 \,]\  \%$}    &      &  \cellcolor{black!25}         &          \\ 
                                &  Diffuse + Iso + Bub + DM    &                                    &                                &   \cellcolor{black!25}        &         \\ \cline{2-3} \cline{5-6} 
                    
                                &  Disk PS + Iso PS + NFW PS         &     \multirow{2}{*} { $\textcolor{violet}{[ \, \textbf{0.0, 0.3} \,]\  \%}$}   &        \multirow{2}{*}{$6\times 10^{19}$ }                          &     \multirow{2}{*}{$9\times 10^2$}     &  \cellcolor{black!25}       \\ 
                                &  Diffuse + Iso + Bub + DM    &                                   &                                  &           &   \cellcolor{black!25}    \\ \cline{2-4} \cline{6-6} 
                                
                                &  Disk PS + Iso PS + NFW PS         &      Fixed at injection         &        \cellcolor{black!25}      &             &    \multirow{2}{*}{$6\times10^{16}$}  \\ 
                                &  Diffuse + Iso + Bub + Fixed DM    &         value ($2.4 \%$)                            &          \cellcolor{black!25}    &           &     \\ \hline
\end{tabular}
\caption{Bayes factors for analyses of the real \textit{Fermi} data injected with a DM signal. The DM flux is not correctly recovered when the standard PS templates are considered (shown as purple), similar to the proof-of-principle case detailed in Tabs.~\ref{tab:bayes_sim}~and~\ref{tab:bayes_sim_unmasked}. The data analysis that is favored has the relative Bayes factor appear next to the analysis row, where the two analyses that are being compared are not grayed out.}
\label{tab:bayes_data_unmasked}
\end{table*}

\section{Injection test on real data: Supplementary Figures and Analyses}

\begin{figure*}[t!]
\leavevmode
\centering
\subfigure{\includegraphics[width=0.49\columnwidth]{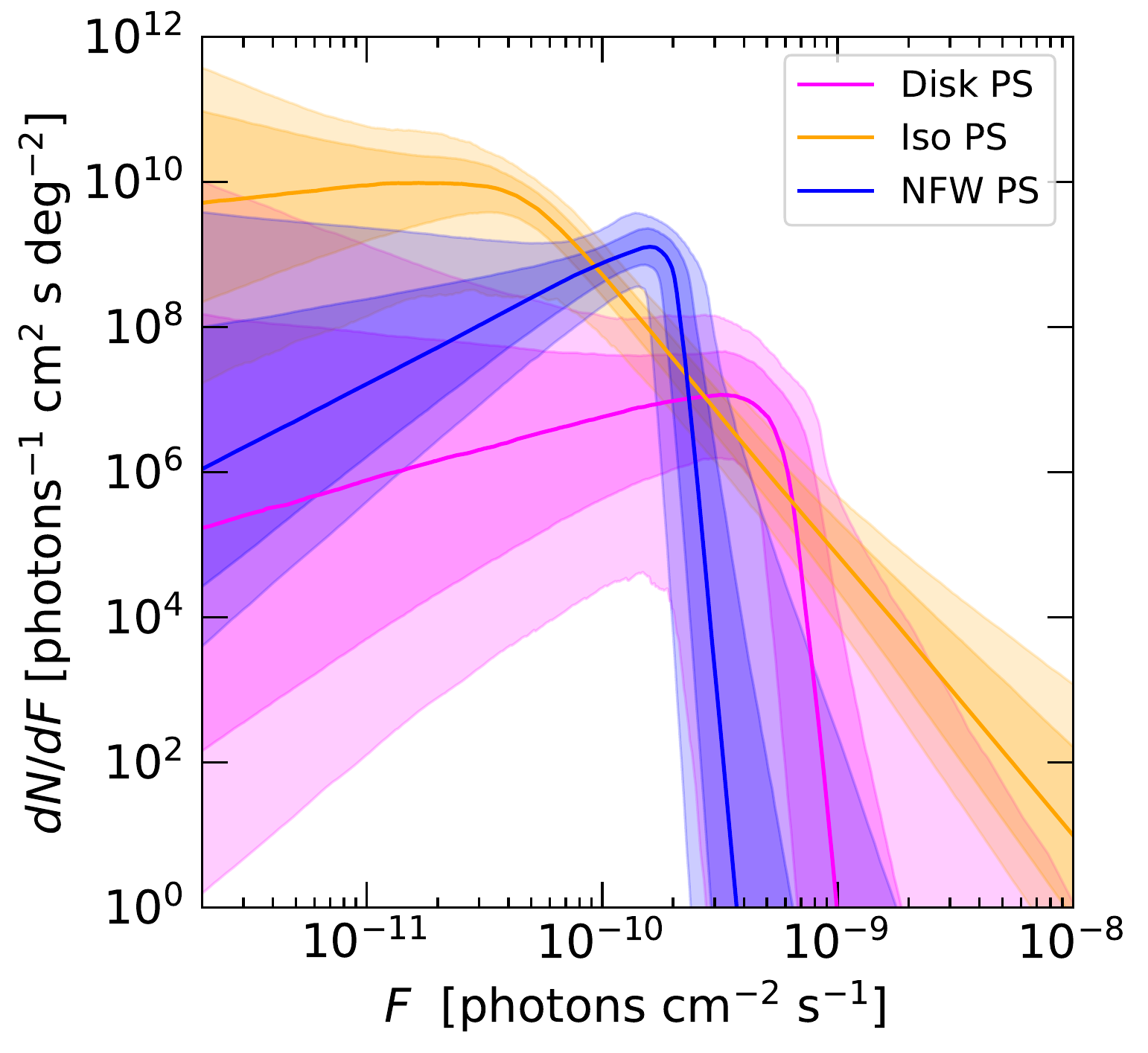}}
\hspace{1mm}
\subfigure{\includegraphics[width=0.49\columnwidth]{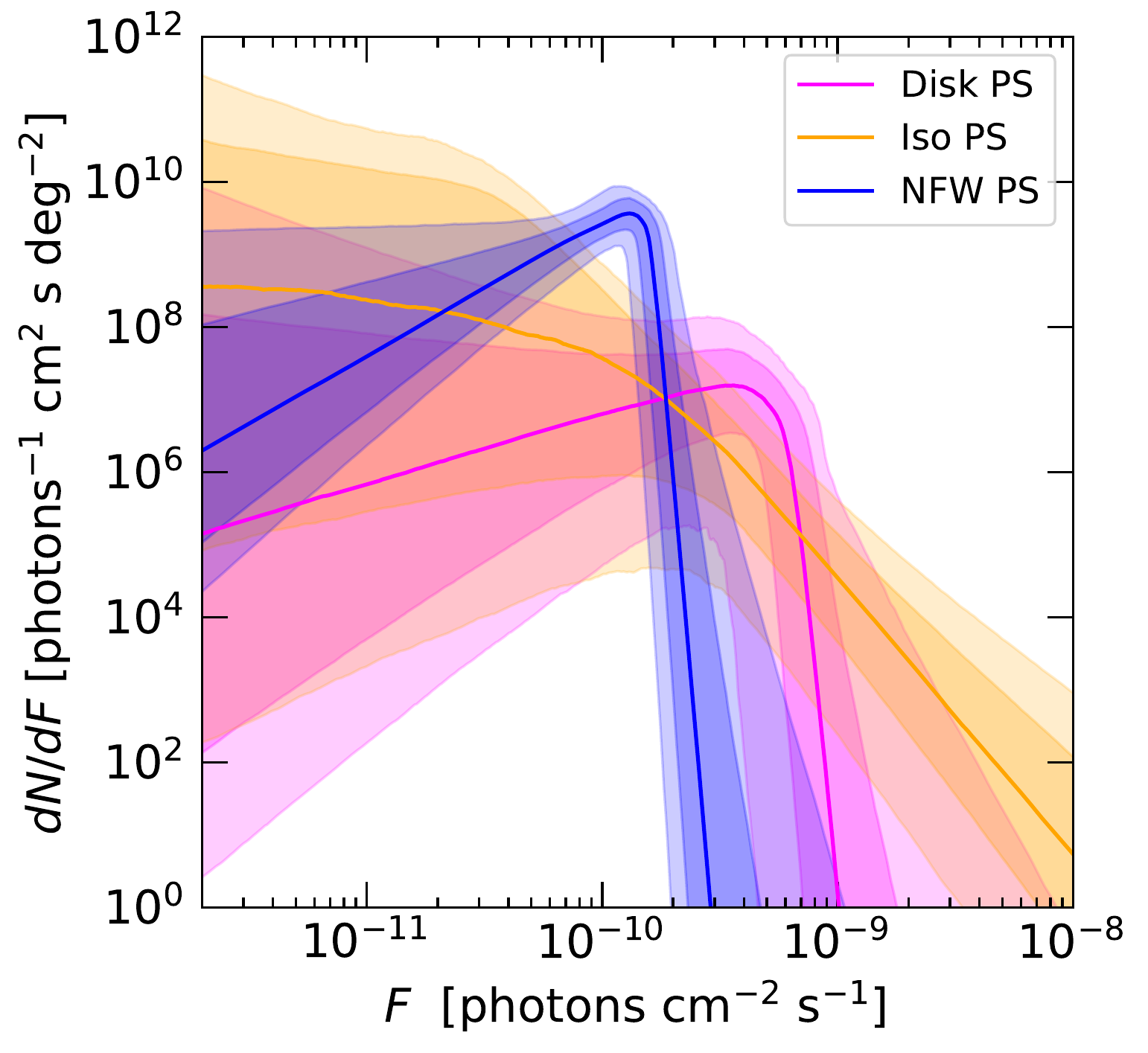}}
\\
\subfigure{\includegraphics[width=0.49\columnwidth]{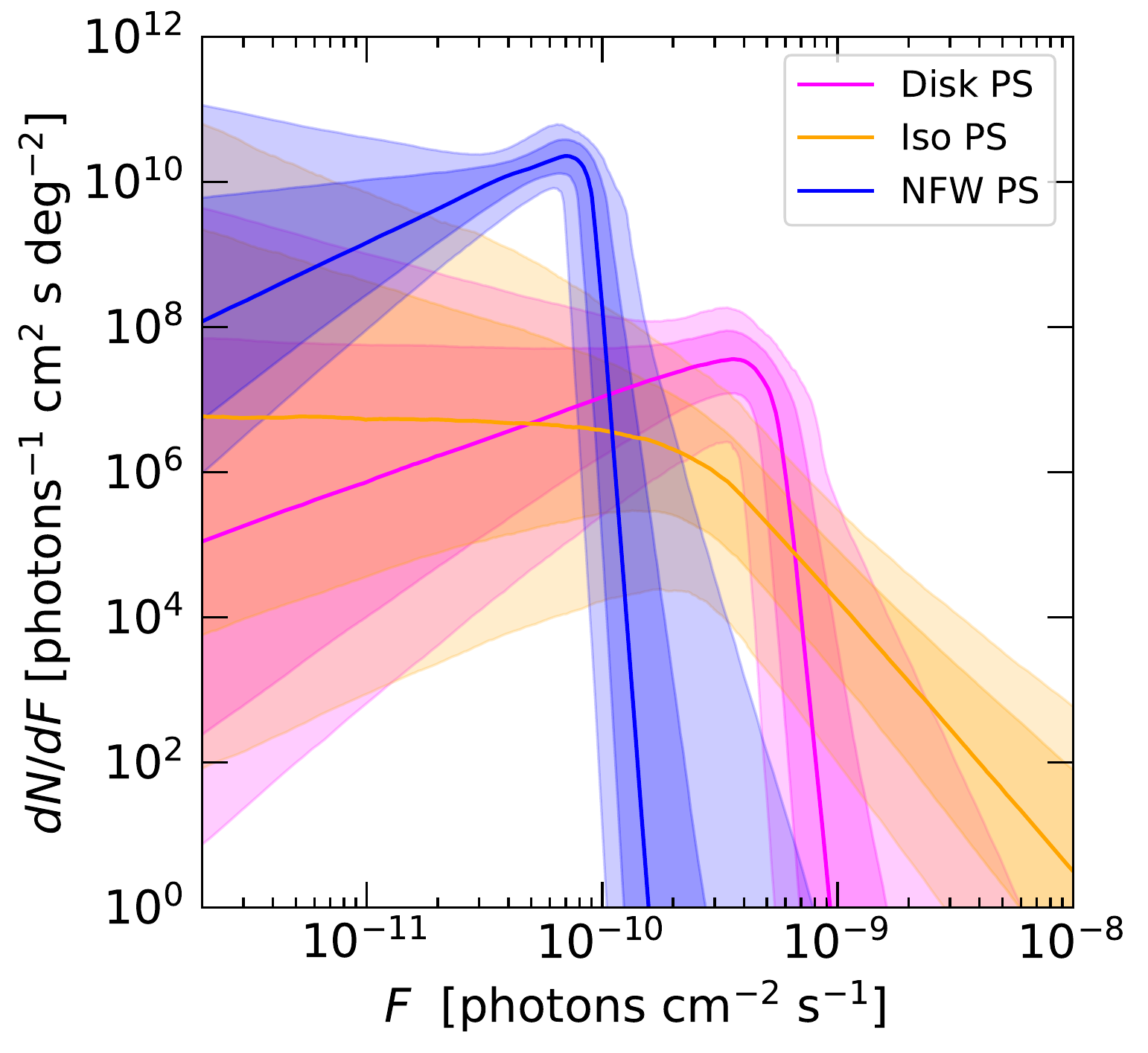}}
\hspace{1mm}
\subfigure{\includegraphics[width=0.49\columnwidth]{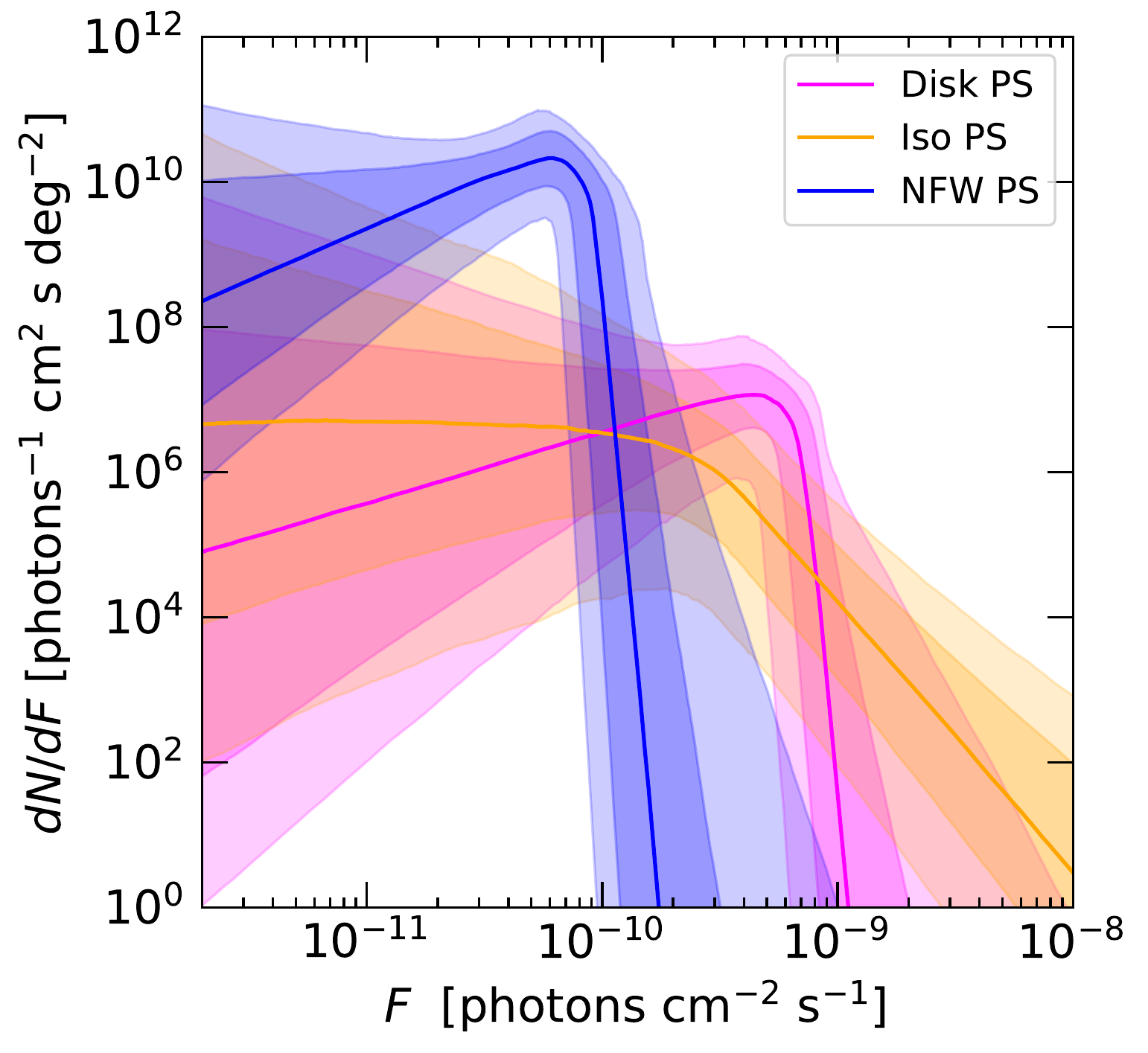}}
\caption{Inner Galaxy (masked) results for the SCF  when increasing amounts of DM are injected into the \textit{Fermi} data, and the ROI is analyzed with NFW PS, Disk PS, and Isotropic PS templates, and smooth DM, Bubbles, Isotropic and Galactic Diffuse templates. (The corresponding flux fractions are shown in Fig.~\ref{fig:bias_data}.) \textbf{Top-Left:} Zero DM injection. \textbf{Top-Right:} $1.8\%$ DM flux injection. No DM is recovered, and DM is instead attributed to PS templates. \textbf{Bottom-Left:} $6.7\%$ DM flux injection. DM is still not recovered. \textbf{Bottom-Right:} $15.2\%$ DM flux injection. Some DM flux is finally identified.}
\label{fig:injscf}
\end{figure*}

\begin{figure*}[t!]
\leavevmode
\centering
\subfigure{\includegraphics[width=0.48\columnwidth]{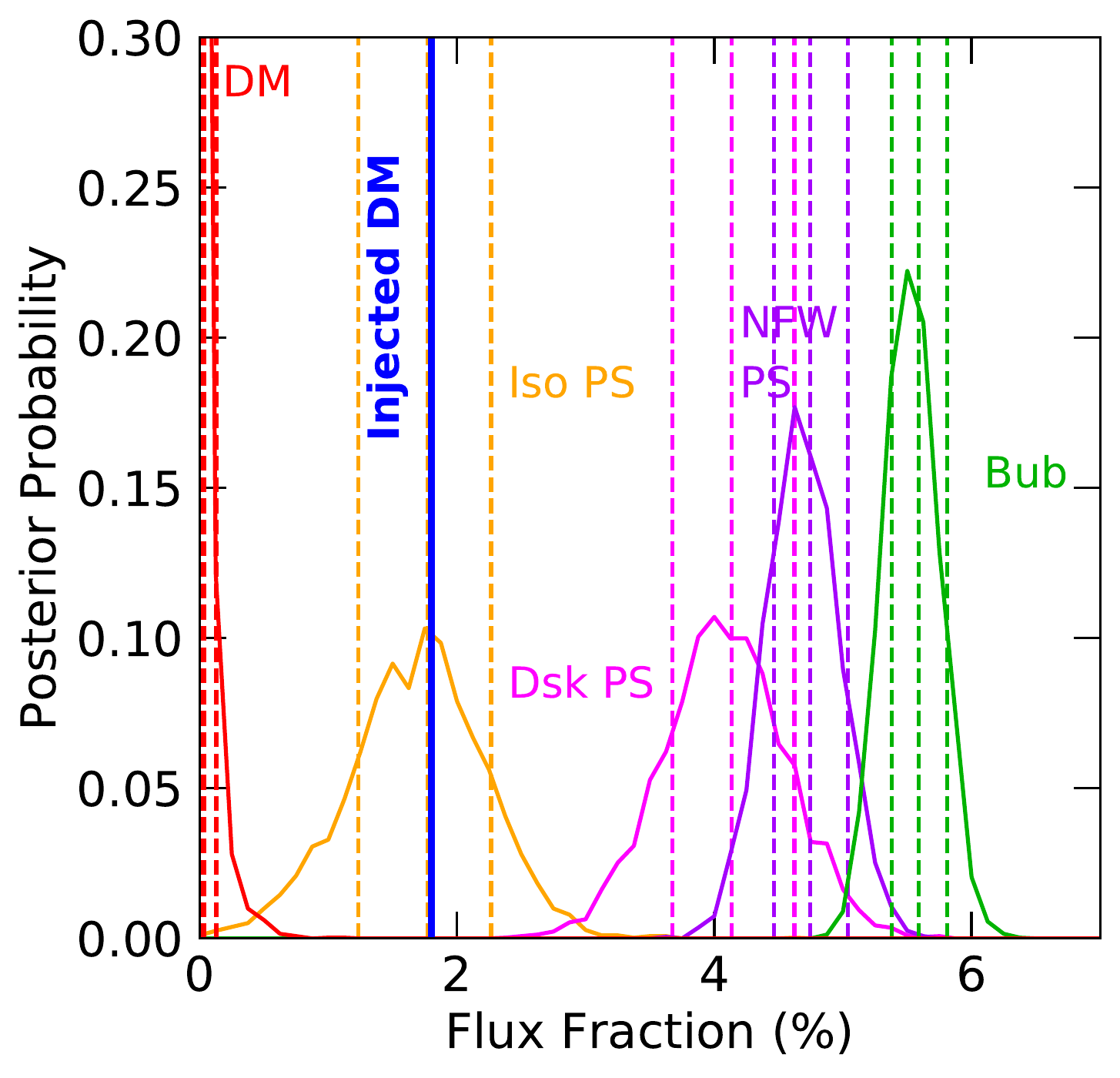}}
\hspace{1mm}
\subfigure{\includegraphics[width=0.49\columnwidth]{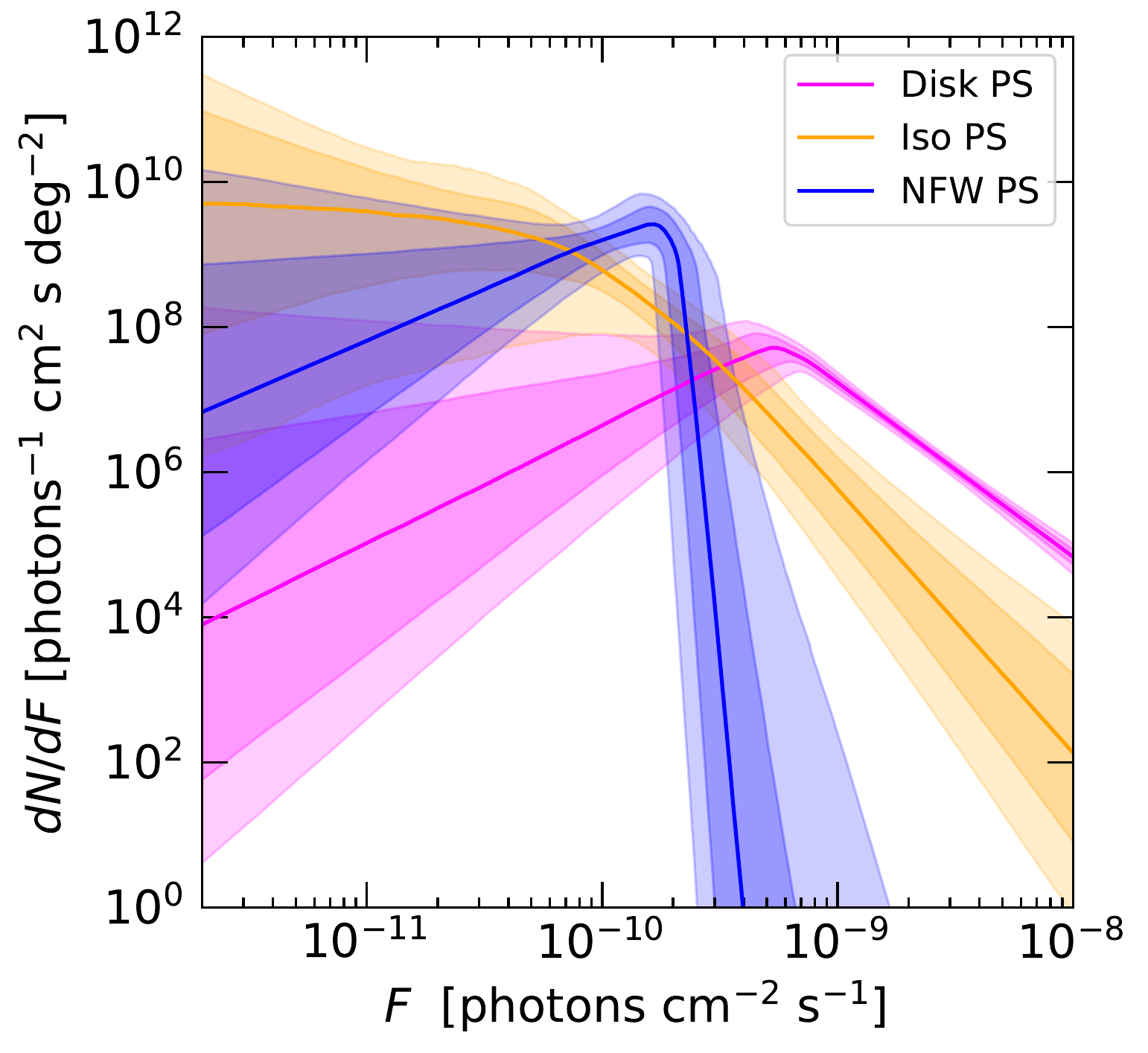}}
\caption{Injection test on real data ($1.8\%$), but with the 3FGL unmasked. All templates are present, but those with fluxes below 0.1\% (except for DM) are not shown for clarity.}
\label{fig:injectDM1_unmasked}
\end{figure*}

\begin{figure*}[t!]
\leavevmode
\centering
\subfigure{\includegraphics[width=0.48\columnwidth]{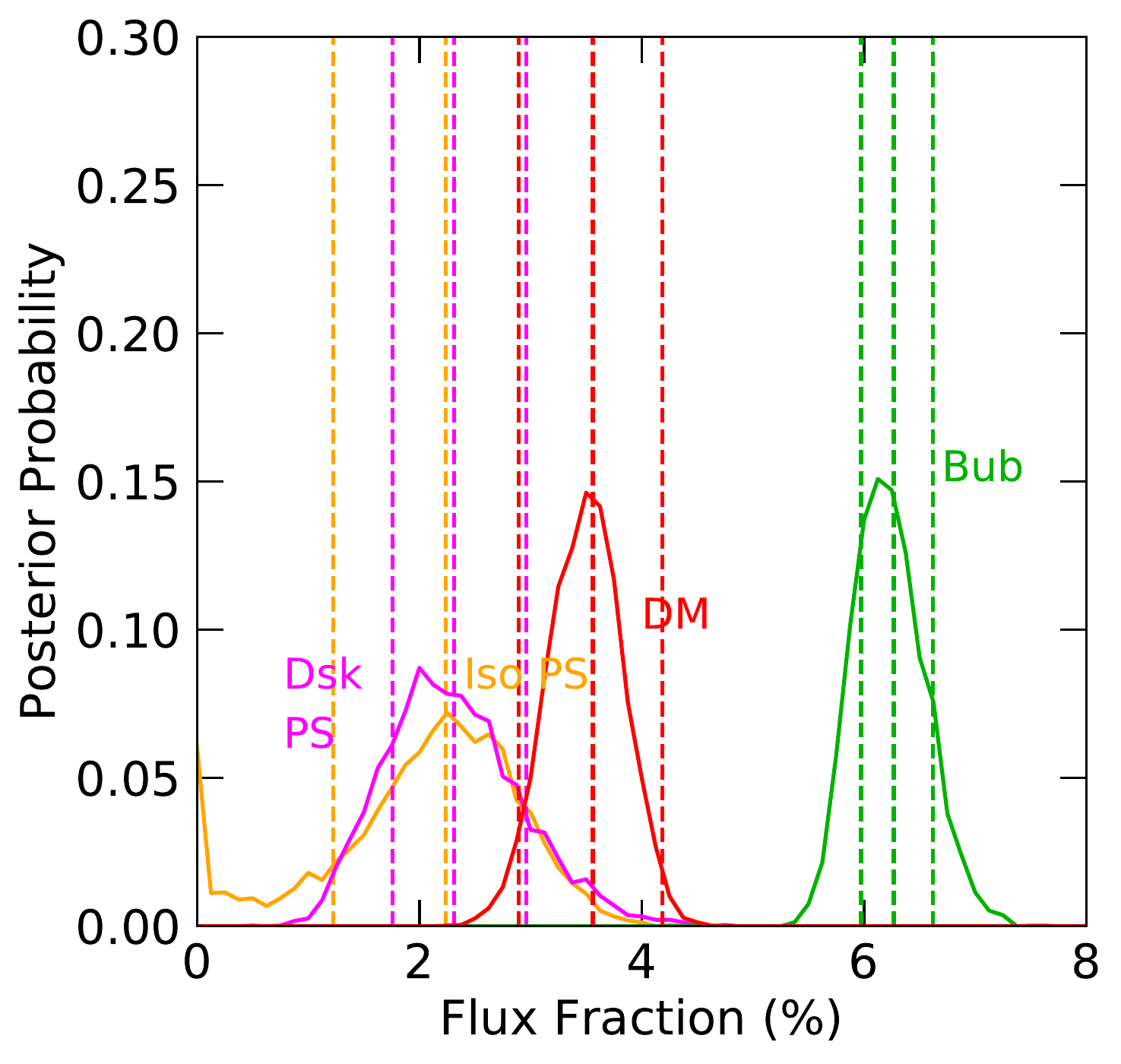}}
\hspace{1mm}
\subfigure{\includegraphics[width=0.49\columnwidth]{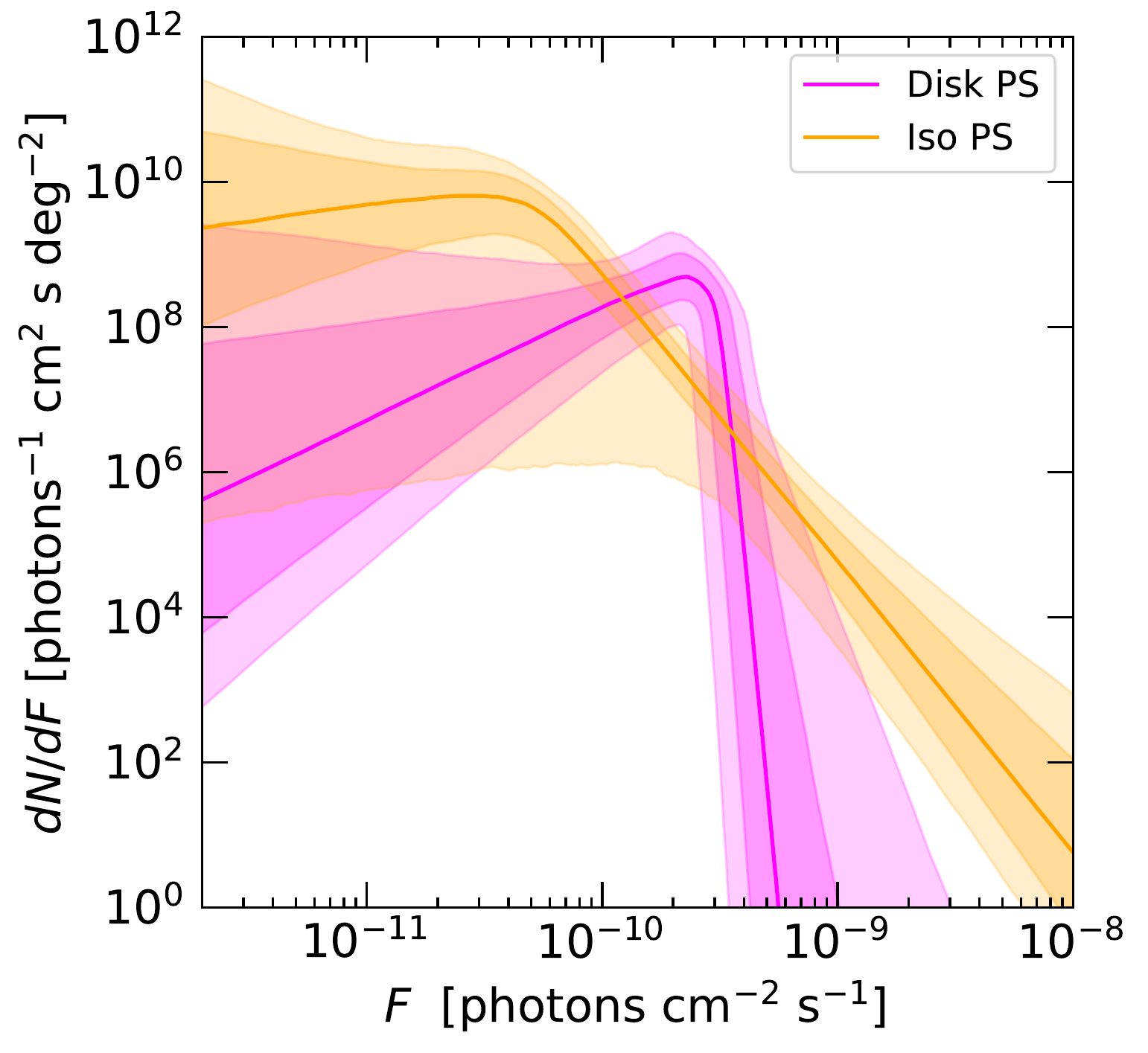}}
\caption{Inner Galaxy (masked) results for the case where an artificial DM signal ($1.8\%$) is injected into the \textit{Fermi} data, and no NFW PSs are included in the analysis. \textbf{Left:} Flux posteriors when analyzed with Disk PS, Isotropic PS, and smooth DM, Bubbles, Isotropic and Galactic Diffuse templates. \textbf{Right:} Luminosity functions for this scenario for Disk PS, and Isotropic PS.}
\label{fig:noPS}
\end{figure*}

Figure~\ref{fig:injscf} shows, as a supplement to the main text, the evolution in the reconstructed source count functions for disk, isotropic and NFW PSs, for no injection as well as 3 different injected DM fluxes (1.8, 6.7, and $15.2\%$ post-injection), in the masked ROI. These results correspond to the flux posteriors shown in Fig.~\ref{fig:bias_data}.

Table~\ref{tab:bayes_data_unmasked} shows the reconstructed DM flux fractions for different combinations of templates, and the relative Bayes factors for these scenarios, in the unmasked ROI, for a single normalization of the injected DM signal (corresponding to an injection of similar size to the GCE). As previously, we see that the DM component is reconstructed to a value consistent with zero and inconsistent with the injected normalization. The scenario where the DM normalization is floated is preferred over the case where it is fixed to the injected value, and also strongly preferred over the case with no NFW PSs. This behavior is again qualitatively consistent with the results we see in the proof-of-principle example using simulated data. 

Figure~\ref{fig:injectDM1_unmasked} shows the posterior probability distributions for the template flux fractions in the case of a $1.8\%$ DM injection when the 3FGL are unmasked, as well as the source count functions for the reconstructed PS populations. 

In Figure~\ref{fig:noPS}, we show for purposes of comparison the recovered flux fractions and intensities when the \textit{Fermi} data, plus an injected DM signal (with normalization equal to the best-fit DM signal under the baseline pipeline), is analyzed with the baseline NPTF pipeline. In this case, there is no NFW PS template available to absorb the GCE, and both the original GCE and the injected signal are attributed to the DM template, which rises in normalization accordingly.

Note to ensure we are not just observing an extreme statistical fluctuation in the simulated DM signal injected into the real data, for the $1.8\%$ injection we also generate 100 sets of simulated DM signals, which are each independently redrawn from a Poisson distribution. We then analyze 100 independent datasets of the \textit{Fermi} data injected with the independent simulated DM signals. We find that the recovered DM flux is always significantly below the true injected value, and is consistent with the values presented in Tab.~\ref{tab:bayes_data}. As such, the DM fluxes recovered and Bayes factors as shown are representative values. (Note the containment bands for Bayes factors for all analyses in Tab.~\ref{tab:bayes_data}, based on 20 sets of independent simulations and analyses, are shown in Fig.~\ref{fig:bayes_compare}).

We also check the impact of injecting a DM signal into an analysis where NFW PSs are interchanged in the fit with Boxy Bulge PSs, to test the possibility that the behavior of the real data under DM injection could simply be due to a mismodeling of PSs associated with the Boxy Bulge. We find that the Boxy Bulge PS template behaves very similarly to the NFW PS template; in particular, it similarly absorbs the injected DM flux.

\subsection{Dark Matter Injection into a Simulated Version of the Best-Fit to the \emph{Fermi} Data}

\begin{table*}
\renewcommand{\arraystretch}{1.3}
\centering
\begin{tabular}{|c|P{2.1cm}|P{6.2cm}|P{3cm}|P{1.0cm}|P{1.0cm}|P{1.0cm}|}
\hline
 \multicolumn{7}{|c|}{\textsc{Simulated Data, 3FGL Masked}} \\
\hline
 \multirow{2}{*}{\textbf{Simulation}}& \textbf{Injected} &  \multirow{2}{*}{\textbf{Analysis Templates}} &  \textbf{DM Flux}                          &  \multicolumn{3}{c|}{\multirow{2}{*}{\textbf{Bayes Factor}} } \\
 & \textbf{DM Flux}   &                                               &   \textbf{(95\%)}                     &     \multicolumn{3}{c|}{}                                 \\ \hline\hline

 & \multirow{4}{*}{None} & Disk PS + Iso PS               &     \multirow{2}{*}{ $[ \, 0.4, 1.9 \,]\  \%$}      &   \multicolumn{3}{c|}{ }          \\ 
                    & & Diffuse + Iso + Bub + DM    &                                      &      \multicolumn{3}{c|}{}        \\ \cline{3-4}
                    & & Disk PS + Iso PS + NFW PS         &      \multirow{2}{*}{ $\textcolor{violet}{[ \, \textbf{0.0, 0.3} \,]\  \%}$}     &      \multicolumn{3}{c|}{\multirow{2}{*}{$8\times10^{6}$}}          \\ 
                    & & Diffuse + Iso + Bub + DM    &                                      &     \multicolumn{3}{c|}{}         \\ \cline{2-7}\cline{2-7}
 
  & \multirow{6}{*}{$1.8 \%$}      &Disk PS + Iso PS                  &     \multirow{2}{*}{ $[ \, 2.4, 3.9 \,]\  \%$}    &      &  \cellcolor{black!25}         &          \\ 
 Disk PS                                &  &Diffuse + Iso + Bub + DM   &                                    &                                &   \cellcolor{black!25}        &         \\ \cline{3-4} \cline{6-7}                     
Iso PS                                &  &Disk PS + Iso PS + NFW PS         &   \multirow{2}{*}   { ${\textcolor{violet}{[ \, \textbf{0.0, 2.1} \,]\  \%}}$}   &        \multirow{2}{*}{$1\times10^{6}$ }                          &     \multirow{2}{*}{$1$}    &  \cellcolor{black!25} \\ 
 NFW PS                                &  &Diffuse + Iso + Bub + DM    &                                   &                                  &           &   \cellcolor{black!25}    \\ \cline{3-5} \cline{7-7} 
   Diffuse                             & & Disk PS + Iso PS + NFW PS         &      Fixed at injection         &        \cellcolor{black!25}      &             &    \multirow{2}{*}{$1\times10^{6}$}  \\ 
   Iso                            &  &Diffuse + Iso + Bub + Fixed DM    &       value ($1.8 \%$)                            &          \cellcolor{black!25}    &           &   \\ \cline{2-7}\cline{2-7}
 Bub  & \multirow{4}{*}{$6.7 \%$}  & Disk PS + Iso PS   &     \multirow{2}{*}{ $[ \, 7.8, 9.2 \,]\  \%$}                   &   \multicolumn{3}{c|}{ }           \\ 
                &  &Diffuse + Iso + Bub + DM   &                         &      \multicolumn{3}{c|}{}        \\ \cline{3-4}
                    &  &Disk PS + Iso PS + NFW PS   &      \multirow{2}{*}{ ${\textcolor{violet}{[ \, \textbf{5.0,7.6}\,]\  \%}}$}                   &     \multicolumn{3}{c|}{\multirow{2}{*}{$8\times10^{5}$}}         \\ 
                    &  &Diffuse + Iso + Bub + DM   &                        &      \multicolumn{3}{c|}{}       \\ \cline{2-7}\cline{2-7}

 & \multirow{4}{*}{$15.2 \%$}   &Disk PS + Iso PS   &     \multirow{2}{*}{$[ \, 12.8, 15.4 \,]\  \%$}                   &  \multicolumn{3}{c|}{ }         \\ 
                    &  &Diffuse + Iso + Bub + DM   &                         &       \multicolumn{3}{c|}{}       \\ \cline{3-4}
                    &  &Disk PS + Iso PS + NFW PS   &     \multirow{2}{*} {${\textcolor{violet}{[ \, \textbf{10.0, 16.5} \,]\  \%}}$}                   &   \multicolumn{3}{c|}{\multirow{2}{*}{$1\times10^{5}$}}           \\ 
                    &  &Diffuse + Iso + Bub + DM   &                        &       \multicolumn{3}{c|}{}      \\ \hline
\end{tabular}
\caption{Bayes factors for analyses of simulated data injected with a DM signal. This is the simulated data parallel of Tab.~\ref{tab:bayes_data}, assuming the best-fit contains no DM, and injecting DM on top of the best-fit. In this case, the DM flux shows distinctly different behavior to the real data case when the standard PS templates are considered (shown in purple); while it is still not always correctly reconstructed, the discrepancy is much smaller than in the real data. The data analysis that is favored has the relative Bayes factor appear next to the analysis row, where the two analyses that are being compared are not grayed out.}
\label{tab:sim_bayes_data}
\end{table*}

\begin{figure*}[t!]
\leavevmode
\centering
\subfigure{\includegraphics[width=0.48\columnwidth]{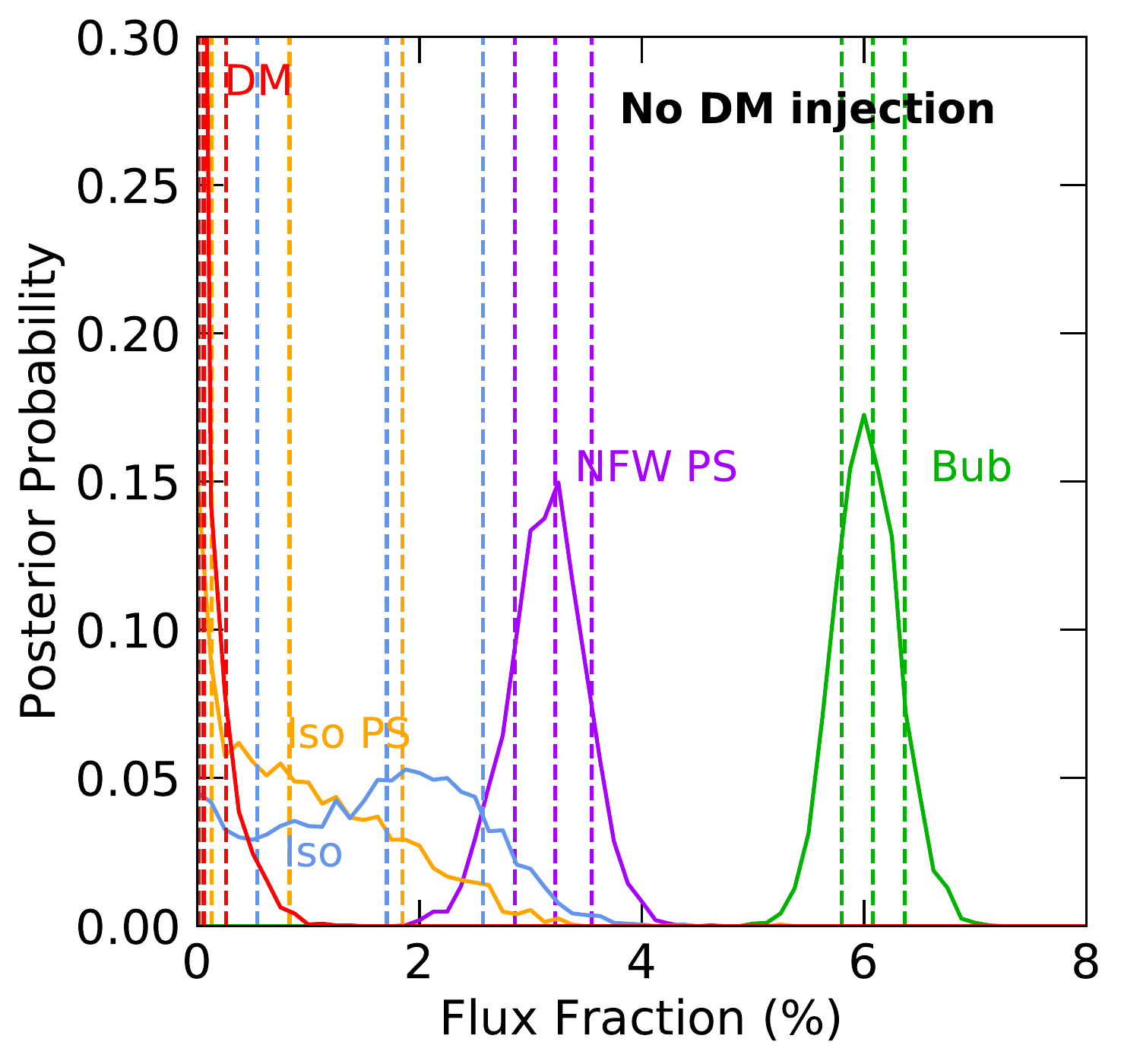}}
\hspace{1mm}
\subfigure{\includegraphics[width=0.48\columnwidth]{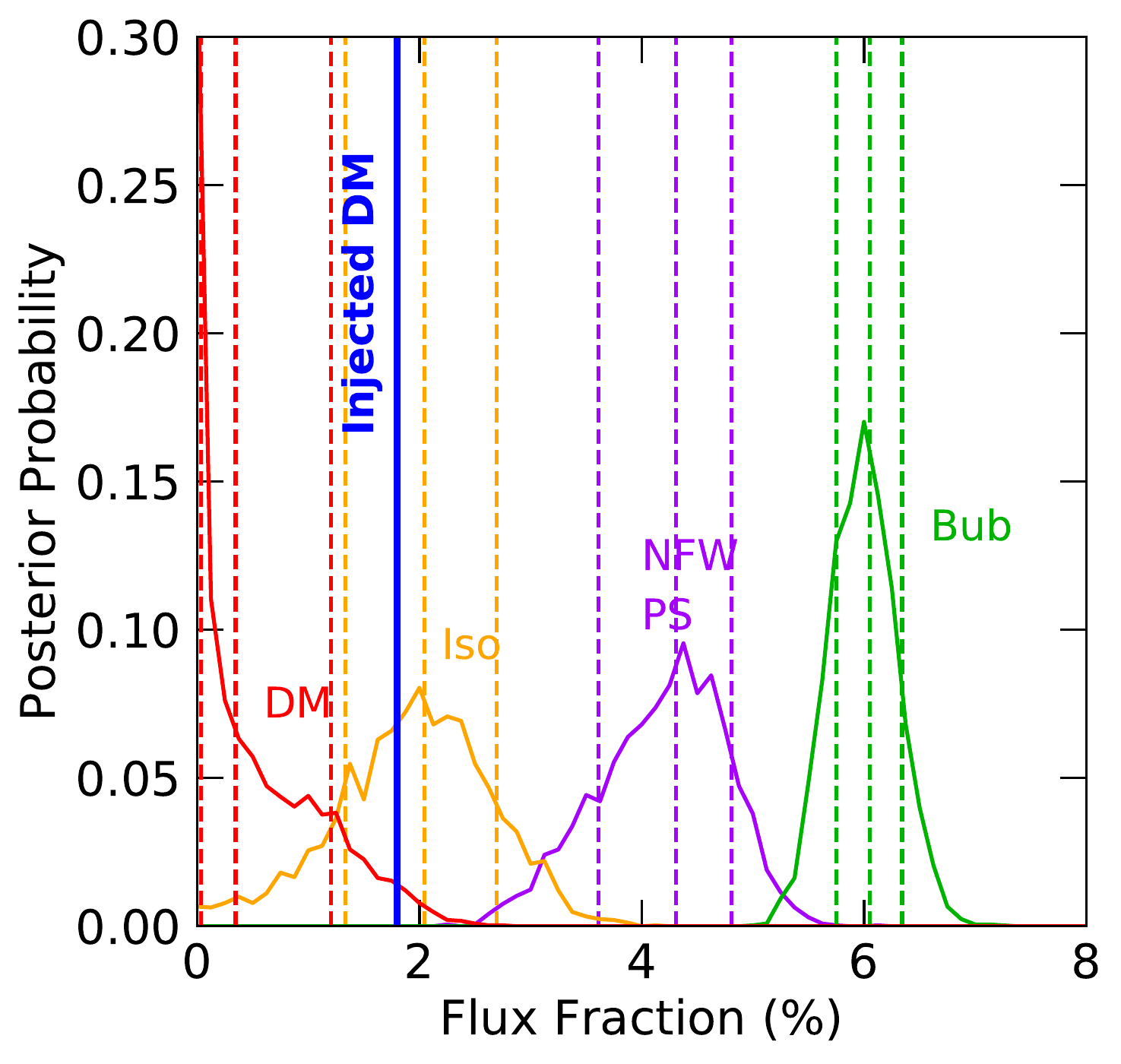}}
\\
\subfigure{\includegraphics[width=0.48\columnwidth]{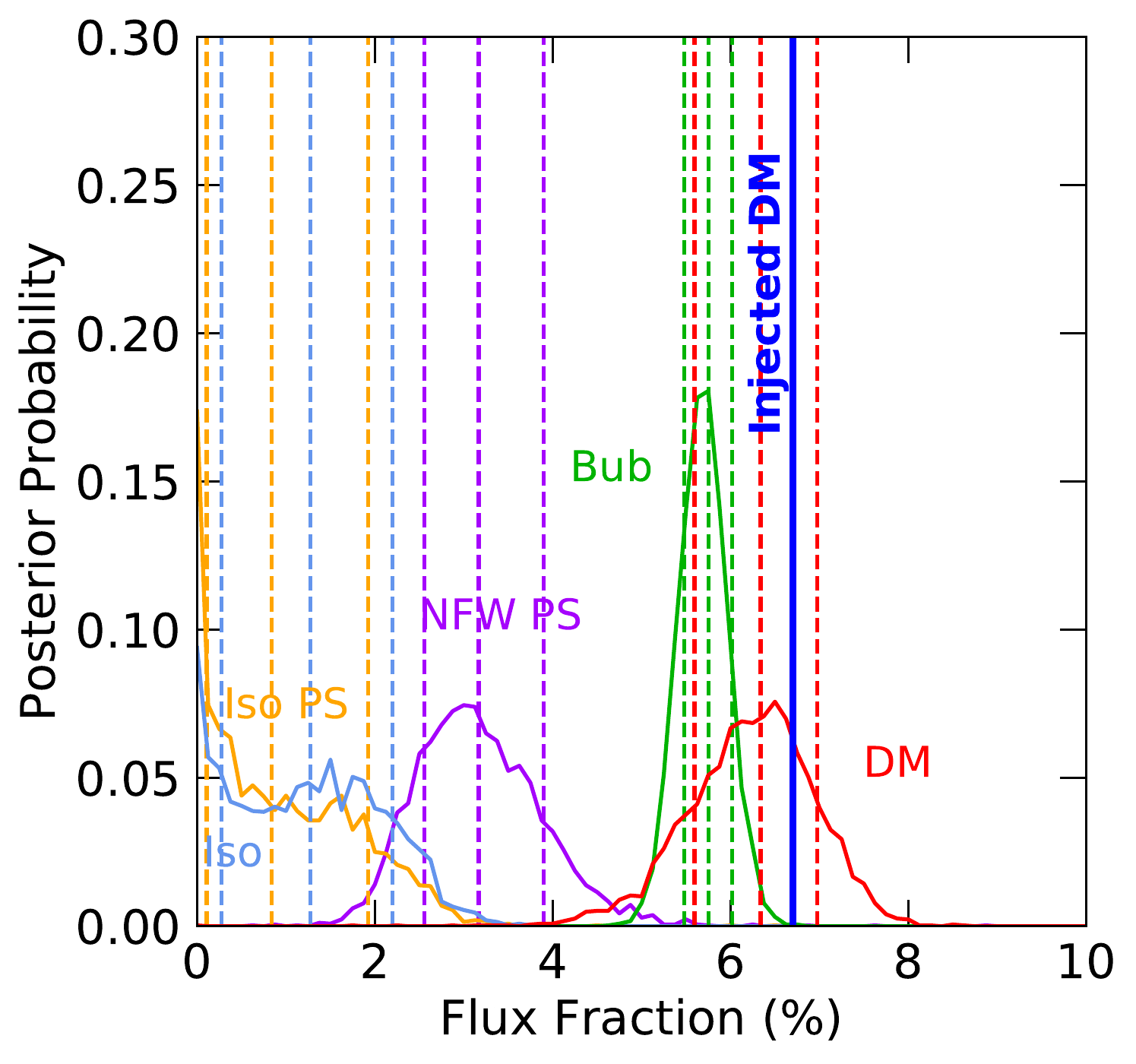}}
\hspace{1mm}
\subfigure{\includegraphics[width=0.48\columnwidth]{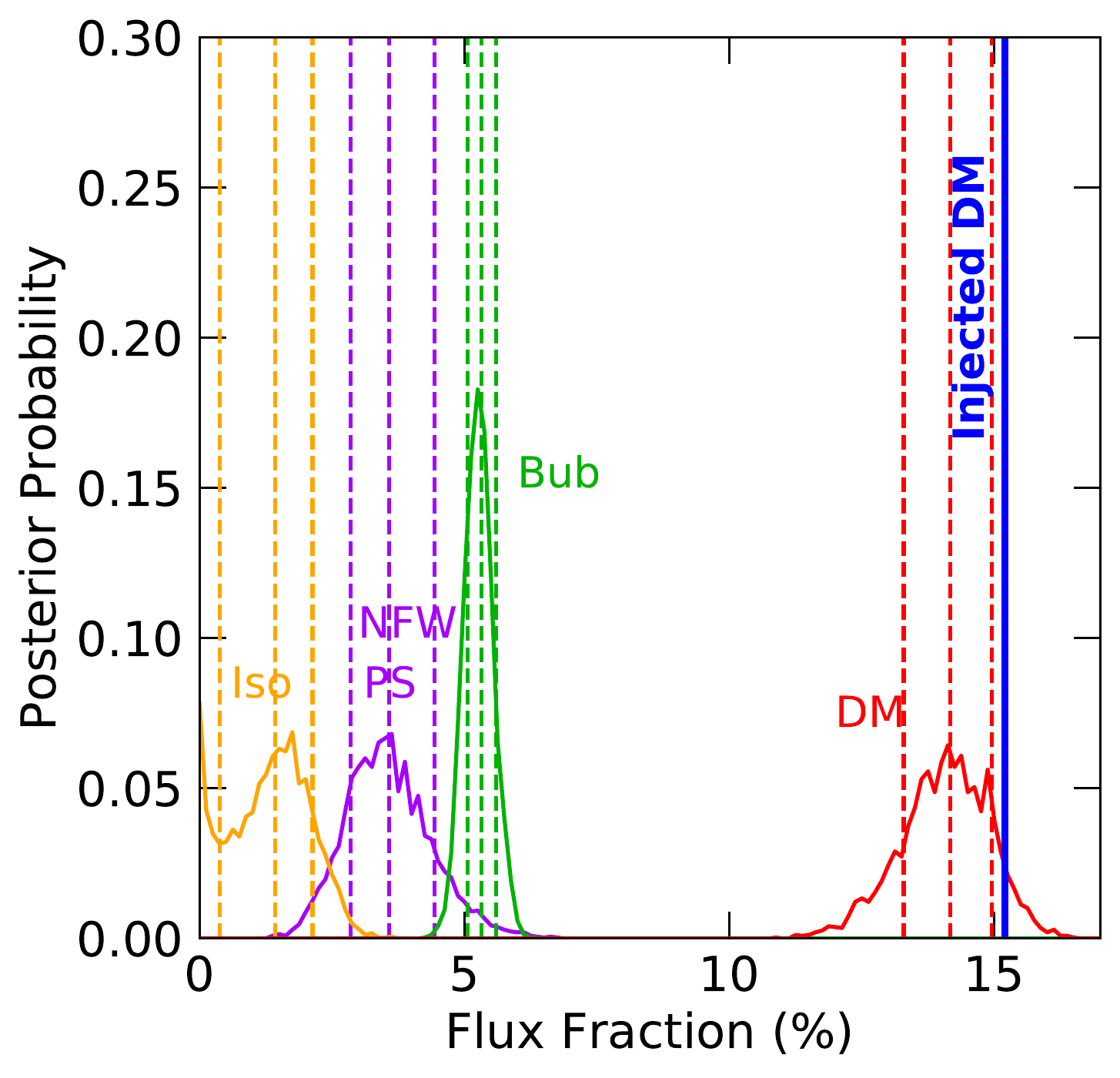}}
\caption{Simulated data comparison for Fig.~\ref{fig:bias_data}, assuming the standard pipeline templates truly describes the data, and injecting the same DM signals on top of the simulated standard pipeline templates. All templates are present, but those peaked at fluxes below $0.1\%$ (except for DM) are not shown for clarity.}
\label{fig:bias_data_nfw}
\end{figure*}

We also consider the case where the best fit values using the standard pipeline on the real data with no injection are used to produce simulated data. We then inject this simulated data with the exact same DM normalizations as Fig.~\ref{fig:bias_data}, and then analyze this data with the exact same templates as simulated. This serves to highlight that the data do not appear to be accurately described by the standard templates -- in this scenario fluxes are approximately recovered within uncertainties.
Specifically, of the 20 realizations, we observe for $1.8\%$ flux injection that 7 out of 20 simulated realizations recover the correct flux within the $68\%$ containment bands, 14 of 20 for $6.7\%$ injection, and 14 out of 20 for $15.2\%$ injection. This is approximately consistent with expectations, although the reconstruction appears to be biased slightly low for the $1.8\%$ injection case. Importantly, in the case where the DM injection is comparable to the GCE itself, there is no statistical preference for the scenario where the DM contribution is allowed to float (and is reconstructed at a low value) over the scenario where the DM contribution is fixed to its injected value, unlike what is found with the injection test on real data.

In the case with no DM injection, the Bayes factor is similar to what we observe in our proof-of-principle example -- although this similarity should be viewed with some caution, as for simplicity, our simulated data in the (masked) proof-of-principle case did not include the isotropic PS population. Nonetheless, this suggests that it could be difficult to distinguish between a case similar to our proof-of-principle scenario, where a real DM signal is incorrectly reconstructed as PSs due to the presence of an unmodeled PS population, and the case where the GCE is entirely PSs, solely via examination of the Bayes factor in favor of PSs. This contrasts with the case where the GCE receives a $50\%$ contribution from DM and $50\%$ from PSs, studied in Ref.~\cite{Lee:2015fea}, where it was found that while the GCE could be incorrectly reconstructed as originating solely from PSs, the Bayes factor in favor of PSs was generically much smaller.

Figure~\ref{fig:bias_data_nfw} shows the recovered flux posteriors in this scenario, in one particular representative realization. Compared to Fig.~\ref{fig:bias_data}, we observe that the injected DM and other simulated templates are much closer to being successfully recovered.

Table~\ref{tab:sim_bayes_data} details the reconstructed DM flux fractions for different combinations of templates, and the relative Bayes factors for these scenarios, in the same realization.

\begin{figure*}[t!]
\leavevmode
\centering
\subfigure{\includegraphics[width=0.48\columnwidth]{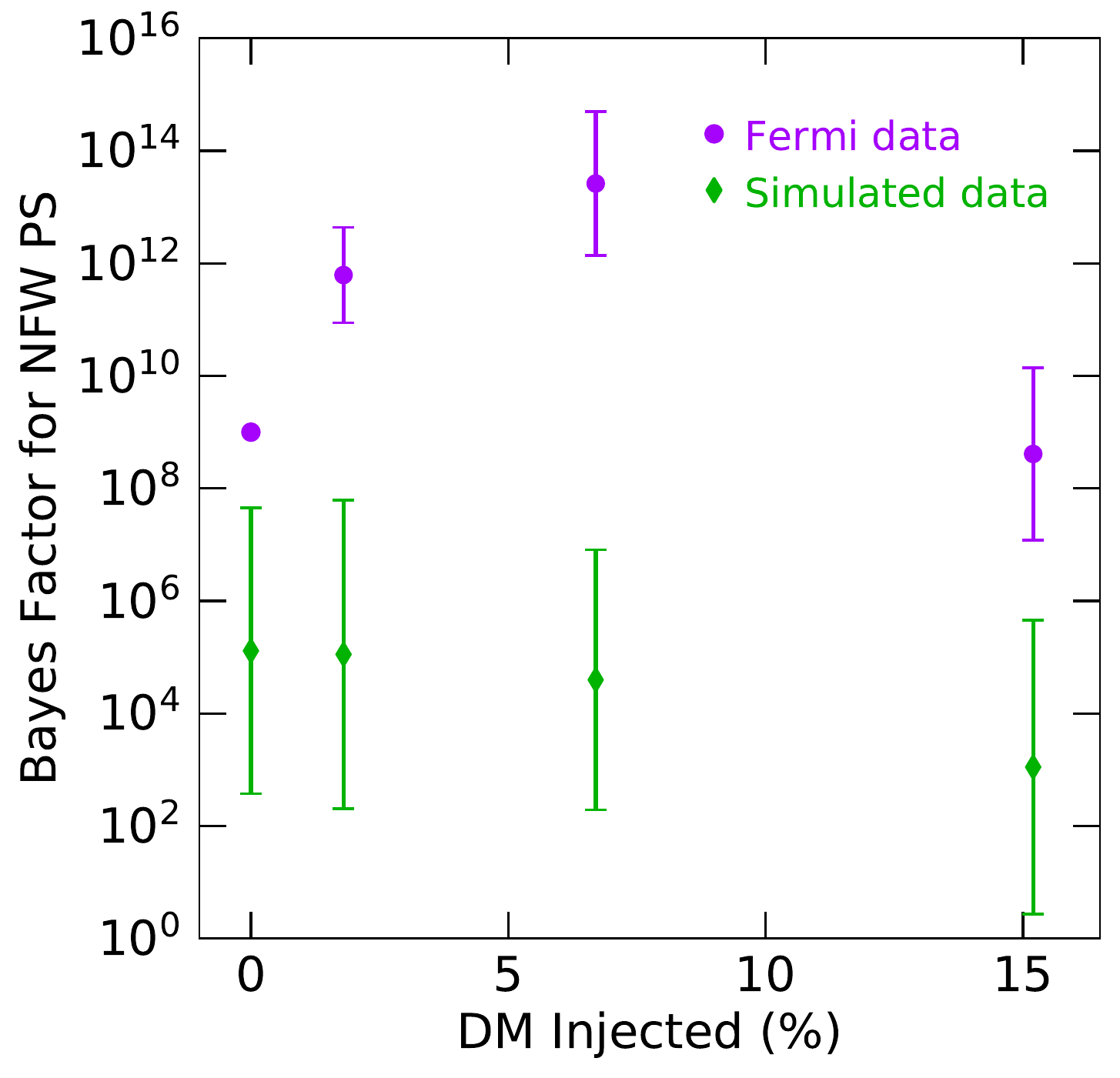}}
\hspace{10mm}
\subfigure{\includegraphics[width=0.228\columnwidth]{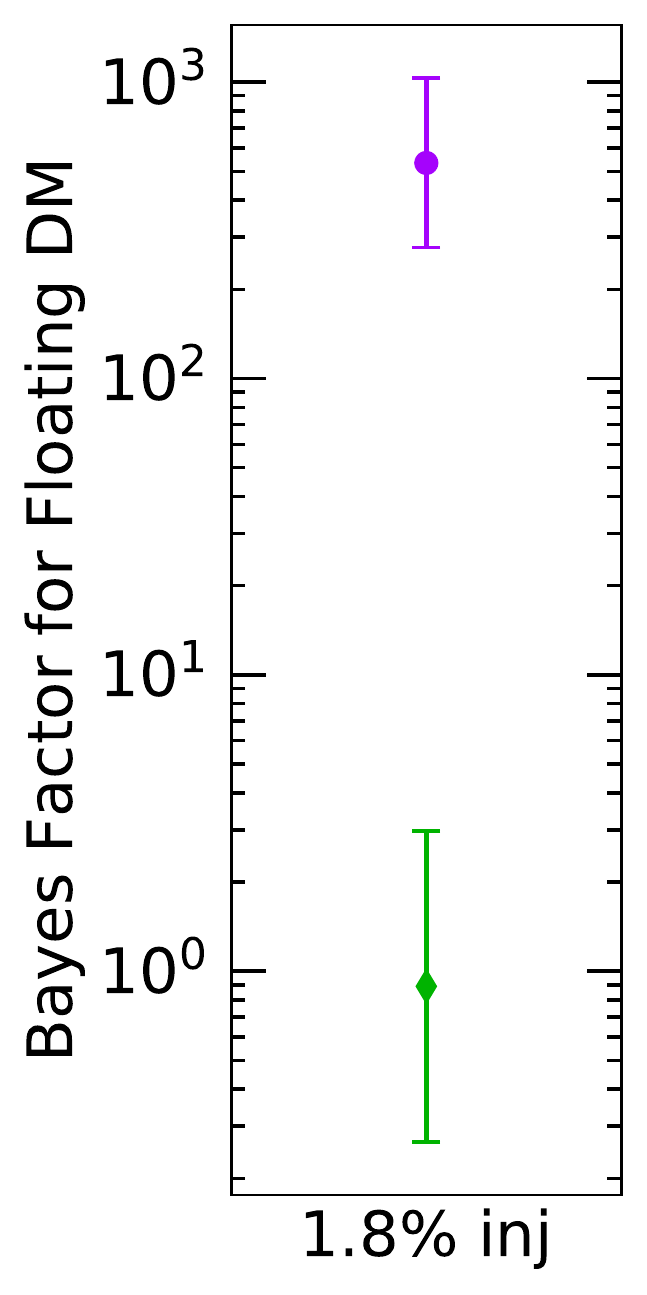}}
\caption{\textbf{Left:} Bayes factors in favor of the standard pipeline (adding NFW PS) with and without DM injection, for both the real \textit{Fermi} data (purple) and simulated data (green) which assumes the standard pipeline templates truly describe the data. \textbf{Right:} Bayes factor in favor of floating the DM template rather than fixing it to its true value (in this case, $1.8\%$ flux injection), when in both cases NFW PS are present. All error bars display $\pm 1$ standard deviation in log(Bayes factor).}
\label{fig:bayes_compare}
\end{figure*}

Figure~\ref{fig:bayes_compare} presents the difference in the spread of Bayes factors between the simulated data based on the best-fit model with the standard pipeline, and the real data with and without injection. This again highlights the data do not appear to be described accurately by the standard templates.

\subsection{Increasing the Number of Breaks in the NFW Source-Count Function}

Very faint PSs approach the limit of statistically behaving just like DM: their probability to emit $n$ photons in a given pixel becomes effectively zero for $n > 1$, as the probability to emit observed photons at all is so low. As such, it is conceivable that the reason the DM flux is not being recovered is because it is being injected into the soft end of an overly-constrained singly-broken power law SCF -- for example, the behavior of the SCF might be almost entirely dictated by the bright sources which are near the single break. As such, we test the effect of enforcing a steep slope right below the 1-photon threshold, as well as allowing a break around this threshold (prior range of the additional break taken as $S_b = [0.1,1]$), with a slope below this threshold that is allowed to float. 

\begin{figure*}[t!]
\leavevmode
\centering
\subfigure{\includegraphics[width=0.48\columnwidth]{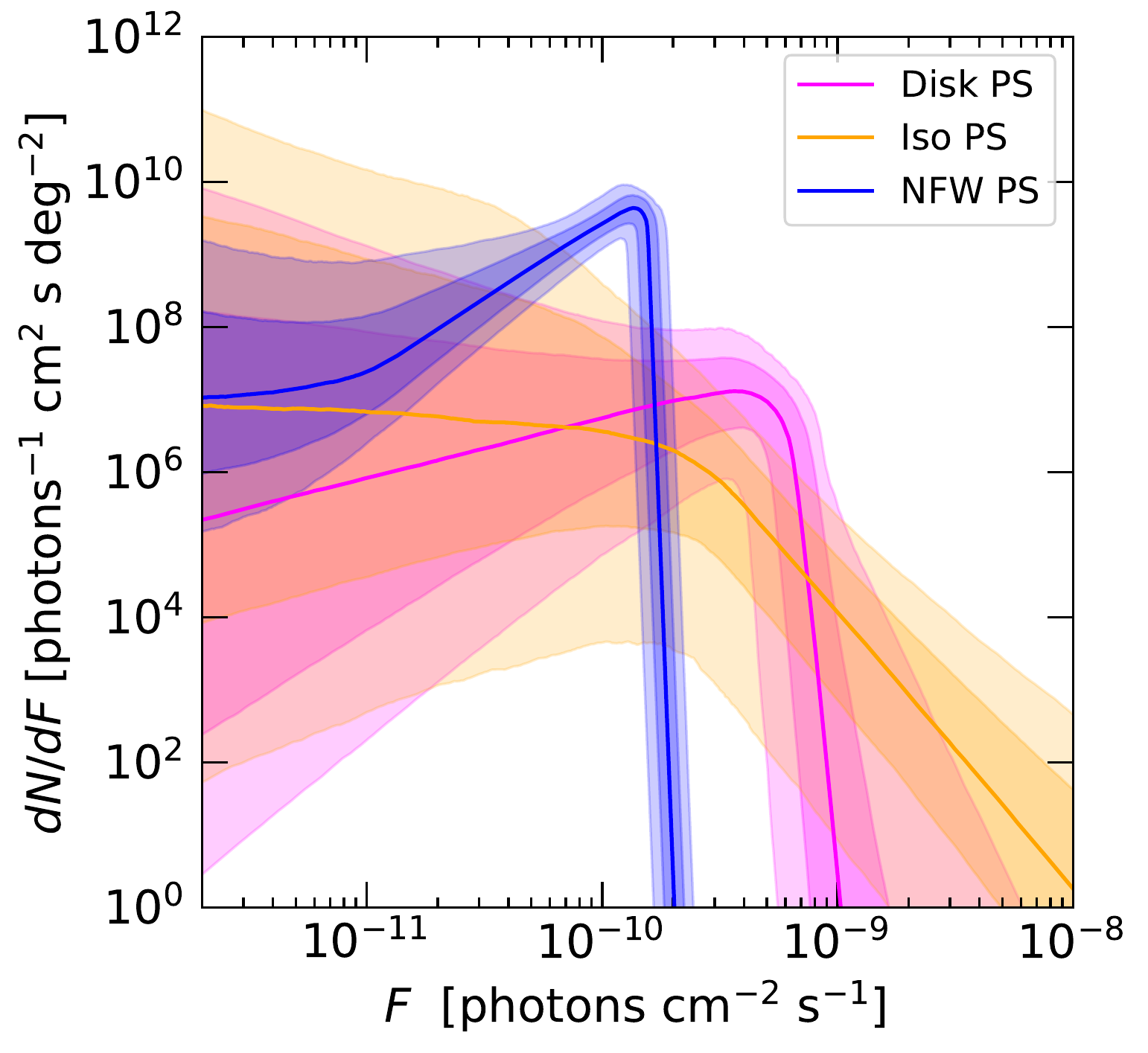}}
\hspace{1mm}
\subfigure{\includegraphics[width=0.48\columnwidth]{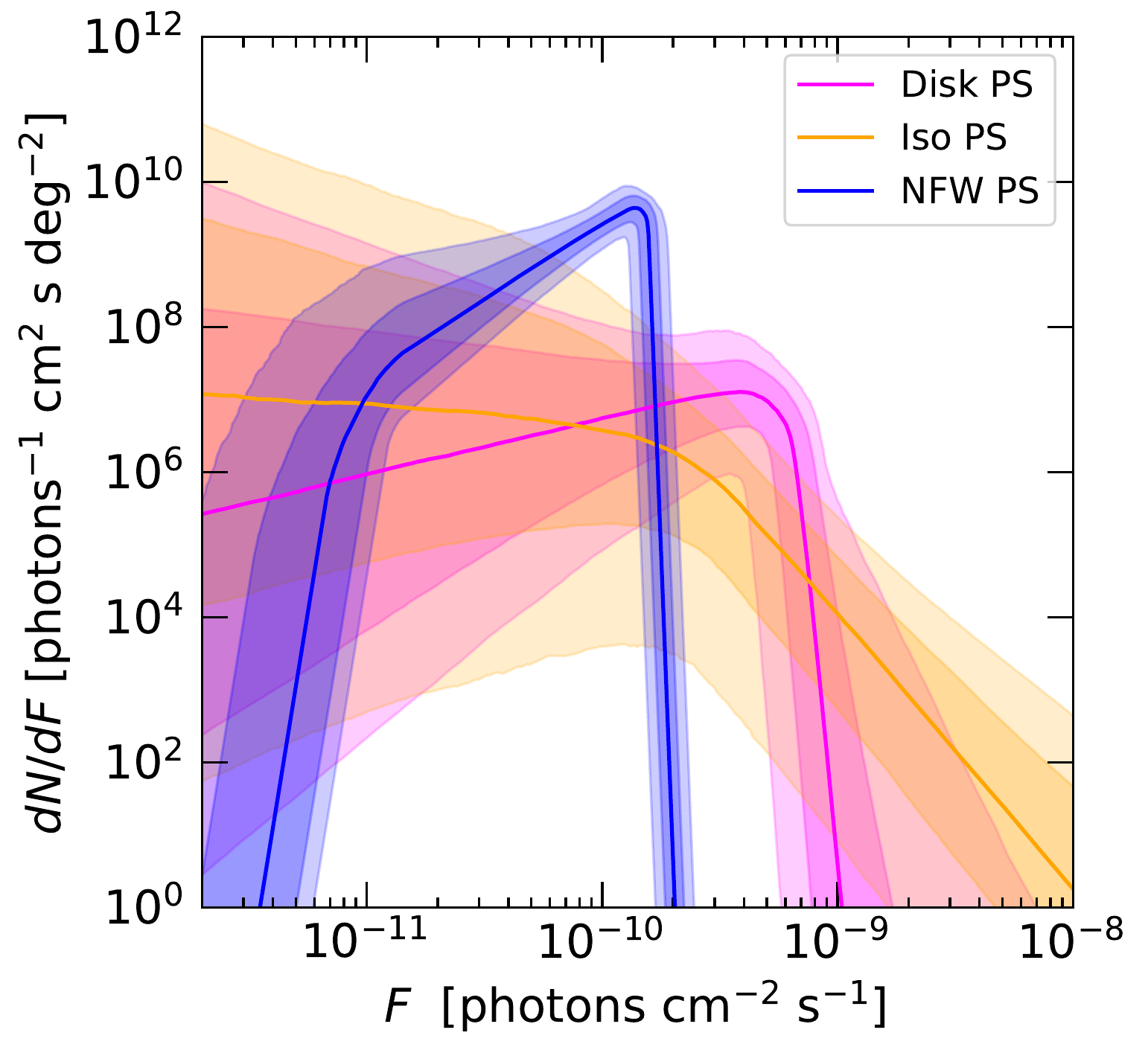}}
\caption{SCF with an additional break in the NFW SCF around the 1-photon threshold. The 3FGL are masked. Neither of these changes impact the results of incorrectly reconstructed DM. \textbf{Left:} The slope below the new break is allowed to float. \textbf{Right:} The slope below the new break is fixed to be very steep.}
\label{fig:scfbreaks}
\end{figure*}

Figure~\ref{fig:scfbreaks} shows how these modifications change the recovered SCF for the case with 3FGL PSs masked and a 1.8\% DM flux injected into the data (though we have also checked we recover the same behavior unmasked).

We find no evidence that such a break is preferred by the data; whether the slope below the break is floated or forced to be steep, we find the change in log evidence relative to the baseline case is within the error bars of the log-evidence, and the flux posteriors for the various components do not change.

\section{Preference For the Dark Matter Normalization to Float Negative}

\begin{figure*}[h]
\leavevmode
\centering
\subfigure{\includegraphics[width=0.48\columnwidth]{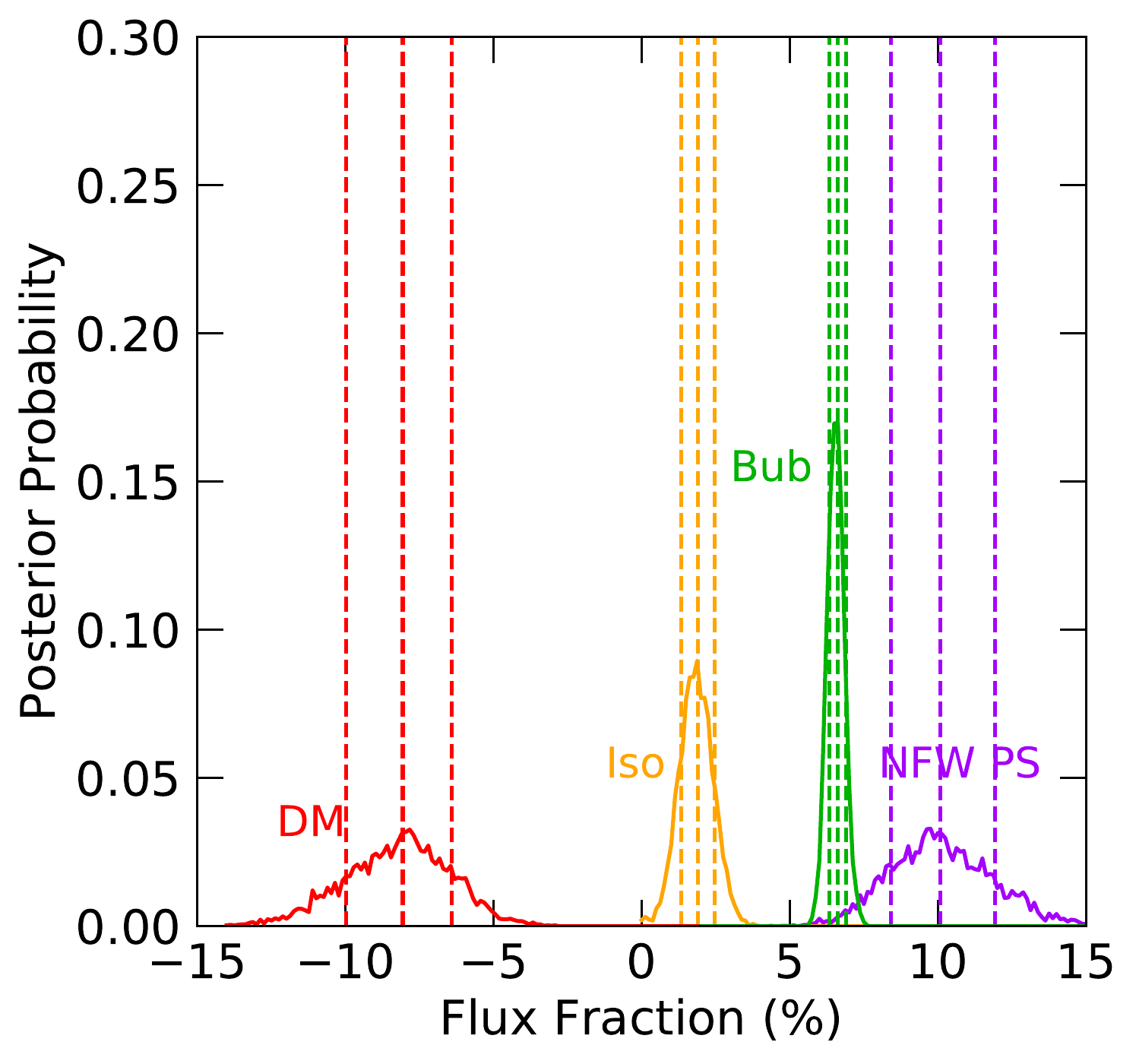}}
\hspace{1mm}
\subfigure{\includegraphics[width=0.48\columnwidth]{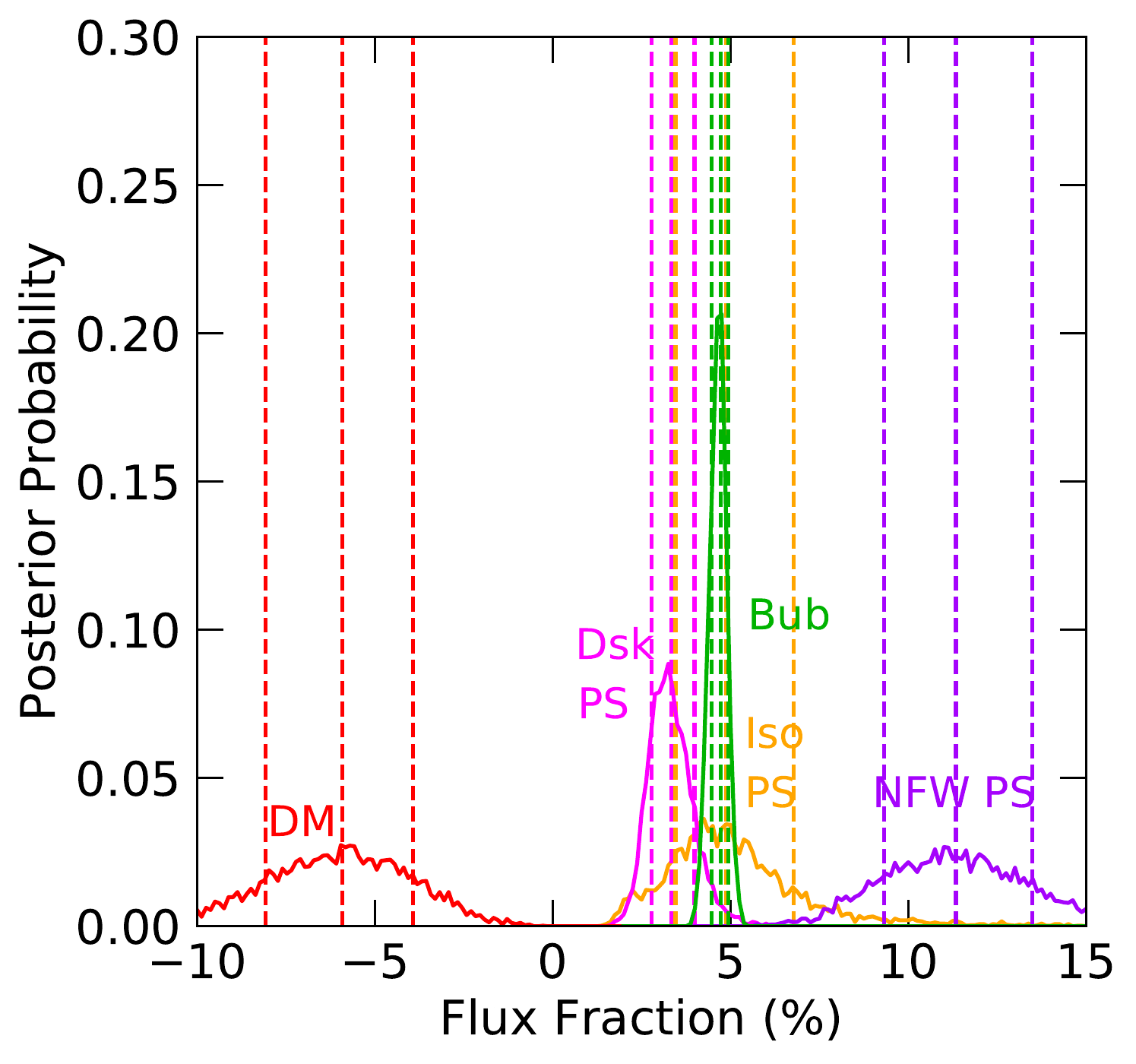}}
\\
\subfigure{\includegraphics[width=0.48\columnwidth]{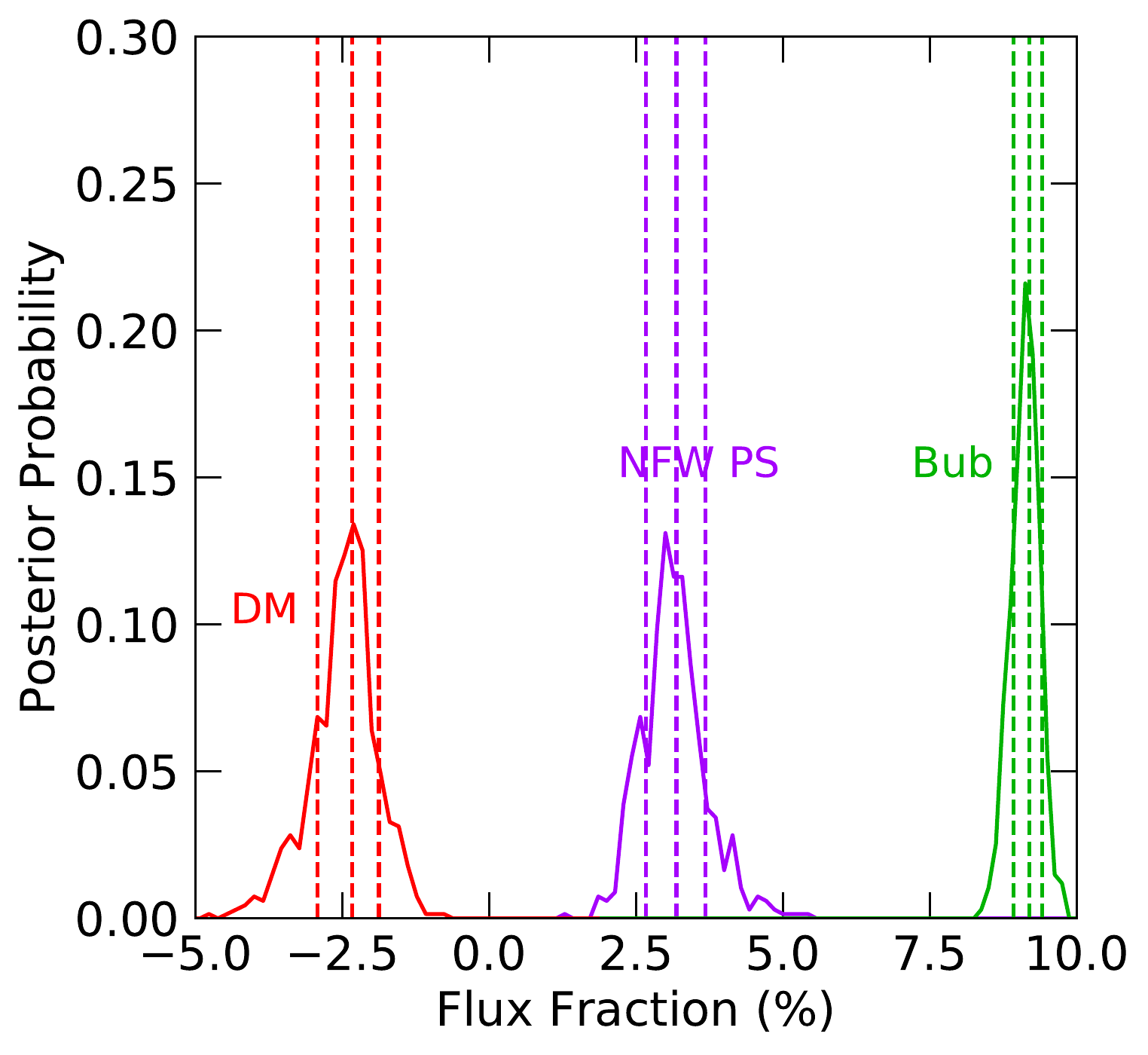}}
\hspace{1mm}
\subfigure{\includegraphics[width=0.48\columnwidth]{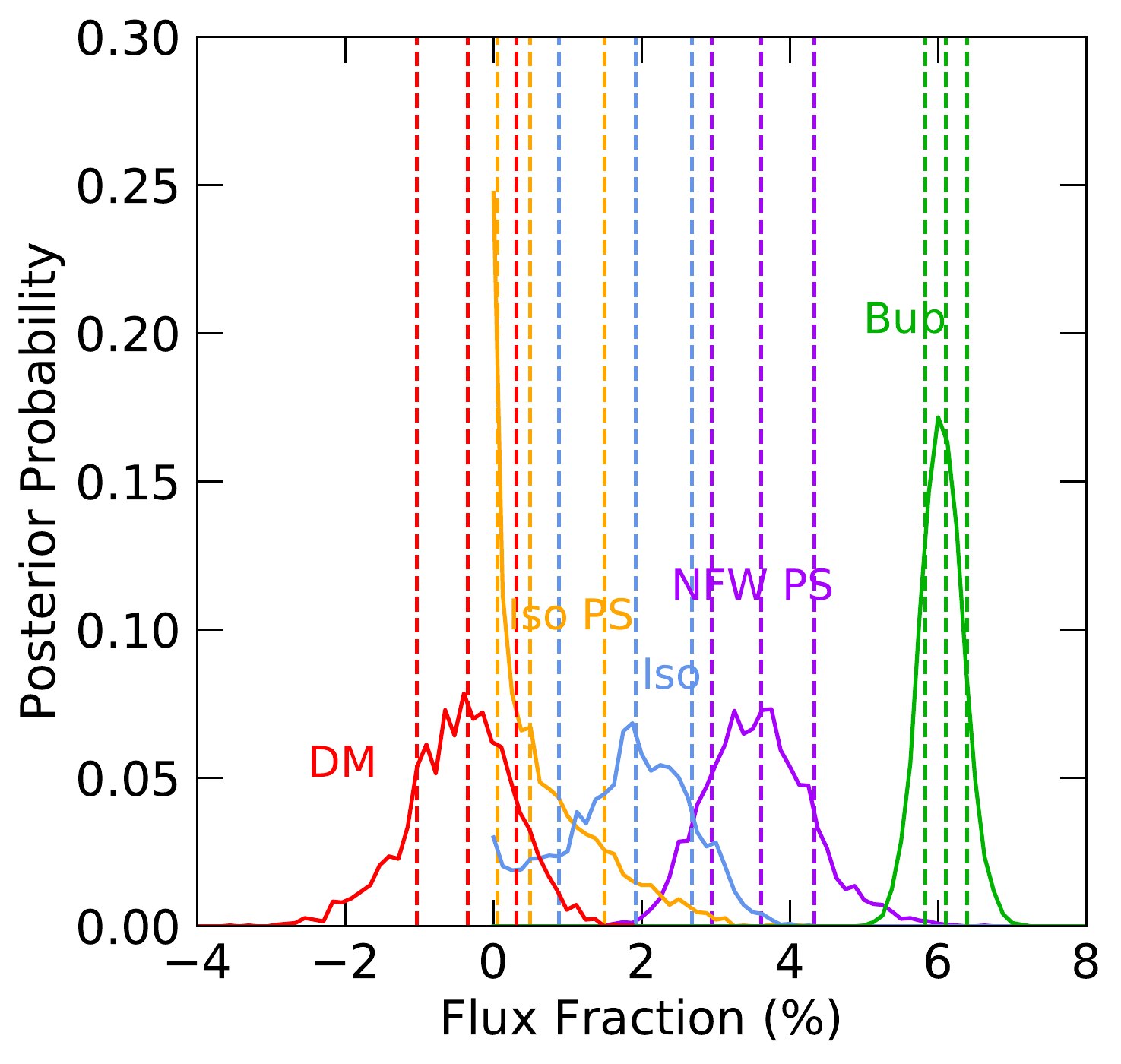}}
\caption{Effects of allowing the DM normalization posterior to float negative. \textbf{Top-Left:} No injection, analyzing the real data, with the \textit{Fermi} \texttt{p6v11} diffuse model. \textbf{Top-Right:} No injection, analyzing the real data, with diffuse Model A. \textbf{Bottom-Left:} Simulated data, using the \textit{Fermi} \texttt{p6v11} diffuse model, in the proof-of-principle case where PSs have been simulated in the \textit{Fermi} Bubbles, but the Bubbles PS template has been replaced with the NFW PS template in the analysis. Similarly to the real data, there is a preference for a negative DM coefficient. \textbf{Bottom-Right:}  Simulated data, using the \textit{Fermi} \texttt{p6v11} diffuse model, in the proof-of-principle case where NFW PSs have been simulated, and the same templates are used at the analysis stage. Over 20 simulated realizations, we observe some bias toward negative DM coefficients (see text), but to a much lesser degree than in the scenario where unmodeled PSs are present.}
\label{fig:floatneg}
\end{figure*}

Prior to this section, we have always forced the normalizations of all templates to be positive, since they represent physical emission mechanisms; we will retain this convention unless explicitly noted otherwise. However, as an alternative way to understand the behavior of the analysis when additional DM is injected, we can expand the prior on the coefficient of the DM template so that it includes negative values. This is clearly not a realistic scenario, but can indicate the presence of systematic effects which drive down the reconstructed DM component. Analyzing the real data (with no additional DM injection), with the \texttt{p6v11} Galactic diffuse emission model, we find that the DM component prefers to float to a very negative value when given the option.

Figure~\ref{fig:floatneg} demonstrates this result; the DM component is reconstructed with a negative flux $\sim 4-5\times$ larger in magnitude than the baseline GCE. The NFW PS template rises in compensation, such that the sum of their fluxes remains approximately constant. This provides an intuitive understanding of the results of the injection test; adding an extra DM signal to the data reduces the degree to which the DM component prefers to run negative, but a positive signal will only be reconstructed once enough simulated DM has been added to completely cancel out the preferred negative flux. 

Furthermore, until this threshold is reached, in fits where the DM component is forced to be non-negative, both the DM coefficient and the amount of power in very faint NFW PSs (degenerate with the DM component) will be forced toward zero. Due to the degeneracy between these components, only the sum of their fluxes is well-constrained, but if their sum prefers a negative value and each component is forced to be non-negative, the unique best-fit solution will set both contributions to zero. Consequently, we expect (as observed) both a near-zero reconstructed DM component and a source count function for the NFW PSs that has very little power at low flux. Adding additional freedom to the source-count function at low flux, as studied above, is not expected to change the results of the analysis, as in this case the degeneracy between DM and low-flux PSs has already been broken (by both components being forced to zero). Only once sufficient DM is injected (or when systematic factors causing the oversubtraction of the smooth component are corrected), and the total preferred power in DM and low-flux sources becomes positive, should the degeneracy between the two components become a potentially important systematic issue.

We can also ask whether this behavior, where the DM coefficient becomes negative, is seen in our proof-of-principle case. We show the results of this test in the bottom left panel of Fig.~\ref{fig:floatneg}, and the DM coefficient indeed prefers a negative value, albeit not to the same degree seen in the real data. In the proof-of-principle example, the DM component prefers a negative flux of magnitude similar to the GCE (rather than several times larger as in the real data); this is consistent with the results of the injection test shown in the main text.  

Finally, we can also test the degree to which the DM coefficient floats negative when we simulate data using the same templates employed in the fit; the results are shown in the bottom right panel of Fig.~\ref{fig:floatneg}. In this case the DM coefficient can still float negative in some realizations, but not to the same extent seen in either the proof-of-principle case or (to a much greater degree) the real data. Specifically, we observe that in 10 out of 20 simulated realizations, the $68\%$ containment bands on the posterior overlap the true DM flux of zero. In the remaining realizations, the upper 68$\%$ containment band on the flux fraction is less than zero, but greater than or equal to $-0.5\%$. In all cases, the degree to which the DM normalization floats negative is representative of how much DM flux needs to be added before DM is recovered in all the analyses.

In all cases, when allowing the DM normalization to float to negative values, all prior ranges remain the same, but the DM normalization prior range is taken to be linear flat, with $A_{\rm NFW}=[-9,9]$.

\section{Varying diffuse background models}
\label{sec:diffuse}

\begin{table*}[t!]
\renewcommand{\arraystretch}{1.5}
\setlength{\tabcolsep}{5.2pt}
\begin{center}
\begin{tabular}{ c  P{3.4cm}  P{3.4cm} P{3.4cm}  P{3.4cm} }
\hline
\multicolumn{5}{c}{\textsc{Prior Ranges, Inner Galaxy, Diffuse Models A and F}}\Tstrut\Bstrut		\\  
\hline Parameter 	&  Model A (masked) & Model A (unmasked) &  Model F (masked) & Model F (unmasked)	\Tstrut\Bstrut \\
\hline 
\hline
$\log_{10}A_\text{iso}$  & $[-3,1]$ & $[-3,1]$  & $[-3,1]$ & $[-3,1]$ \Tstrut\Bstrut \\ 
$\log_{10}A_\text{dif}^\text{ICS}$  & $[-2,1]$ & $[-2,1]$  & $[-2,1]$ & $[-2,1]$  \Tstrut\Bstrut \\
$\log_{10}A_\text{dif}^{\pi^0\text{brem}}$  & $[-1,2]$ & $[-1,2]$  & $[-1,2]$ & $[-1,2]$  \Tstrut\Bstrut \\
$\log_{10}A_\text{bub}$  & $[-3, 1]$ & $[-3, 1]$ & $[-3, 1]$ & $[-3, 1]$  \Tstrut\Bstrut \\ 
$\log_{10} A_\text{NFW}$ & $[-3, 1]$ & $[-3, 1]$ & $[-3, 1]$ & $[-3, 1]$  \Tstrut\Bstrut	\\
$\log_{10}A_\text{PS}^\text{NFW}$ & $[-6, 1]$ & $[-6, 1]$  & $[-6, 1]$ & $[-6, 1]$ \Tstrut\\
$S_b^\text{NFW}$  & $[0.05 ,60]$      & $[0.05 ,30]$  & $[5 ,60]$      & $[0.05 ,60]$   \\
$n_1^\text{NFW}$   & $[50, 95]$     & $[5, 45]$ & $[2.05, 60]$     & $[5, 60]$    \\
$n_2^\text{NFW}$   & $[ -3 ,1.95]$    & $[ -3 ,1.95]$  & $[ -3 ,1.95]$    & $[ -3 ,1.95]$ \\
$\log_{10}A_\text{PS}^\text{disk}$ & $[-6, 1]$ & $[-6, 1]$   & $[-6, 1]$ & $[-6, 1]$ \Tstrut\\
$S_b^\text{disk}$ & $[0.05 ,60]$      & $[0.05 ,50]$  & $[0.05 ,60]$      & $[0.05 ,60]$     \\
$n_1^\text{disk}$  & $[25, 80]$     & $[2.05, 5]$  & $[25, 80]$     & $[2.05, 65]$    \\
$n_2^\text{disk}$   & $[ -3 ,1.95]$    & $[ -3 ,1.95]$  & $[ -3 ,1.95]$    & $[ -3 ,1.95]$ \\
$\log_{10}A_\text{PS}^\text{iso}$  & $[-6, 1]$ & $[-6, 1]$  & $[-6, 1]$ & $[-6, 1]$  \Tstrut\\
$S_{b_1}^\text{iso}$  & $[10, 50]$ & $[1,40]$ & $[1,40]$ & $[1,40]$   \\
$S_{b_2}^\text{iso}$  & $[0.05,30]$ & $[0.05,30]$  & $[0.05,30]$ & $[0.05,30]$   \\
$n_1^\text{iso}$ & $[2.05, 5]$& $[2.05, 5]$ & $[2.05, 5]$& $[2.05, 5]$  \\
$n_2^\text{iso}$ & $[1.5, 4.5]$  & $[0.5, 4.5]$ & $[0.5, 4.5]$  & $[0.5, 4.5]$  \\
$n_3^\text{iso}$  & $[-1.95, 1.95]$    & $[-1.95, 1.95]$ & $[-1.95, 1.95]$    & $[-1.95, 1.95]$ \Bstrut\\
\hline
\end{tabular}
\end{center}
\caption{Parameters and associated prior ranges for the Inner Galaxy analyses, when using diffuse models A and F, where the inverse Compton and $\pi^0+$bremsstrahlung contributions are considered independently with their own templates.}
\label{tab:priors_ccw}
\end{table*}

As Galactic diffuse emission is the dominant contribution to the gamma-ray sky and could be a large source of systematic effects, in this section we check that the injection test on the real data behaves similarly to the description in the main text for the alternative diffuse models A and F. These models were identified by Ref.~\cite{Calore:2014xka} as providing either the best fit to the data among a wide range of \texttt{GALPROP}-based models (model F) or a comparable fit with physical values for various floated parameters (model A). Both models are expected to provide a better fit to the gamma-ray sky than our default \texttt{p6v11} diffuse model in the energy band of interest ($2-20$ GeV) \cite{Calore:2014xka}.

Table~\ref{tab:priors_ccw} details the prior ranges used for analyses involving diffuse models A and F. A key difference to the previous prior ranges considered in Tab.~\ref{tab:priors} is that the inverse-Compton and $\pi^0$-bremsstrahlung (gas-correlated) contributions are allowed to float separately, whereas the relative normalizations of the $\pi^0$ and bremsstrahlung contributions are held fixed in the main analysis with the \texttt{p6v11} \textit{Fermi} diffuse model. Note that to speed up the runs, for some parameters we first perform a broad preliminary scan, and then in our final runs we restrict the priors to a narrower range around the converged posterior values. We always use identical priors when directly comparing analyses on real data to those on simulated data.

\begin{table*}[t!]
\renewcommand{\arraystretch}{1.3}
\centering
\begin{tabular}{|P{2.4cm}|P{7.2cm}|P{3cm}|P{1.4cm}|P{1.4cm}|P{1.4cm}|}
\hline
\multicolumn{6}{|c|}{\textsc{Diffuse Model A, Real Data, 3FGL Unmasked}} \\
\hline
 \textbf{Injected} &  \multirow{2}{*}{\textbf{Analysis Templates}} &  \textbf{DM Flux}                          &  \multicolumn{3}{c|}{\multirow{2}{*}{\textbf{Bayes Factor}} } \\
\textbf{DM Flux}   &                                               &   \textbf{(95\%)}                     &     \multicolumn{3}{c|}{}                                 \\ \hline\hline
 
 \multirow{4}{*}{None} &  Disk PS + Iso PS               &     \multirow{2}{*}{ $[ \, 4.0, 5.0 \,]\  \%$}      &   \multicolumn{3}{c|}{ }          \\ 
                    &  Diff IC + Diff $\pi^0$brem + Iso + Bub + DM    &                                      &      \multicolumn{3}{c|}{}        \\ \cline{2-3}
                    &  Disk PS + Iso PS + NFW PS         &      \multirow{2}{*}{ $\textcolor{violet}{[ \, \textbf{0.0, 0.7} \,]\  \%}$}     &      \multicolumn{3}{c|}{\multirow{2}{*}{$3\times10^{9}$}}          \\ 
                    &  Diff IC + Diff $\pi^0$brem + Iso + Bub+ DM    &                                      &     \multicolumn{3}{c|}{}         \\ \hline\hline
                    
\multirow{6}{*}{$2 \%$}    &  Disk PS + Iso PS                  &     \multirow{2}{*}{ $[ \, 5.7, 6.8 \,]\  \%$}    &      &  \cellcolor{black!25}         &          \\ 
                                &  Diff IC + Diff $\pi^0$brem + Iso + Bub + DM    &                                    &                                &   \cellcolor{black!25}        &         \\ \cline{2-3} \cline{5-6} 
                    
                                &  Disk PS + Iso PS + NFW PS         &      \multirow{2}{*}{  $\textcolor{violet}{[ \, \textbf{0.0, 0.8} \,]\  \%}$}   &        \multirow{2}{*}{$3\times10^9$  }                          &     \multirow{2}{*}{$10$}     &  \cellcolor{black!25}       \\ 
                                &  Diff IC + Diff $\pi^0$brem + Iso + Bub + DM    &                                  &                                  &           &   \cellcolor{black!25}    \\ \cline{2-4} \cline{6-6} 
                                
                                &  Disk PS + Iso PS + NFW PS         &      Fixed at injection         &        \cellcolor{black!25}      &             &    \multirow{2}{*}{$5\times10^8$}  \\ 
                                &  Diff IC + Diff $\pi^0$brem + Iso + Bub + Fixed DM    &        value ($2\%$)                            &          \cellcolor{black!25}    &           &     \\ \hline
\end{tabular}
\caption{Bayes factors for analyses of the real \textit{Fermi} data injected with a DM signal, in the Inner Galaxy ROI (unmasked), using the diffuse model A. The standard-pipeline analysis (modulo the replacement of the diffuse model) is shown in purple. The data analysis that is favored has the relative Bayes factor appear next to the analysis row, where the two analyses that are being compared are not grayed out.}
\label{tab:ccw_diffa_real_unmsk}
\end{table*}

\begin{table*}
\renewcommand{\arraystretch}{1.3}
\centering
\begin{tabular}{|P{2.4cm}|P{7.2cm}|P{3cm}|P{1.4cm}|P{1.4cm}|P{1.4cm}|}
\hline
\multicolumn{6}{|c|}{\textsc{Diffuse Model F, Real Data, 3FGL Unmasked}} \\
\hline
 \textbf{Injected} &  \multirow{2}{*}{\textbf{Analysis Templates}} &  \textbf{DM Flux}                          &  \multicolumn{3}{c|}{\multirow{2}{*}{\textbf{Bayes Factor}} } \\
\textbf{DM Flux}   &                                               &   \textbf{(95\%)}                     &     \multicolumn{3}{c|}{}                                 \\ \hline\hline
 
  \multirow{4}{*}{None} &  Disk PS + Iso PS               &     \multirow{2}{*}{ $[ \, 2.6, 3.7 \,]\  \%$}      &   \multicolumn{3}{c|}{ }          \\ 
                    &  Diff IC + Diff $\pi^0$brem + Iso + Bub + DM    &                                      &      \multicolumn{3}{c|}{}        \\ \cline{2-3}
                    &  Disk PS + Iso PS + NFW PS         &      \multirow{2}{*}{ $\textcolor{violet}{[ \, \textbf{0.0, 0.6} \,]\  \%}$}     &      \multicolumn{3}{c|}{\multirow{2}{*}{$6\times10^{3}$}}          \\ 
                    &  Diff IC + Diff $\pi^0$brem + Iso + Bub + DM    &                                      &     \multicolumn{3}{c|}{}         \\ \hline\hline
                    
\multirow{6}{*}{$2 \%$}    &  Disk PS + Iso PS                  &     \multirow{2}{*}{ $[ \,4.2,5.8 \,]\  \%$}    &      &  \cellcolor{black!25}         &          \\ 
                                &  Diff IC + Diff $\pi^0$brem + Iso + Bub + DM    &                                    &                                &   \cellcolor{black!25}        &         \\ \cline{2-3} \cline{5-6} 
                    
                                &  Disk PS + Iso PS + NFW PS         &     \multirow{2}{*} { $\textcolor{violet}{[ \, \textbf{0.0, 0.9}\,]\  \%}$}   &        \multirow{2}{*}{ $1\times10^5$ }                          &     \multirow{2}{*}{$10$}     &  \cellcolor{black!25}       \\ 
                                &  Diff IC + Diff $\pi^0$brem + Iso + Bub + DM    &                                 &                                  &           &   \cellcolor{black!25}    \\ \cline{2-4} \cline{6-6} 
                                
                                &  Disk PS + Iso PS + NFW PS         &      Fixed at injection         &        \cellcolor{black!25}      &             &    \multirow{2}{*}{$1\times 10^{4}$}  \\ 
                                &  Diff IC + Diff $\pi^0$brem + Iso + Bub + Fixed DM    &      value ($2\%$)                            &          \cellcolor{black!25}    &           &     \\ \hline
\end{tabular}
\caption{Bayes factors for analyses of the real \textit{Fermi} data injected with a DM signal, in the Inner Galaxy ROI (unmasked), using the diffuse model F. The standard-pipeline analysis (modulo the replacement of the diffuse model) is shown in purple. The data analysis that is favored has the relative Bayes factor appear next to the analysis row, where the two analyses that are being compared are not grayed out.}
\label{tab:ccw_difff_real_unmsk}
\end{table*}

Tables~\ref{tab:ccw_diffa_real_unmsk} and \ref{tab:ccw_difff_real_unmsk} show results for the unmasked case for models A and F respectively, both with and without an additional injected DM signal. In the baseline case without injection, the Bayes factor preference for PSs varies strongly between model F, model A, and the \texttt{p6v11} Galactic diffuse emission model used in the main text, with the smallest value for model F and the largest for the \texttt{p6v11} model. This suggests that variations in the diffuse modeling can markedly reduce or enhance the apparent significance of a PS population, and consequently that errors in the diffuse modeling could significantly modify the Bayes factors obtained from the real data.

\begin{figure*}[t!]
\leavevmode
\centering
\subfigure{\includegraphics[width=0.48\columnwidth]{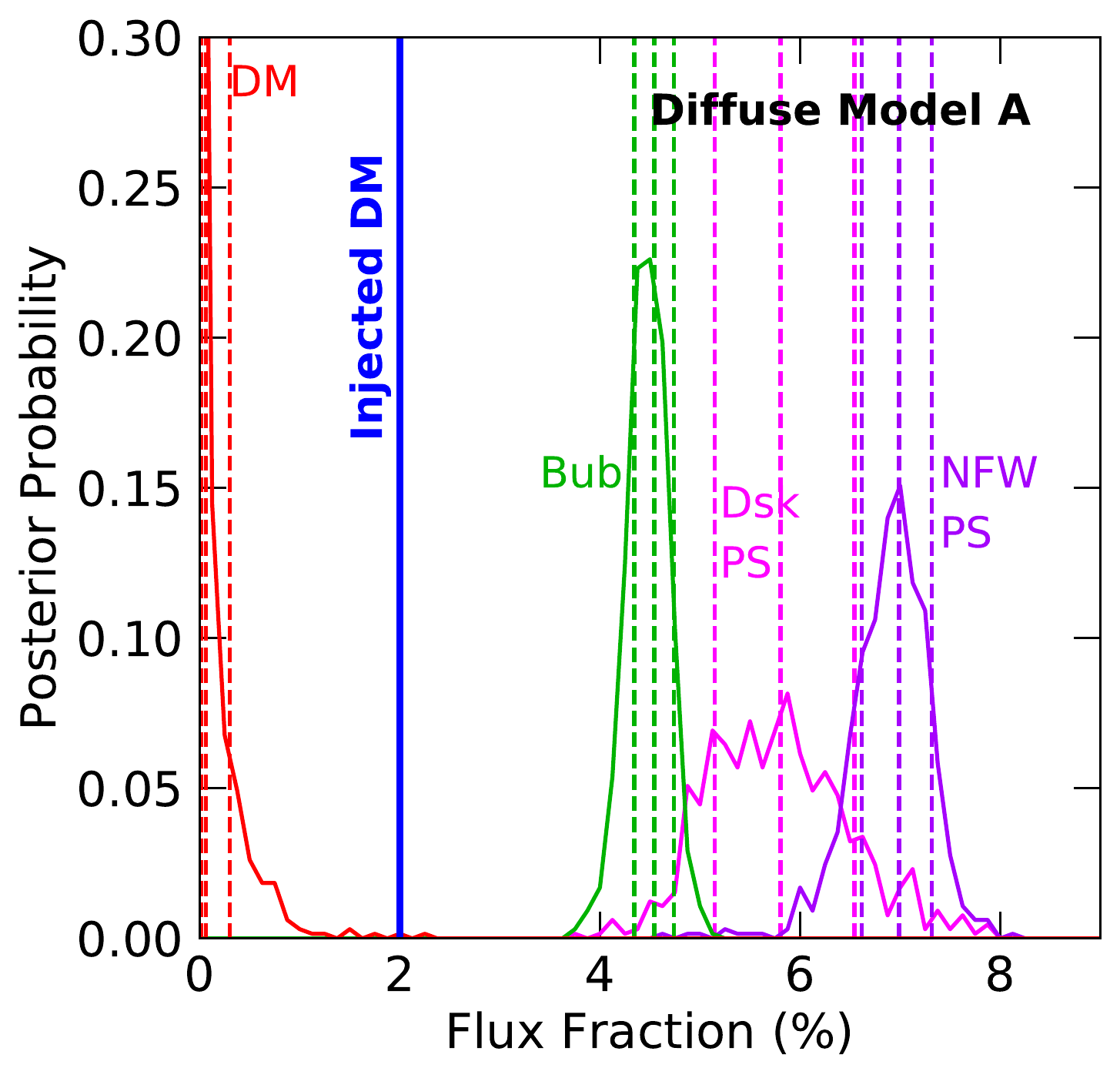}}
\hspace{1mm}
\subfigure{\includegraphics[width=0.49\columnwidth]{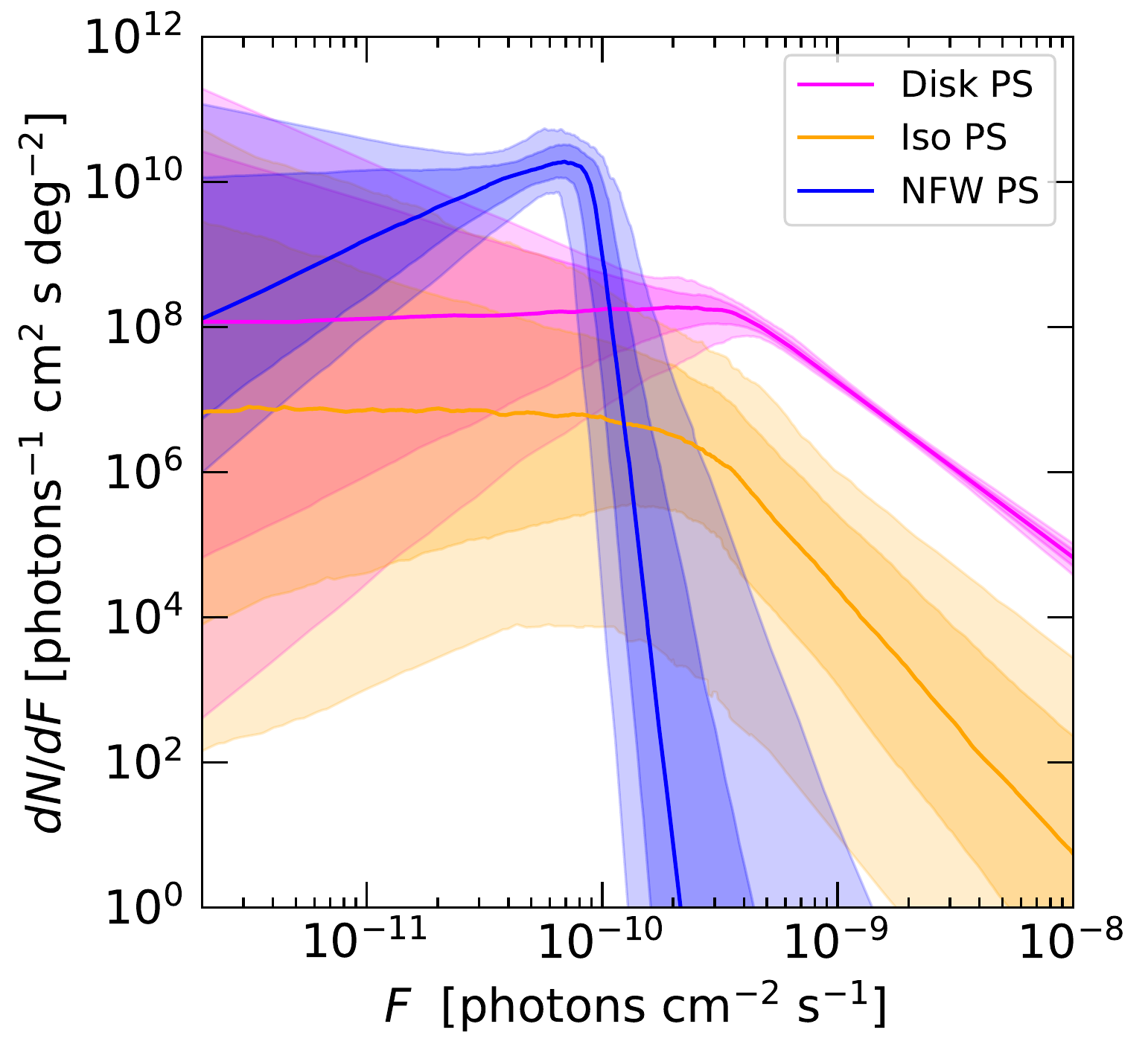}}
\\
\subfigure{\includegraphics[width=0.48\columnwidth]{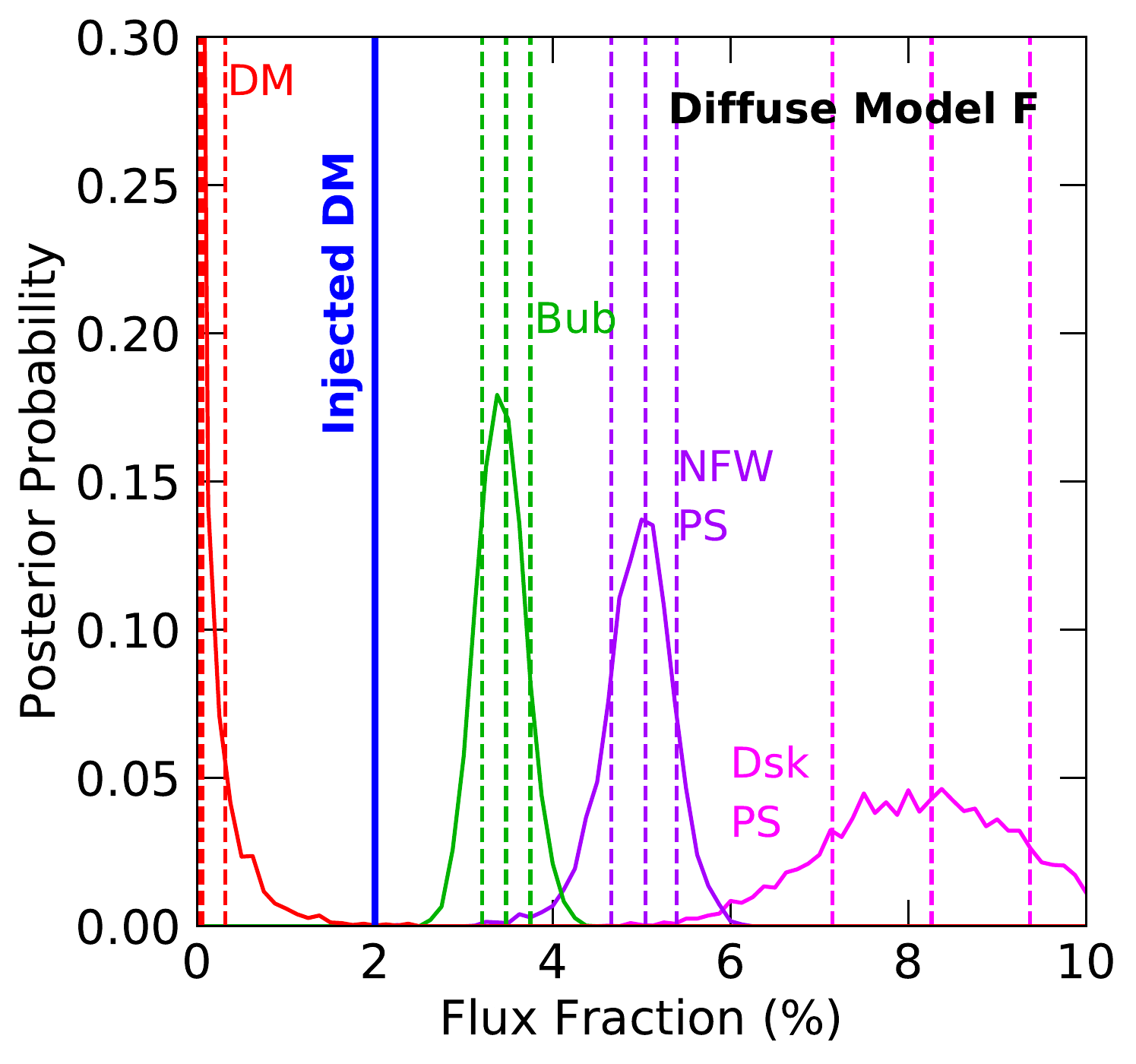}}
\hspace{1mm}
\subfigure{\includegraphics[width=0.49\columnwidth]{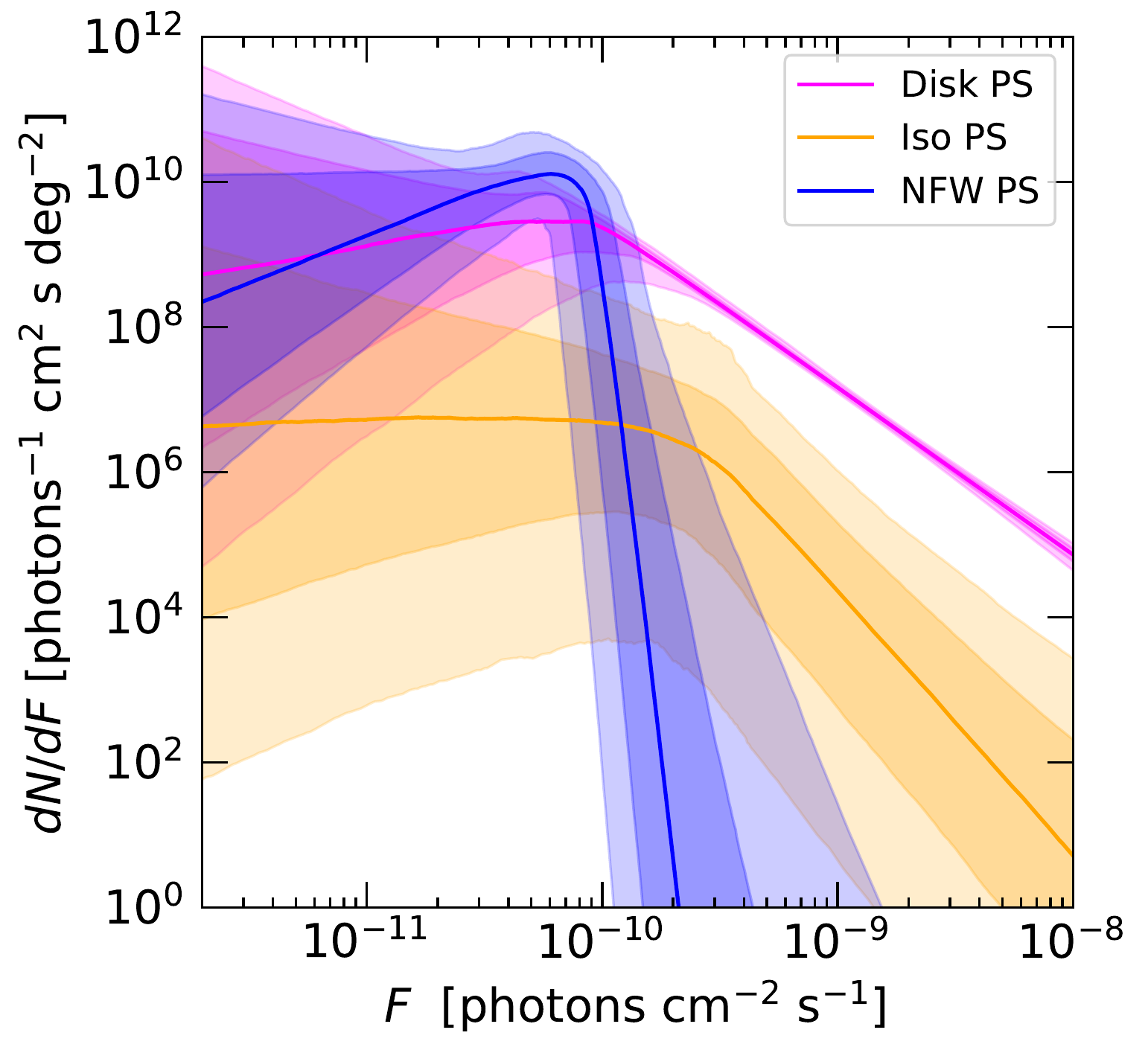}}
\caption{Injection test replacing the \mbox{\texttt{Pass 6}} \textit{Fermi} diffuse model with the diffuse Model A (top row) or F (bottom row) from Ref.~\cite{Calore:2014xka}, and injecting a DM flux making up $\sim 2\%$ of the ROI. All templates are present, but those with fluxes peaked below 0.1\% (except for DM) are not shown for clarity. 3FGL are unmasked.}
\label{fig:Run_15_dif_unmsk}
\end{figure*}

Figure~\ref{fig:Run_15_dif_unmsk} shows the flux posteriors for the unmasked study using diffuse models A and F.

In both Model A and Model F, the injected 2\% DM-like signal is excluded by the reconstructed value, as for the \texttt{p6v11} diffuse model. However, in both cases, the Bayes factor preference for the model with free DM fraction, over the model with the DM component fixed to its injected value, is modest, only $\sim 10$. In this sense model A and model F perform better than the default \texttt{p6v11} model, in the unmasked case; we note that their Bayes factors in favor of PSs are much smaller than in the \texttt{p6v11} case, in the data without injection (and much smaller for Model F than Model A).

Tables~\ref{tab:ccw_diffa_real} and \ref{tab:ccw_difff_real} show results for the masked case for models A and F respectively. In this scenario model F completely loses any preference for point sources, even without any DM injection. Model A prefers PSs (prior to injection) with a Bayes factor of $\sim 10^3$; when extra DM is injected, the Bayes factor in favor of PSs stays fairly constant, but the DM component is not reconstructed correctly. 

For diffuse model A we perform simulations for both the masked and unmasked cases, and find that the recovered Bayes factor in the real data is broadly consistent with expectations, albeit on the low side in the masked case (a detailed study of consistency would require a larger number of realizations). For model F, we have performed one simulation in the masked case, and find that there is no strong preference for PSs in the simulated data (consistent with the real data). 

\begin{figure*}[t!]
\leavevmode
\centering
\subfigure{\includegraphics[width=0.48\columnwidth]{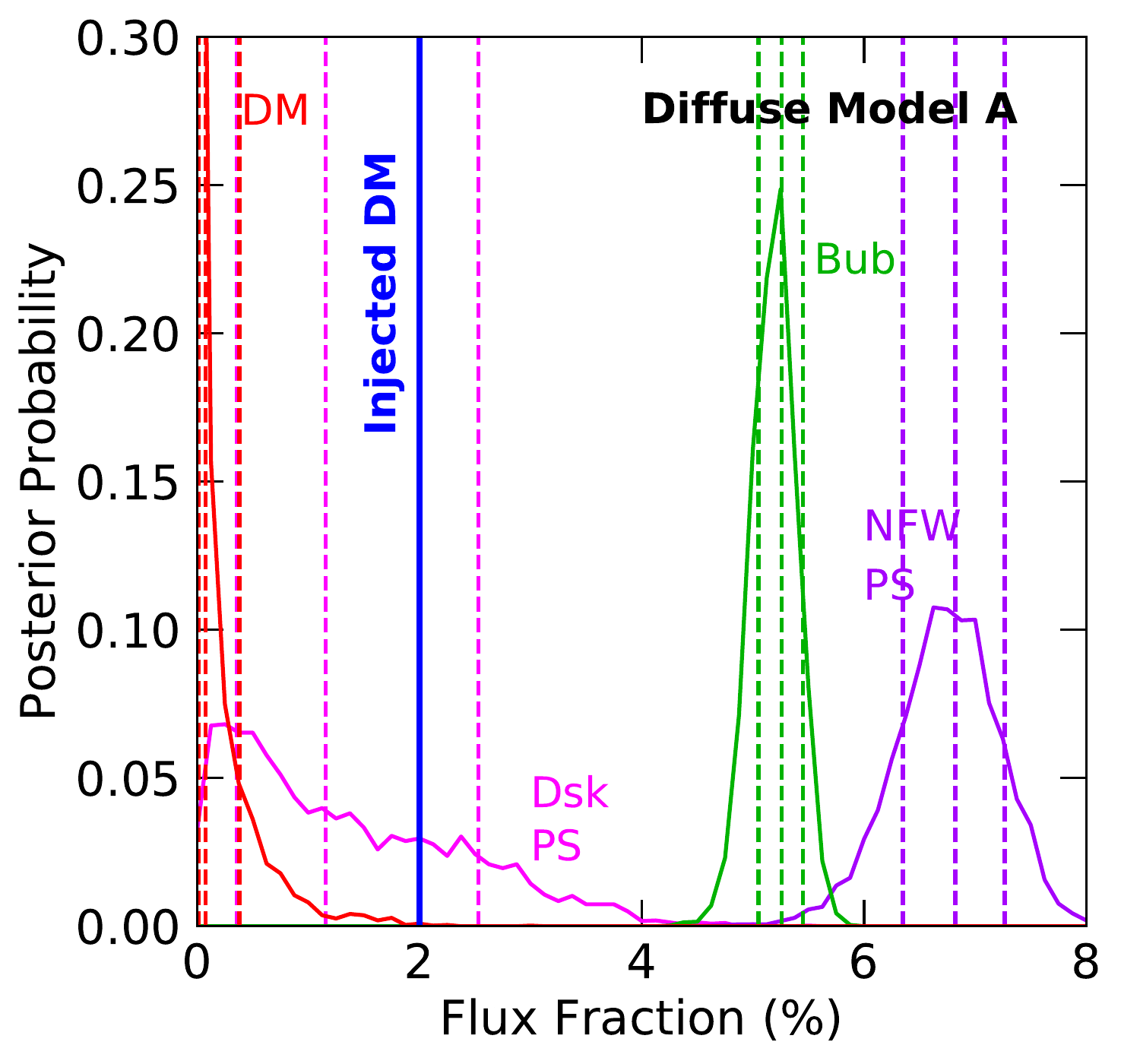}}
\hspace{1mm}
\subfigure{\includegraphics[width=0.49\columnwidth]{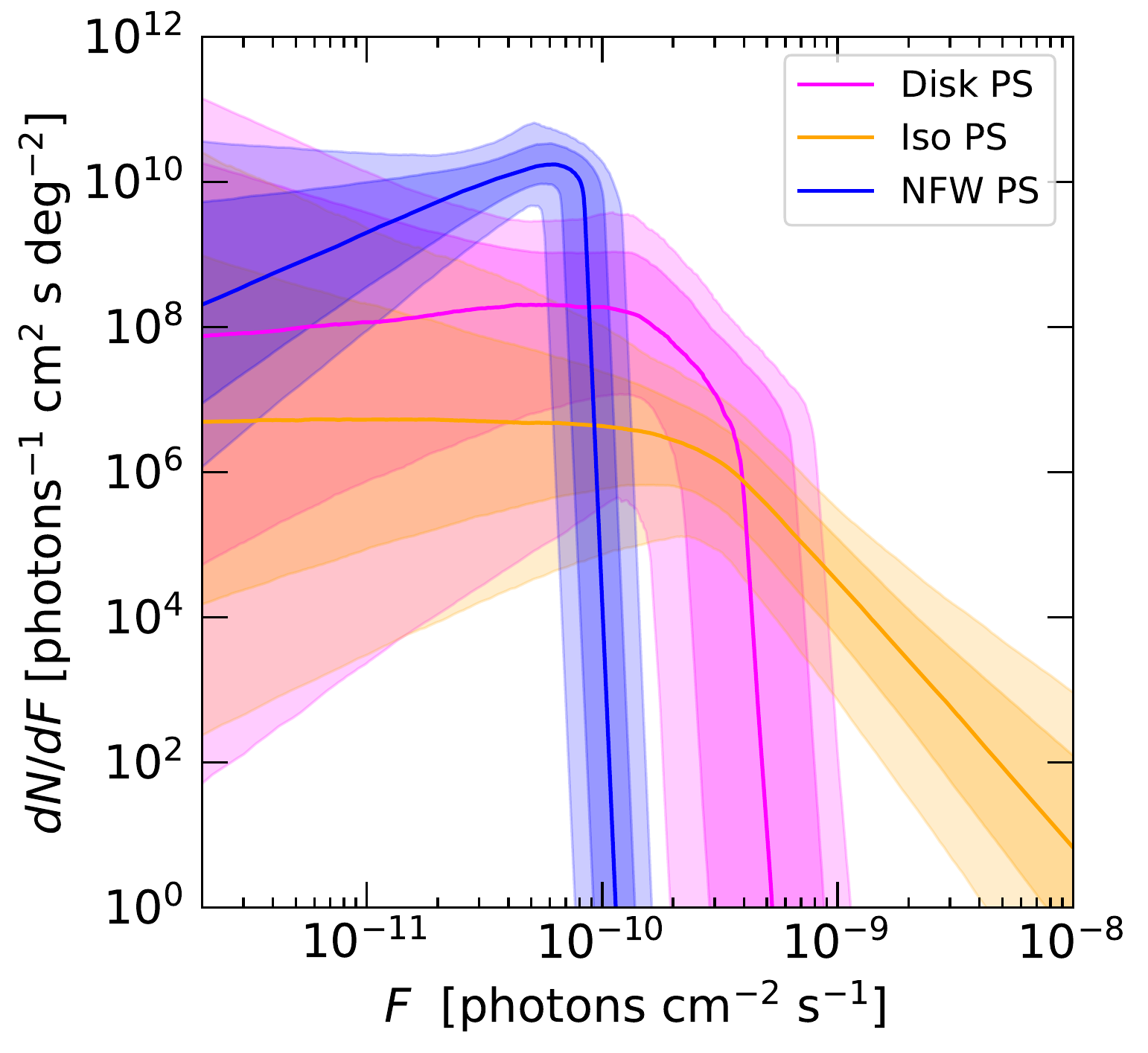}}
\\
\subfigure{\includegraphics[width=0.48\columnwidth]{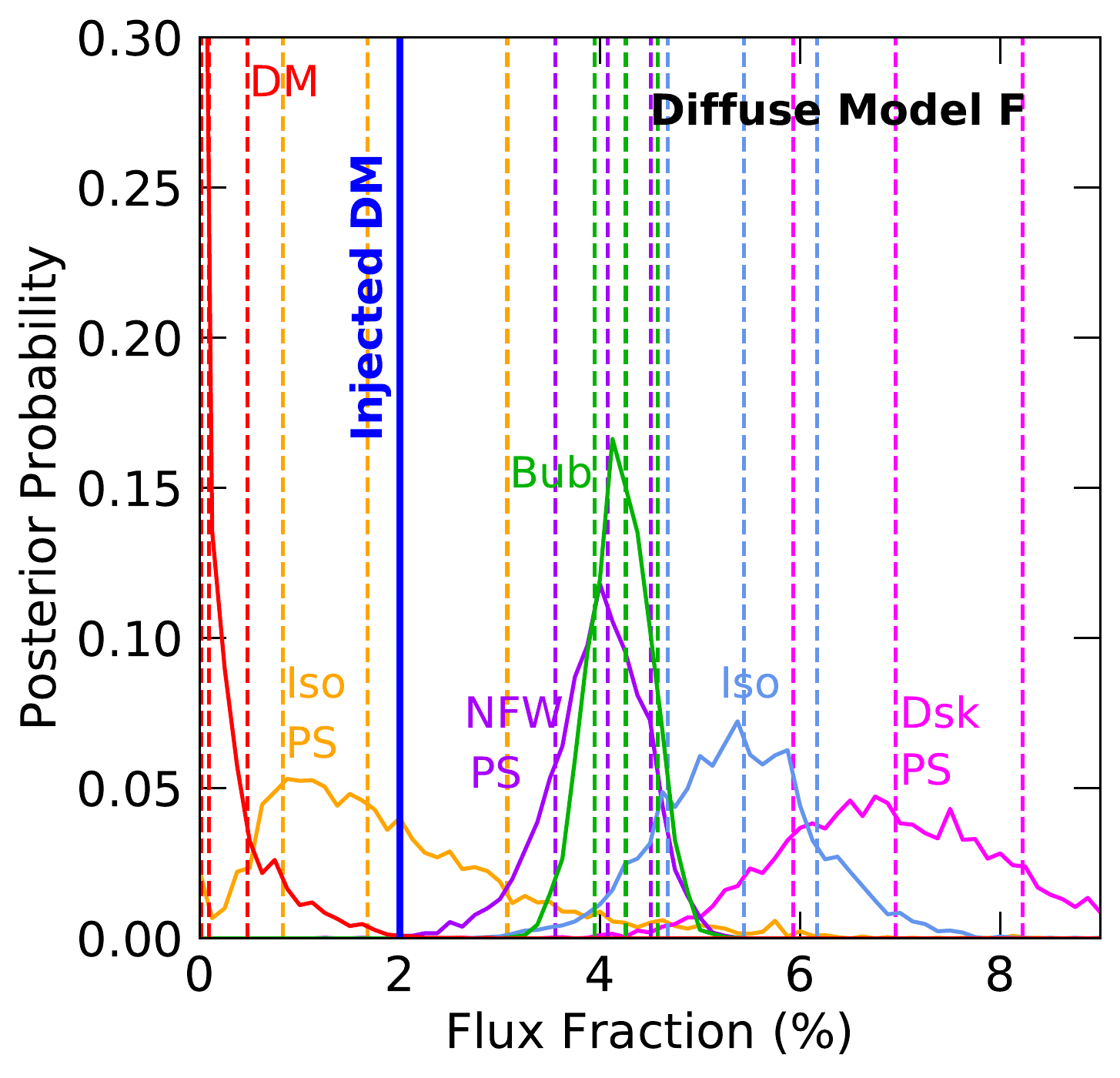}}
\hspace{1mm}
\subfigure{\includegraphics[width=0.49\columnwidth]{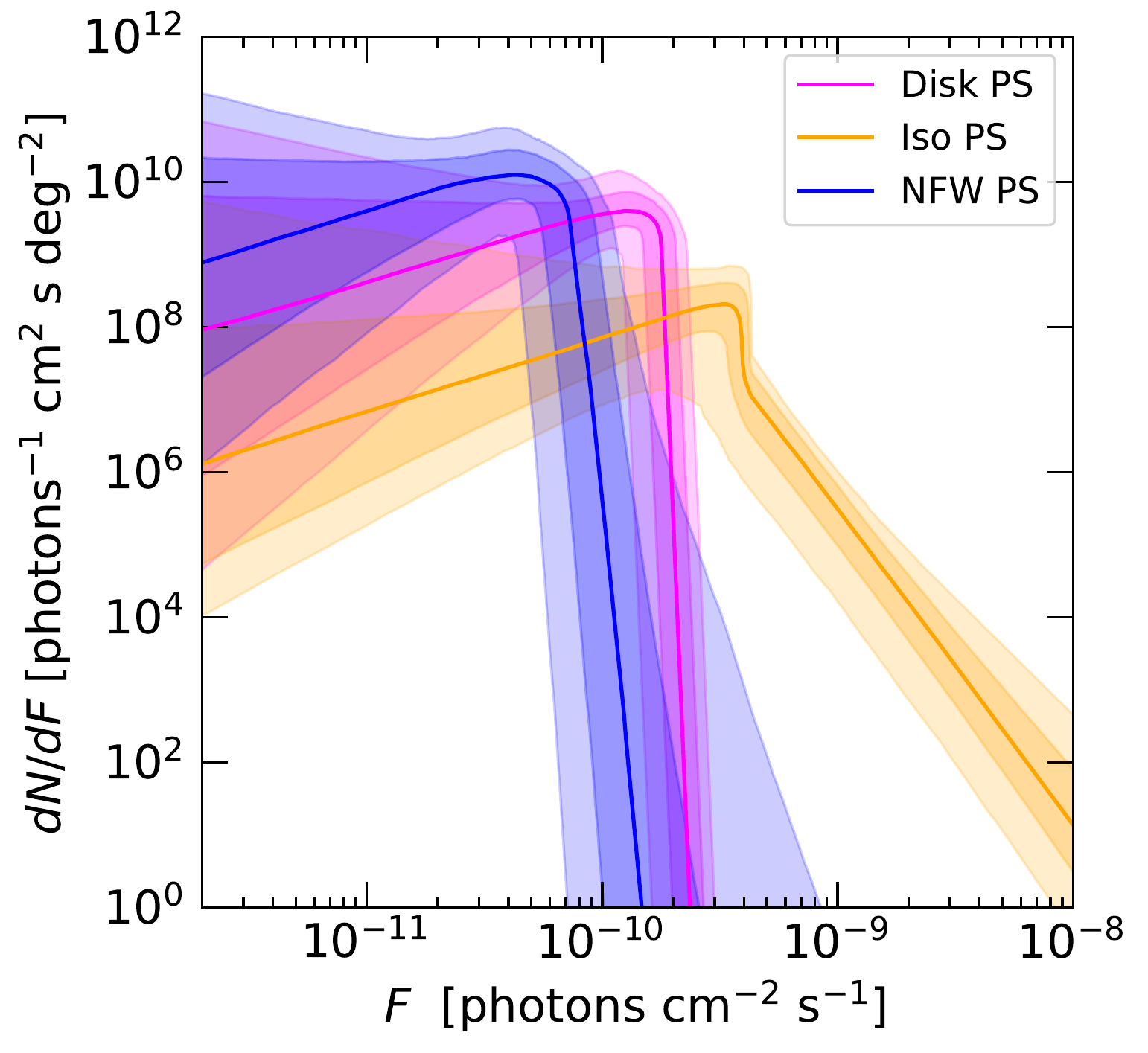}}
\caption{Injection test replacing the \mbox{\texttt{Pass 6}} \textit{Fermi} diffuse model with the diffuse Model A (top row) or F (bottom row) from Ref.~\cite{Calore:2014xka}, and injecting a DM flux making up $\sim 2\%$ of the ROI. All templates are present, but those with fluxes peaked below 0.1\% (except for DM) are not shown for clarity. 3FGL are masked. Diffuse model F masked does not differentiate between models with and without NFW PS.}
\label{fig:Run_15_dif_msk}
\end{figure*}

Figure~\ref{fig:Run_15_dif_msk} shows the flux posteriors for the masked study using diffuse models A and F.

\begin{table*}[t!]
\renewcommand{\arraystretch}{1.3}
\centering
\begin{tabular}{|P{2.4cm}|P{7.2cm}|P{3cm}|P{1.4cm}|P{1.4cm}|P{1.4cm}|}
\hline
\multicolumn{6}{|c|}{\textsc{Diffuse Model A, Real Data, 3FGL Masked}} \\
\hline
 \textbf{Injected} &  \multirow{2}{*}{\textbf{Analysis Templates}} &  \textbf{DM Flux}                          &  \multicolumn{3}{c|}{\multirow{2}{*}{\textbf{Bayes Factor}} } \\
\textbf{DM Flux}   &                                               &   \textbf{(95\%)}                     &     \multicolumn{3}{c|}{}                                 \\ \hline\hline
 
 \multirow{4}{*}{None} &  Disk PS + Iso PS               &     \multirow{2}{*}{ $[ \, 4.0, 5.0 \,]\  \%$}      &   \multicolumn{3}{c|}{ }          \\ 
                    &  Diff IC + Diff $\pi^0$brem + Iso + Bub + DM    &                                      &      \multicolumn{3}{c|}{}        \\ \cline{2-3}
                    &  Disk PS + Iso PS + NFW PS         &      \multirow{2}{*}{ $\textcolor{violet}{[ \, \textbf{0.0, 0.1} \,]\  \%}$}     &      \multicolumn{3}{c|}{\multirow{2}{*}{$1\times10^{3}$ }}          \\ 
                    &  Diff IC + Diff $\pi^0$brem + Iso + Bub + DM    &                                      &     \multicolumn{3}{c|}{}         \\ \hline\hline
                    
 \multirow{4}{*}{$2 \%$} &  Disk PS + Iso PS               &     \multirow{2}{*}{ $[ \, 5.5, 6.7 \,]\  \%$}      &   \multicolumn{3}{c|}{ }          \\ 
                    &  Diff IC + Diff $\pi^0$brem + Iso + Bub + DM    &                                      &      \multicolumn{3}{c|}{}        \\ \cline{2-3}
                    &  Disk PS + Iso PS + NFW PS         &      \multirow{2}{*}{ $\textcolor{violet}{[ \, \textbf{0.0, 1.0} \,]\  \%}$}     &      \multicolumn{3}{c|}{\multirow{2}{*}{$2\times10^3$ }}          \\ 
                    &  Diff IC + Diff $\pi^0$brem + Iso + Bub + DM    &                                      &     \multicolumn{3}{c|}{}         \\ \hline
\end{tabular}
\caption{Bayes factors for analyses of the real \textit{Fermi} data injected with a DM signal, in the Inner Galaxy ROI (masked), using the diffuse model A. The standard-pipeline analysis (modulo the replacement of the diffuse model) is shown in purple. The data analysis that is favored has the relative Bayes factor appear next to the analysis row, where the two analyses that are being compared are not grayed out.}
\label{tab:ccw_diffa_real}
\end{table*}

\begin{table*}
\renewcommand{\arraystretch}{1.3}
\centering
\begin{tabular}{|P{2.4cm}|P{7.2cm}|P{3cm}|P{1.4cm}|P{1.4cm}|P{1.4cm}|}
\hline
\multicolumn{6}{|c|}{\textsc{Diffuse Model A, Simulated Data, 3FGL Masked}} \\
\hline
 \textbf{Injected} &  \multirow{2}{*}{\textbf{Analysis Templates}} &  \textbf{DM Flux}                          &  \multicolumn{3}{c|}{\multirow{2}{*}{\textbf{Bayes Factor}} } \\
\textbf{DM Flux}   &                                               &   \textbf{(95\%)}                     &     \multicolumn{3}{c|}{}                                 \\ \hline\hline
 
 \multirow{4}{*}{None} &  Disk PS + Iso PS               &     \multirow{2}{*}{ $[ \, 4.2, 5.6 \,]\  \%$}      &   \multicolumn{3}{c|}{ }          \\ 
                    &  Diff IC + Diff $\pi^0$brem + Iso + Bub + DM    &                                      &      \multicolumn{3}{c|}{}        \\ \cline{2-3}
                    &  Disk PS + Iso PS + NFW PS         &      \multirow{2}{*}{ $\textcolor{violet}{[ \, \textbf{0.0, 0.4} \,]\  \%}$}     &      \multicolumn{3}{c|}{\multirow{2}{*}{$5\times10^{4}$}}          \\ 
                    &  Diff IC + Diff $\pi^0$brem + Iso + Bub + DM    &                                      &     \multicolumn{3}{c|}{}         \\ \hline\hline
                    
 \multirow{4}{*}{$2 \%$} &  Disk PS + Iso PS               &     \multirow{2}{*}{ $[ \, 6.0, 7.5\,]\  \%$}      &   \multicolumn{3}{c|}{ }          \\ 
                    &  Diff IC + Diff $\pi^0$brem + Iso + Bub + DM    &                                      &      \multicolumn{3}{c|}{}        \\ \cline{2-3}
                    &  Disk PS + Iso PS + NFW PS         &      \multirow{2}{*}{ $\textcolor{violet}{[ \, \textbf{0.0, 0.5} \,]\  \%}$}     &      \multicolumn{3}{c|}{\multirow{2}{*}{$2\times10^{4}$}}          \\ 
                    &  Diff IC + Diff $\pi^0$brem + Iso + Bub + DM    &                                      &     \multicolumn{3}{c|}{}         \\ \hline
\end{tabular}
\caption{Bayes factors for analyses of the simulated data (based on the best fit with NFW PS explaining the GCE) injected with a DM signal, in the Inner Galaxy ROI (masked), using the diffuse model A. The standard-pipeline analysis (modulo the replacement of the diffuse model) is shown in purple. The data analysis that is favored has the relative Bayes factor appear next to the analysis row, where the two analyses that are being compared are not grayed out.}
\label{tab:ccw_diffa_sim}
\end{table*}

\begin{table*}
\renewcommand{\arraystretch}{1.3}
\centering
\begin{tabular}{|P{2.4cm}|P{7.2cm}|P{3cm}|P{1.4cm}|P{1.4cm}|P{1.4cm}|}
\hline
\multicolumn{6}{|c|}{\textsc{Diffuse Model F, Real Data, 3FGL Masked}} \\
\hline
 \textbf{Injected} &  \multirow{2}{*}{\textbf{Analysis Templates}} &  \textbf{DM Flux}                          &  \multicolumn{3}{c|}{\multirow{2}{*}{\textbf{Bayes Factor}} } \\
\textbf{DM Flux}   &                                               &   \textbf{(95\%)}                     &     \multicolumn{3}{c|}{}                                 \\ \hline\hline
 
  \multirow{4}{*}{None} &  Disk PS + Iso PS               &     \multirow{2}{*}{ $[ \, 1.2, 3.0 \,]\  \%$}      &   \multicolumn{3}{c|}{ }          \\ 
                    &  Diff IC + Diff $\pi^0$brem + Iso + Bub + DM    &                                      &      \multicolumn{3}{c|}{}        \\ \cline{2-3}
                    &  Disk PS + Iso PS + NFW PS         &      \multirow{2}{*}{ $\textcolor{violet}{[ \, \textbf{0.0, 1.0} \,]\  \%}$ }     &      \multicolumn{3}{c|}{\multirow{2}{*}{$2$}}          \\ 
                    &  Diff IC + Diff $\pi^0$brem + Iso + Bub + DM    &                                      &     \multicolumn{3}{c|}{}         \\ \hline\hline
                    
  \multirow{4}{*}{$2 \%$} &  Disk PS + Iso PS               &     \multirow{2}{*}{ $[ \, 3.5, 4.7 \,]\  \%$}      &   \multicolumn{3}{c|}{ }          \\ 
                    &  Diff IC + Diff $\pi^0$brem + Iso + Bub + DM    &                                      &      \multicolumn{3}{c|}{}        \\ \cline{2-3}
                    &  Disk PS + Iso PS + NFW PS         &      \multirow{2}{*}{ $\textcolor{violet}{[ \, \textbf{0.0, 1.2} \,]\  \%}$}     &      \multicolumn{3}{c|}{\multirow{2}{*}{$1$}}          \\ 
                    &  Diff IC + Diff $\pi^0$brem + Iso + Bub + DM    &                                      &     \multicolumn{3}{c|}{}         \\ \hline
\end{tabular}
\caption{Bayes factors for analyses of the real \textit{Fermi} data injected with a DM signal, in the Inner Galaxy ROI (masked), using the diffuse model F. The standard-pipeline analysis (modulo the replacement of the diffuse model) is shown in purple.The data analysis that is favored has the relative Bayes factor appear next to the analysis row, where the two analyses that are being compared are not grayed out.}
\label{tab:ccw_difff_real}
\end{table*}

As discussed above, allowing the DM normalization to float negative provides an alternate form of the injection test. The top right panel of Fig.~\ref{fig:floatneg} shows the impact of allowing the DM normalization to float negative in model A, when the 3FGL sources are not masked; we find almost identical results for model F. For the unmasked ROI, the DM normalization floats significantly negative in both cases, similarly to the results obtained with the \texttt{p6v11} Galactic diffuse model. 

When the 3FGL sources are masked, and the Galactic diffuse emission is modeled with model A, the DM component also floats negative (consistent with our earlier finding that the injection test is also failed for this combination of model and ROI). When model F is used instead, there is no preference for the DM to float negative, but this is a corollary of the fact that in this case there is no significant preference for the GCE to be comprised of PSs rather than DM.

We note that because we simulate based on the posterior medians, the simulated flux and properties of the NFW PSs are different between different diffuse models; it is thus not straightforward to use these simulations to answer how the sensitivity of the fit to a fixed PS population differs between different diffuse models. It would be interesting to understand whether combinations of diffuse model + ROI that pass the injection test (correctly recovering the injected DM, and with the DM coefficient not floating significantly negative in the absence of injection) retain a robust preference for a dominant PS contribution to the GCE. At present, the example of model F in the masked ROI suggests this is not the case, but this could be either an outlier or simply a reflection of specific features of model F that cause the fit to lose sensitivity to a dominant PS contribution in the masked ROI. We leave an in-depth analysis of this question to future work.

\section{Survival function analysis}
\label{sec:survival}

\begin{figure*}[t!]
\leavevmode
\centering
\subfigure{\includegraphics[width=0.43\columnwidth]{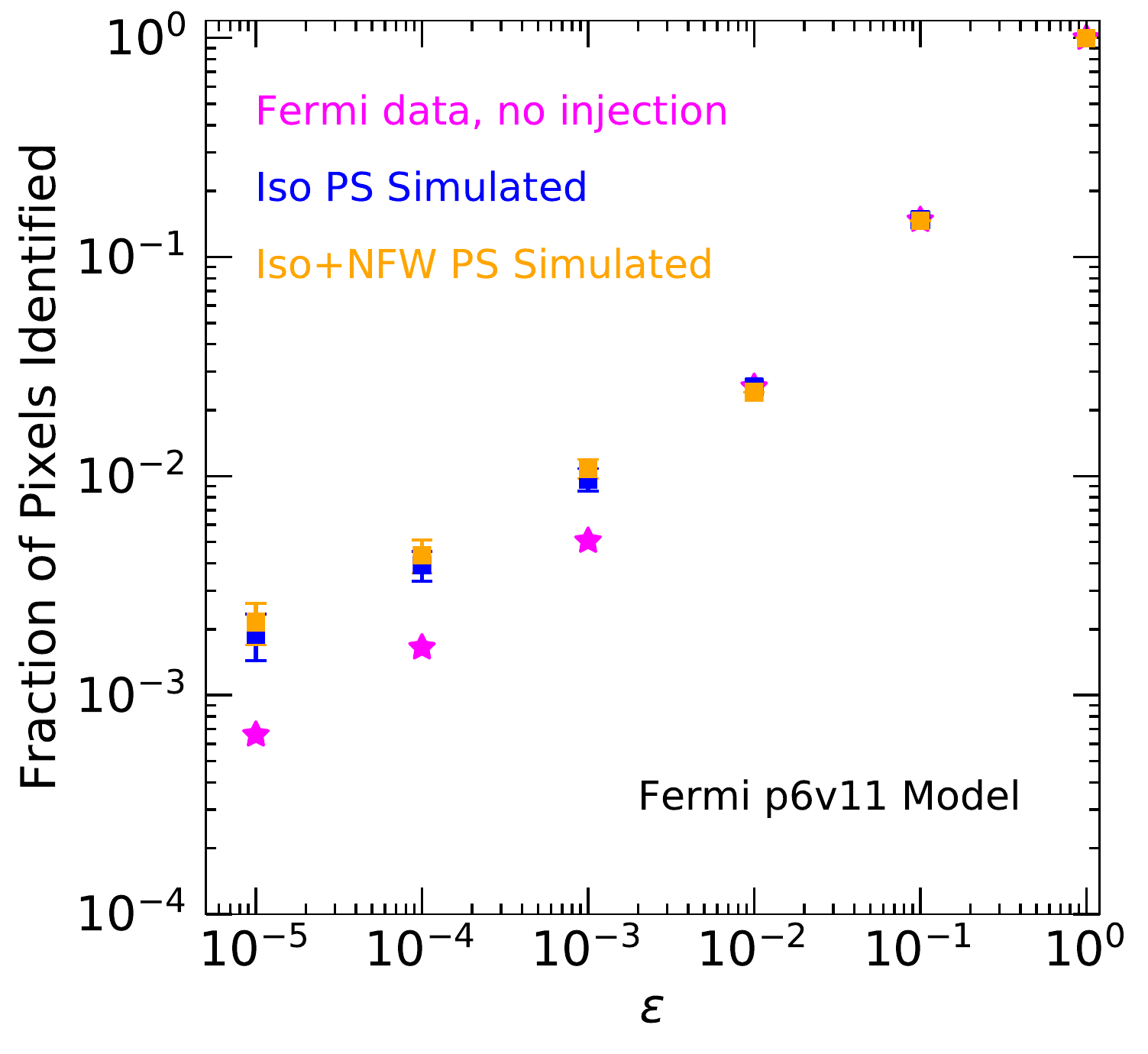}}
\hspace{1mm}
\subfigure{\includegraphics[width=0.43\columnwidth]{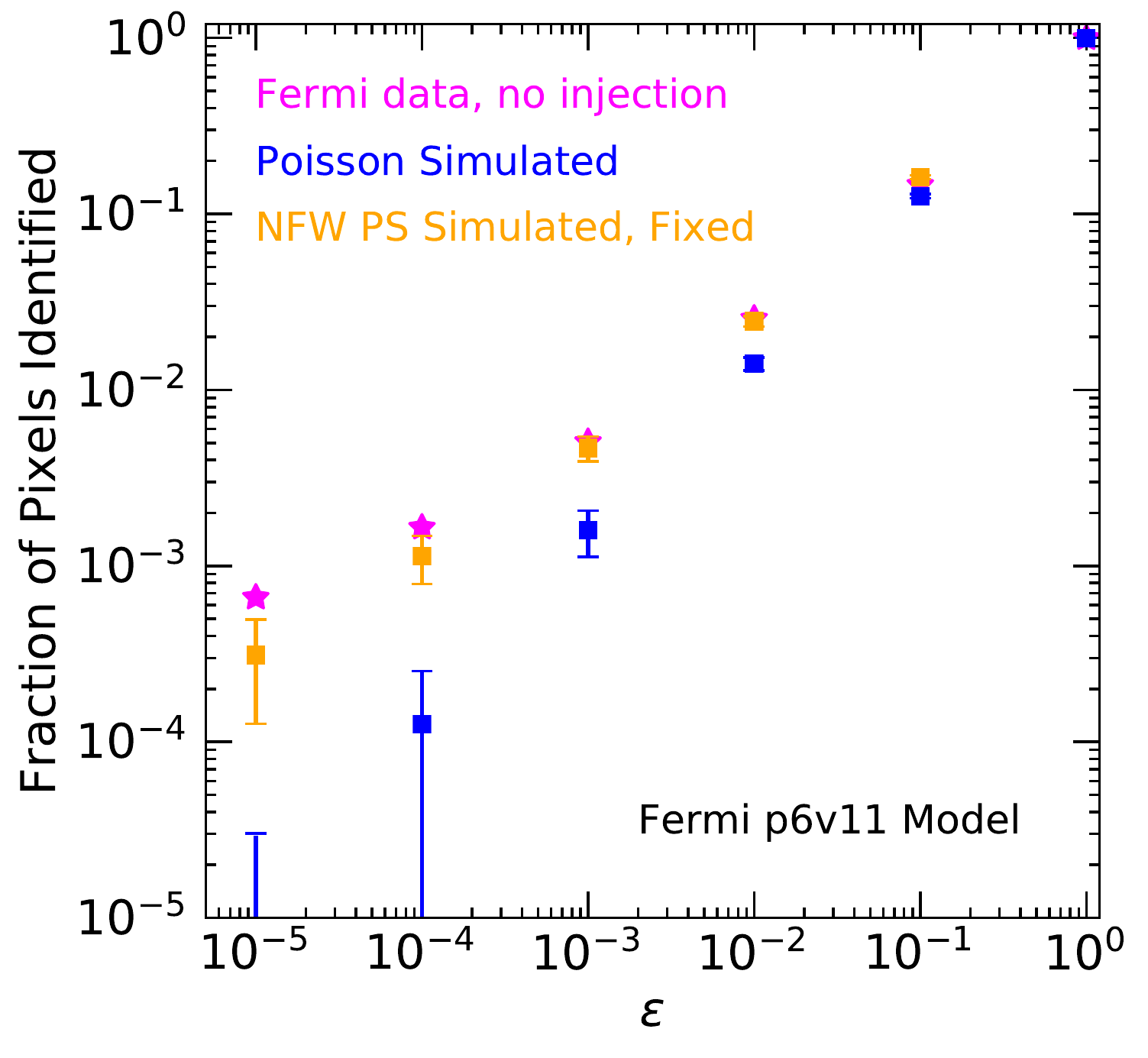}}
\\
\subfigure{\includegraphics[width=0.43\columnwidth]{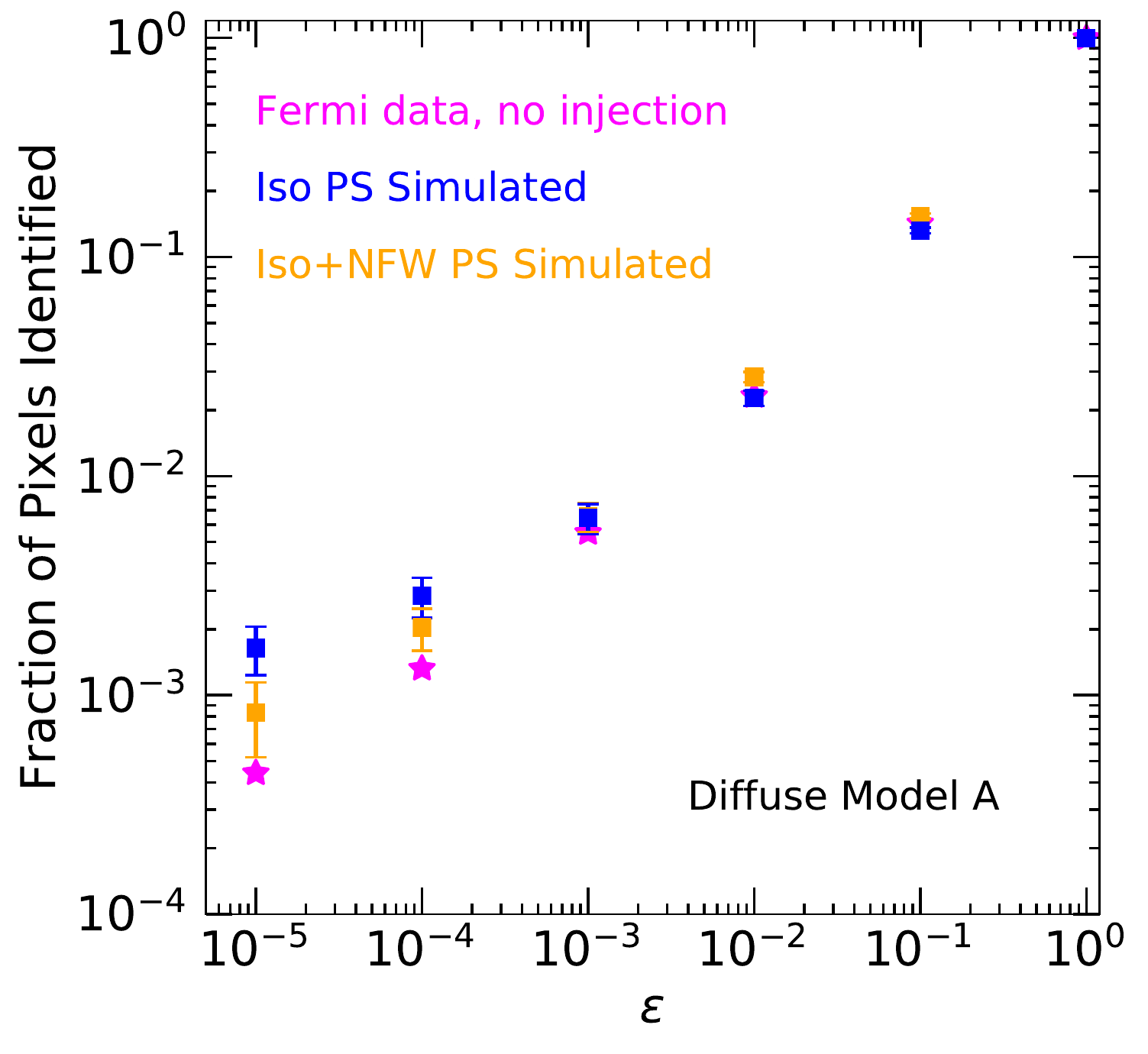}}
\hspace{1mm}
\subfigure{\includegraphics[width=0.43\columnwidth]{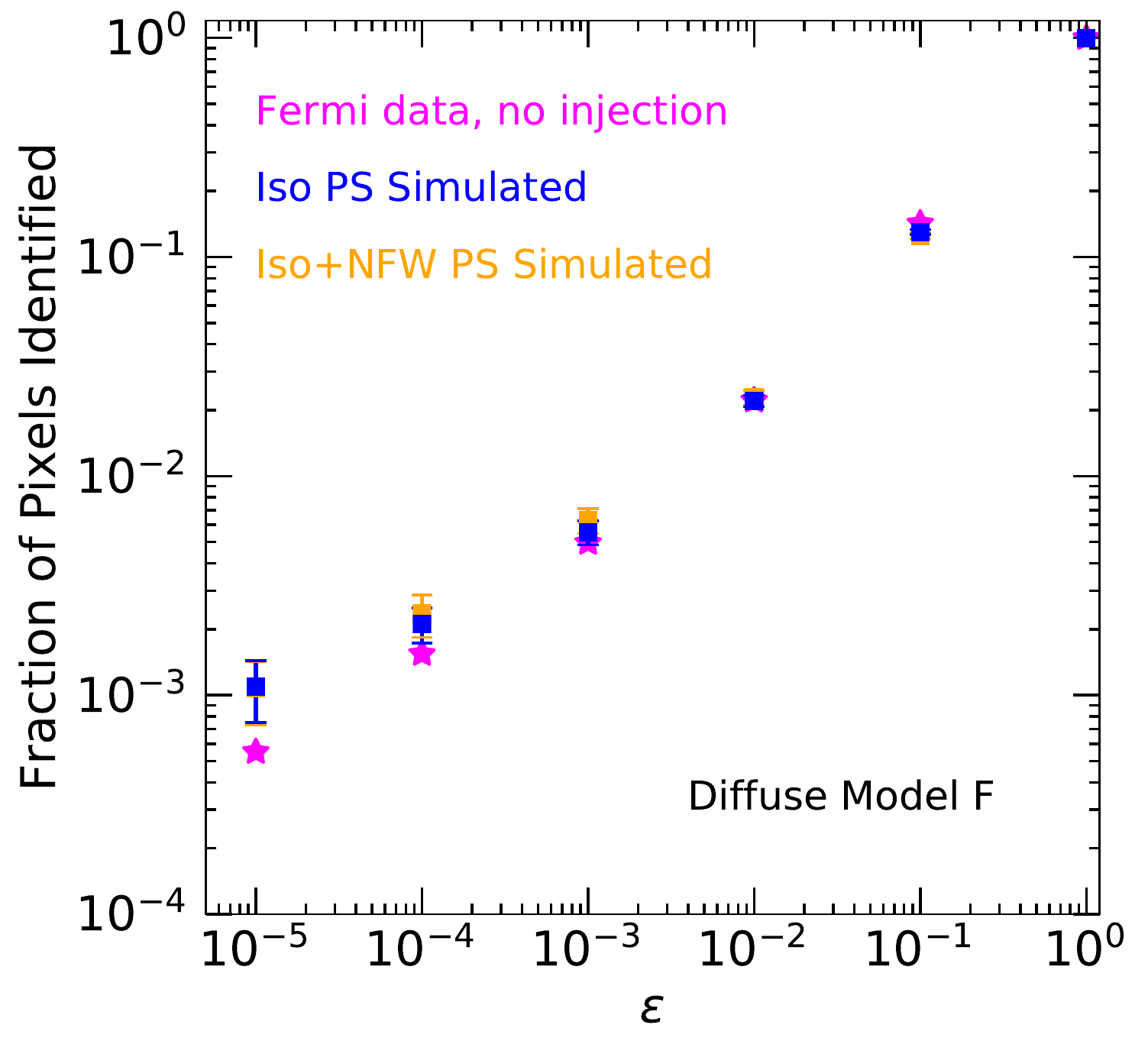}}
\\
\subfigure{\includegraphics[width=0.43\columnwidth]{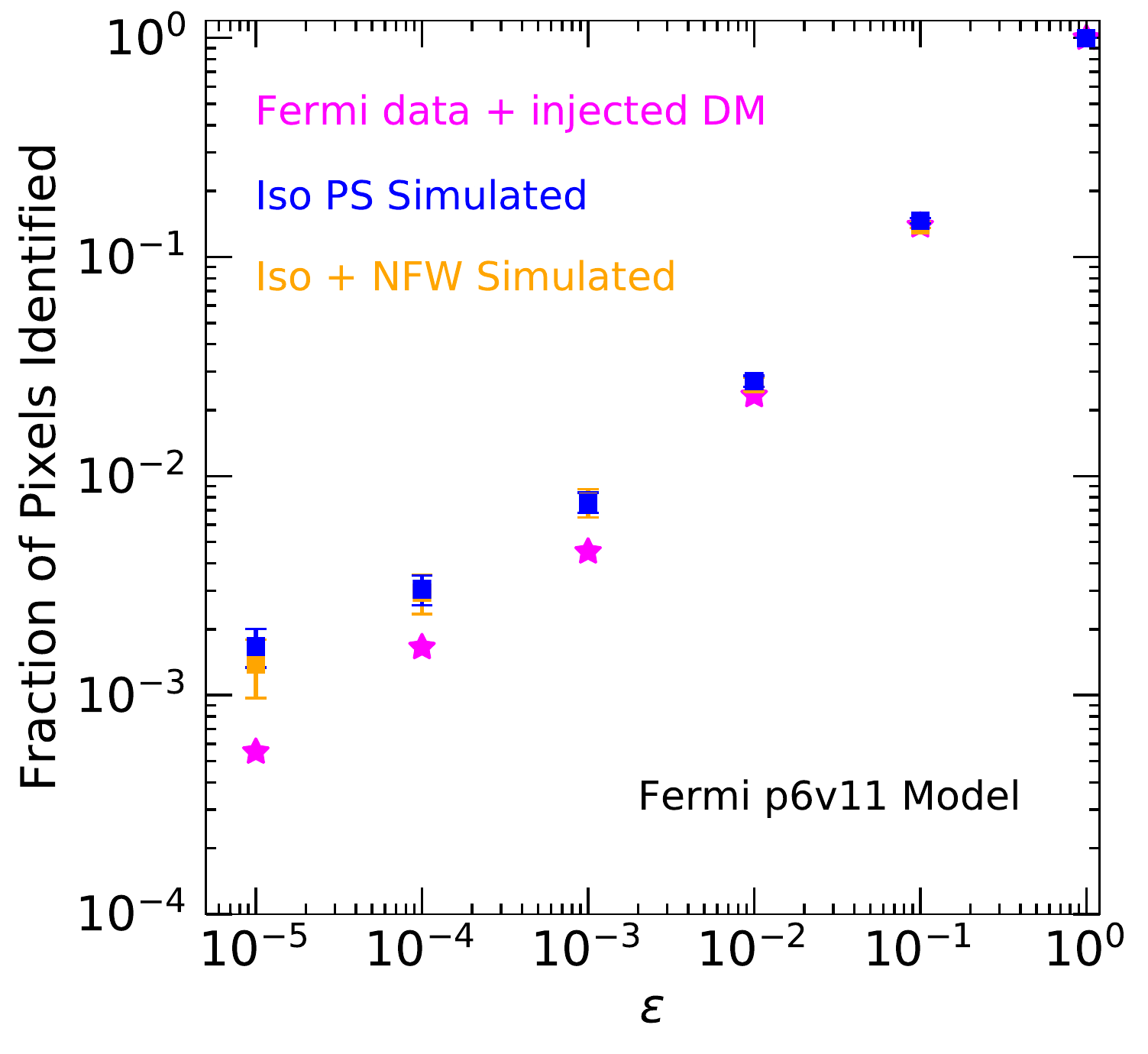}}
\hspace{1mm}
\subfigure{\includegraphics[width=0.43\columnwidth]{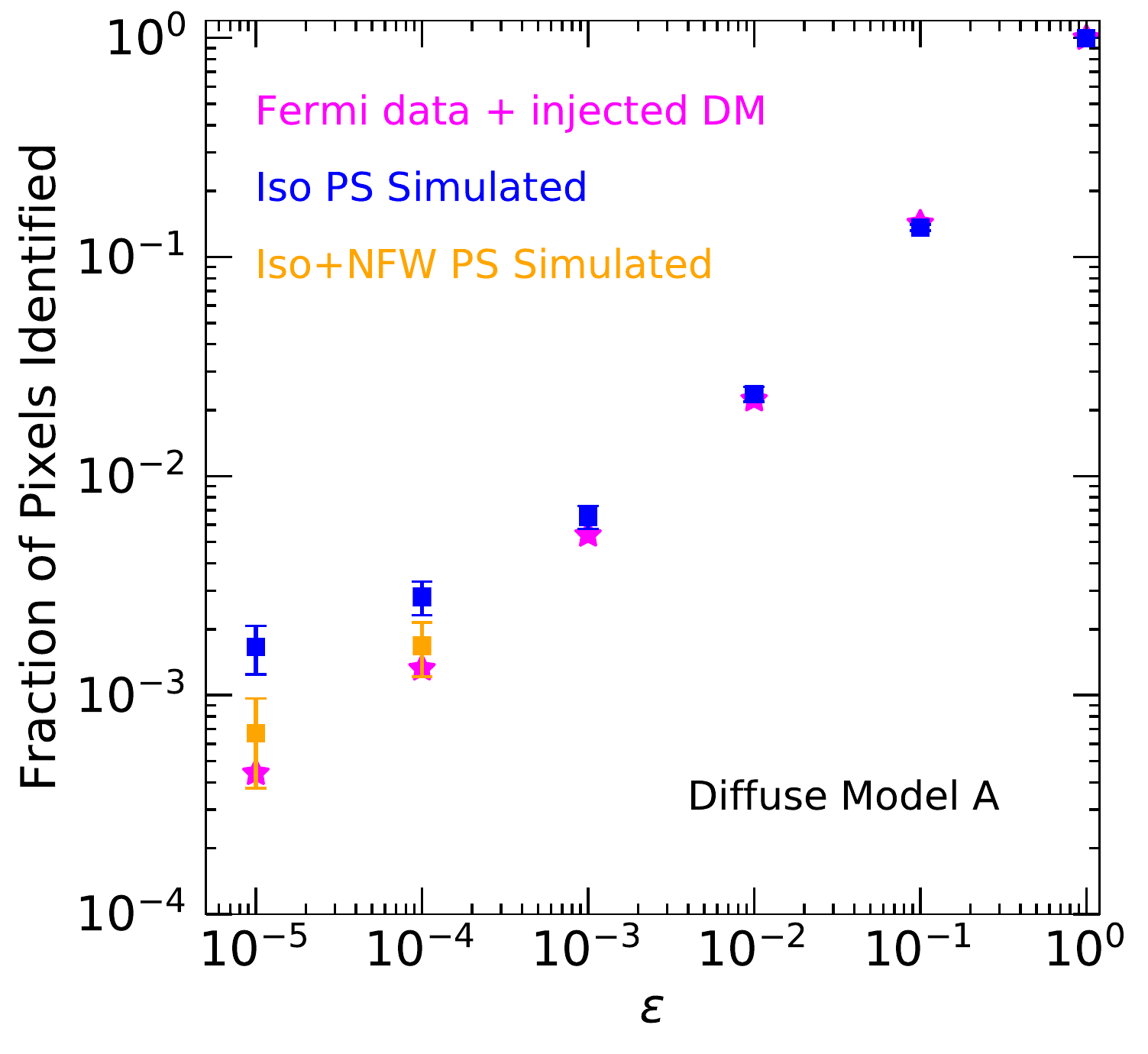}}
\caption{Fraction of pixels in the dataset consisting of the \textit{Fermi} data with or without an additional injected DM signal (pink stars), which satisfy  $\epsilon_p < \epsilon$, where $\epsilon_p$ is as defined in Eq.~(\ref{eq:survival}). Various diffuse models are tested, as labelled. We also show the fraction of pixels satisfying this criterion in simulated data maps with and without NFW PS, with $68\%$ confidence interval bands shown. The ROI is the Inner Galaxy with a 1 degree 3FGL mask.}
\label{fig:survival}
\end{figure*}

NPTF methods do not by default reconstruct the likely positions of the PSs. As a proxy, however, we can ask which pixels have photon count values that are particularly unlikely in the absence of PSs -- so-called ``hot pixels'' -- following Ref.~\cite{Lee:2015fea}. Quantitatively, we first perform a Poissonian template fit in the ROI, omitting the PS templates from the standard NPTF pipeline. We then define a reference model as the sum of these Poissonian templates, with normalizations given by the posterior medians from the fit. Suppose the reference model gives an expected mean number of counts $\mu_p$ in pixel $p$, and the real data contain $n_p$ photon counts in the same pixel. Then we can define the survival function as:
\begin{equation}
 \epsilon_p \equiv 1 - \textrm{CDF}\,[\mu_p, \, n_p],
 \label{eq:survival}
\end{equation}
where CDF$\,[\mu_p, \, n_p]$ is the cumulative probabilty of observing $n_p$ counts for a Poisson function with mean $\mu_p$. The smaller the value of $\epsilon_p$ in a pixel, the more unlikely the observed number of counts are in the absence of PSs, and thus the more likely the pixel is to contain an unresolved PS. Even in the absence of any PSs, there will still be pixels with small $\epsilon_p$, purely due to statistical fluctuations; however, the number of pixels with $\epsilon_p$ at a given level can discriminate between populations with different photon statistics. 

We compute the cumulative fraction of pixels with $\epsilon_p < \epsilon$, as a function of $\epsilon$, in the real data; we then repeat this test in simulated data constructed from fits (to the real data) with and without the NFW PS template, and from the Poissonian reference model. Following Ref.~\cite{Lee:2015fea}, by default we omit disk PSs from both the fits and the resulting simulated data; thus the three cases are (1) Poissonian templates only, (2) Poissonian templates + isotropic PSs, (3) Poissonian templates + isotropic PSs + NFW PSs. For ease of comparison to Ref.~\cite{Lee:2015fea}, we also use a $1^\circ$ radius mask for this analysis, slightly larger than our standard mask; we have tested somewhat smaller masks on a limited number of realizations, and found little effect on the results. 

For each case we take the true values of the parameters to be the medians of their posteriors in the relevant fit, and simulate 100 different datasets; the PS populations and Poissonian components are independently redrawn from the source-count functions or template normalizations respectively, for each of the 100 simulations. From each set of 100 simulations, we extract the median fraction of pixels with $\epsilon_p < \epsilon$ (as a function of $\epsilon$), and also the standard deviation of this value. We then further repeat this pipeline, but adding a $2\%$ DM flux injection to the real data at all points.

Figure~\ref{fig:survival} shows the results of this analysis for three Galactic diffuse models: the \textit{Fermi} \texttt{p6v11} diffuse model, and two models of Galactic diffuse emission generated using \texttt{GALPROP}~\cite{Strong:1998pw}, labeled models A and F in Ref.~\cite{Calore:2014xka}. 

We find that in all cases, the fraction of pixels with small $\epsilon_p$ in real data exceeds that for the Poissonian-only reference model, indicating (as expected) the presence of non-Poissonian fluctuations away from the reference model in the data. However, with the \texttt{p6v11} diffuse model, the simulated data including PSs -- either both NFW PSs and isotropic PSs, or isotropic PSs alone -- overproduces the fraction of pixels with small $\epsilon_p$. This could be due to a mismodeling of the spatial distribution of the PSs, the diffuse background, or both; the pixel $\epsilon_p$ depends strongly on where the PS is located, not only its brightness, as even a relatively faint source may be fairly significant if located in a region of low background. For example, suppose a non-isotropic population of PSs lying preferentially in regions of high background brightness is absorbed by the isotropic PS template; these PSs will then be simulated as purely isotropic, without the correlation with bright background regions, and the simulated data will have more PSs lying in regions of low background brightness (and hence more hot pixels) than the real data. 

This effect appears to be largely due to the isotropic PS population, which is detected with a significant positive flux in our analysis, unlike the analysis of Ref.~\cite{Lee:2015fea} which found it to be consistent with zero. That analysis used an earlier dataset for this CDF test (\texttt{Pass 7} vs \texttt{Pass 8}), and a slightly different energy range. We have tested changing the energy range in our version of this analysis, to match the results of Ref.~\cite{Lee:2015fea}, and continue to find that the simulated data overproduces the pixel fraction at small $\epsilon_p$ compared to the real data.

We have tested adding disk PSs to the fit and the simulated data, but the fit generally prefers to assign very little flux to disk PSs in our masked ROI, and this change does not qualitatively alter our results. In the other direction, we have tested forcing the isotropic PS population to zero, and fitting purely with NFW PSs; however, we then find the best-fit NFW PS population becomes brighter, with a higher cutoff in flux (presumably because the NFW PS template absorbs sources that were previously attributed to isotropic PSs). In this case the number of small-$\epsilon_p$ pixels is again overproduced. However, if we force the NFW PS population to have the same source-count-function parameters as extracted in Ref.~\cite{Lee:2015fea} -- with the exception of the overall normalization, which is allowed to float -- then we recover similar results to Ref.~\cite{Lee:2015fea}, with the number of small-$\epsilon_p$ pixels agreeing fairly well between the real and simulated data.

Replacing the \texttt{p6v11} diffuse model with models A and F (which give an overall better fit to the data), the mismatch between simulated and real data for small-$\epsilon_p$ pixels is significantly mitigated in models A and F. It is interesting to note that in model A, the simulated data containing isotropic PSs but no NFW PSs yields more low-$\epsilon_p$ pixels than the case with both isotropic and NFW PSs. This may again be because of the spatial mismodeling issue discussed above; PSs spatially coincident with the GCE, in a high-background region, will generally have larger $\epsilon_p$ (lower significance) than PSs with the same brightness distributed isotropically. If the isotropic PS template absorbs such sources, when they are re-simulated, they are likely to be placed in regions with lower background than in the real data, where their significance is accordingly higher.

Using Model A to describe the Galactic diffuse emission, when the artificial DM component is injected into the data, the fraction of pixels with the smallest $\epsilon_p$ values remains essentially unchanged. However, the fraction of small-$\epsilon_p$ pixels in the simulated data (where the NFW PS template rises to absorb the injected DM component) slightly decreases, becoming more similar to the real data. This is counterintuitive -- increasing the number of NFW PSs decreases the fraction of hot pixels -- but appears to be due to a shift in the median source-count function for the NFW PSs toward lower fluxes, although this shift is within the uncertainties. This highlights that the error bars for simulated data, in these survival-function analyses, do not include the scatter from the uncertainty in the simulated template parameters.

\section{Two Dimensional Posterior Values of Floated Parameters}
\label{sec:triangle}

In this section we show the triangle plots for several key analyses. These are the one and two dimensional posteriors of the parameters floated in the analyses.

\begin{figure*}[p!]
	\leavevmode
	\begin{center}
	\includegraphics[width=\textwidth]{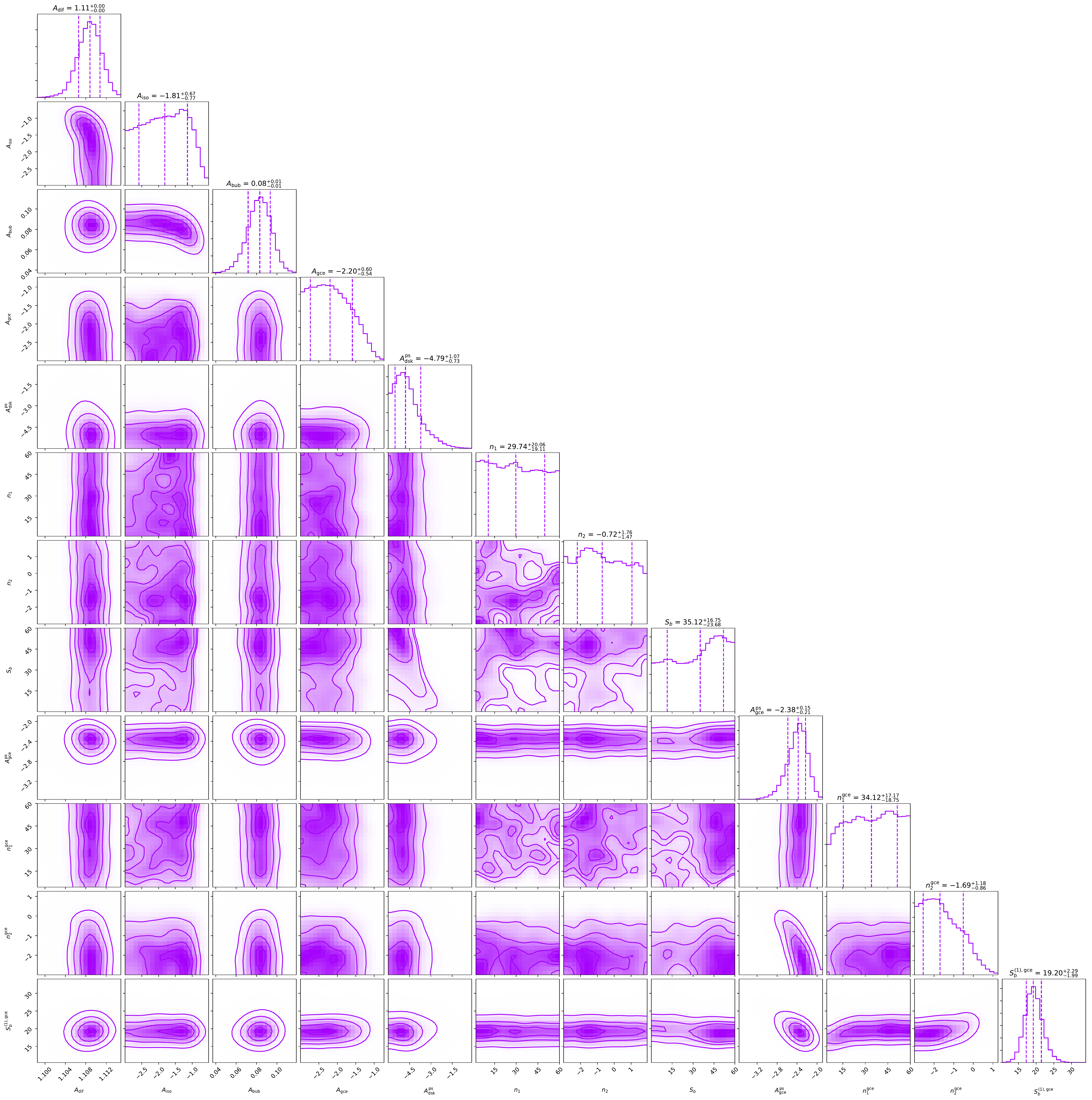} 
	\end{center}
	\vspace{-.50cm}
	\caption{Triangle plot for the simulated NP Bubbles, NP Disk, and Poissonian Bubbles, DM, Diffuse and Isotropic backgrounds, when they are analyzed with the same templates but the Bubbles PS are interchanged with NFW PS. The DM flux is not recovered.}     
	\vspace{3in}
	\label{fig:Run_6_tri}
\end{figure*}

Figure~\ref{fig:Run_6_tri} shows the triangle plot corresponding to the scenario where we have simulated NP Bubbles, NP Disk, and Poissonian Bubbles, DM, Diffuse and Isotropic backgrounds, when they are analyzed with the same templates but the Bubbles PS are interchanged with NFW PS. The DM flux is not recovered.

\begin{figure*}[p!]
	\leavevmode
	\begin{center}
	\includegraphics[width=\textwidth]{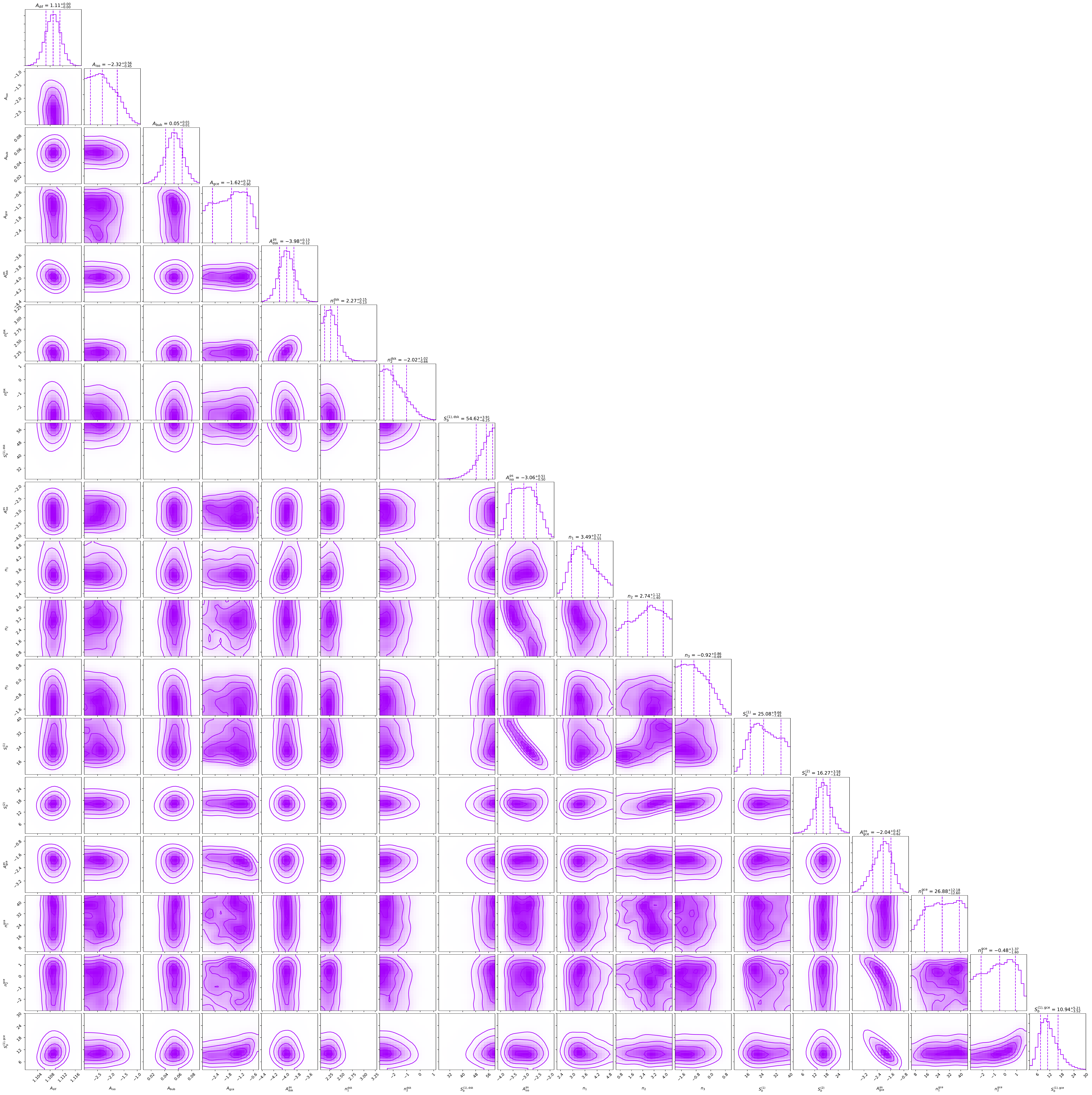} 
	\end{center}
	\vspace{-.50cm}
	\caption{Triangle plot for the simulated NP Bubbles, NP Disk, NP isotropic, and Poissonian Bubbles, DM, Diffuse and Isotropic backgrounds, when they are analyzed with the same templates but the Bubbles PS are interchanged with NFW PS, and the 3FGL are unmasked. The DM flux is not recovered.}     
	\vspace{3in}
	\label{fig:Run_6_tri_unmasked}
\end{figure*}

Figure~\ref{fig:Run_6_tri_unmasked} shows the triangle plot corresponding to the scenario where we have simulated NP Bubbles, NP Disk, NP isotropic, and Poissonian Bubbles, DM, Diffuse and Isotropic backgrounds, when they are analyzed unmasked with the same templates but the Bubbles PS are interchanged with NFW PS. The DM flux is not recovered.

\begin{figure*}[p!]
	\leavevmode
	\begin{center}
	\includegraphics[width=\textwidth]{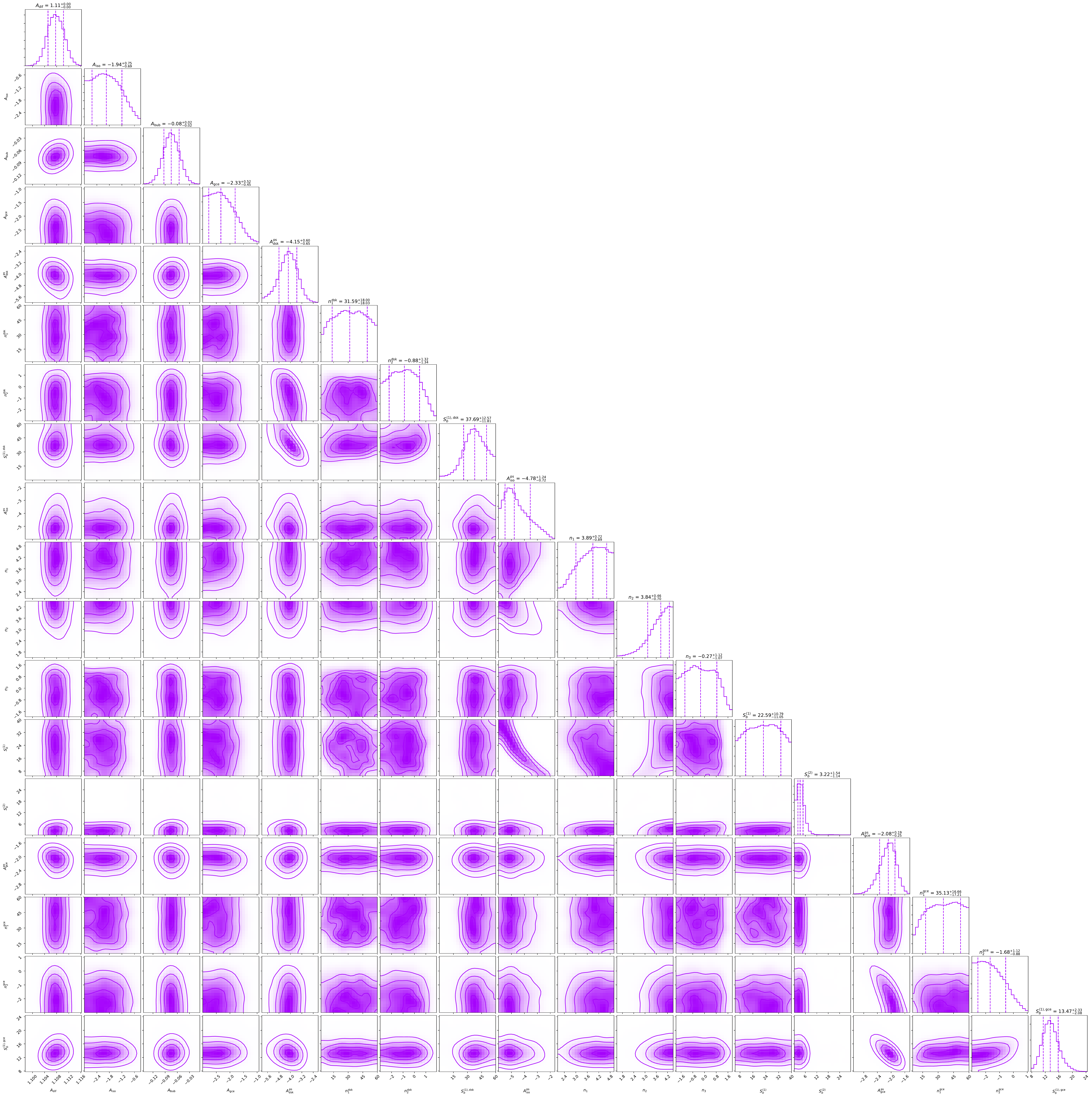} 
	\end{center}
	\vspace{-.50cm}
	\caption{Triangle plot for analysis of the \textit{Fermi} data without any injected DM flux, in the Inner Galaxy ROI. These posteriors for the templates are used as parameter values in the simulated data, except for the DM normalization which is taken from the same run but without NFW PS.}     
	\vspace{3in}
	\label{fig:DMinject_none_tri}
\end{figure*}

Figure~\ref{fig:DMinject_none_tri} shows the triangle plot corresponding to the case where the \textit{Fermi} data are analyzed without any injected DM flux, in the Inner Galaxy ROI. These posteriors for the templates are used as parameter values in the simulated data, except for the DM normalization which is taken from the same run but without NFW PS.

\begin{figure*}[p!]
	\leavevmode
	\begin{center}
	\includegraphics[width=\textwidth]{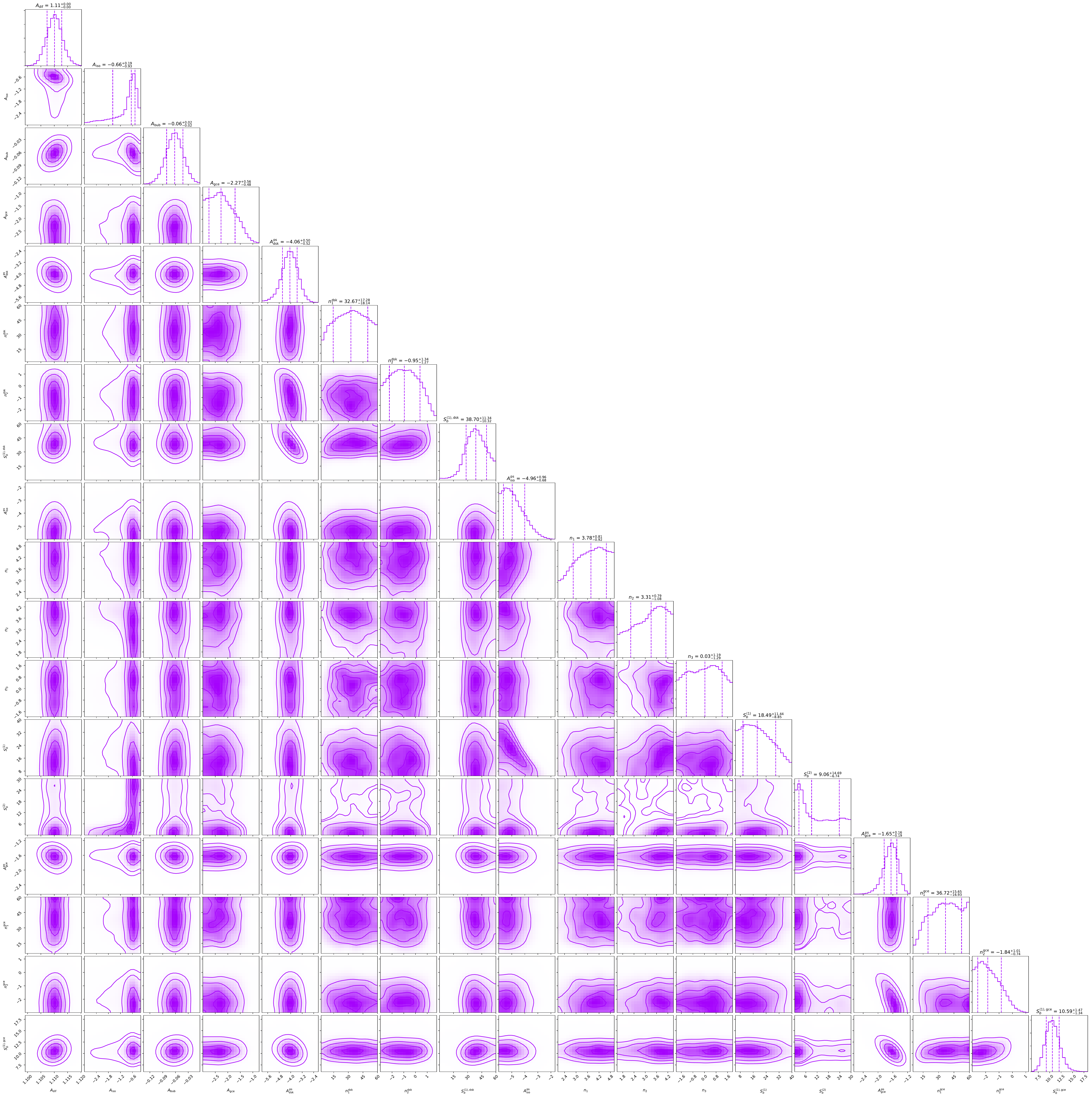} 
	\end{center}
	\vspace{-.50cm}
	\caption{Triangle plot for analysis of an artificial DM signal ($1.8\%$ flux post-injection) inserted into the \textit{Fermi} data. The DM flux is not recovered.}     
	\vspace{3in}
	\label{fig:Run_15_tri}
\end{figure*}

Figure~\ref{fig:Run_15_tri} shows the triangle plot corresponding to the analysis of an artificial DM signal inserted into the \textit{Fermi} data (1.8$\%$ DM injection). The DM flux is not recovered.

\begin{figure*}[p!]
	\leavevmode
	\begin{center}
	\includegraphics[width=\textwidth]{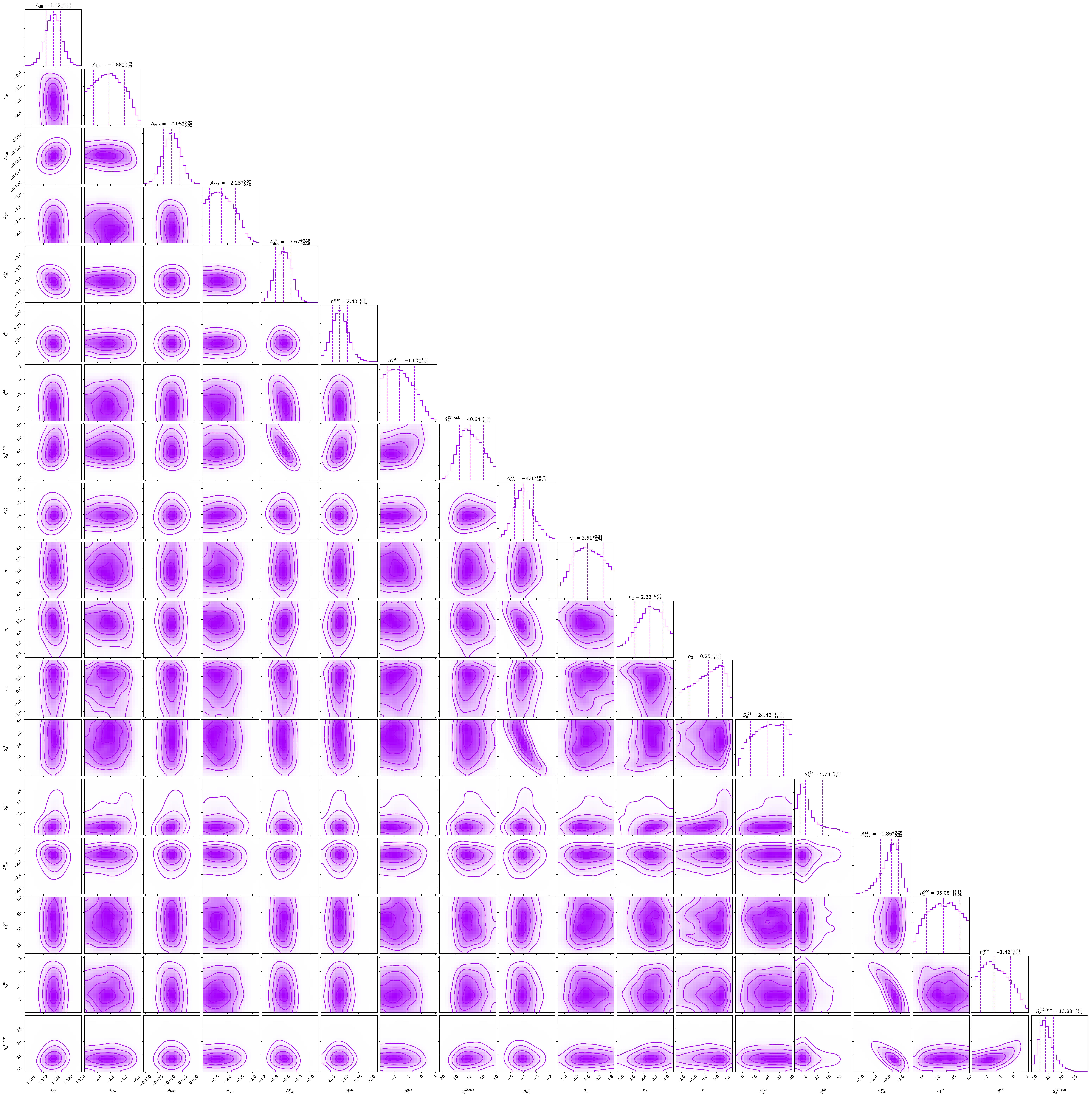} 
	\end{center}
	\vspace{-.50cm}
	\caption{Triangle plot for analysis of an artificial DM signal ($2.4\%$ flux post-injection) inserted into the \textit{Fermi} data, with the 3FGL unmasked. The DM flux is not recovered.}     
	\vspace{3in}
	\label{fig:Run_15_tri_unmasked}
\end{figure*}

Figure~\ref{fig:Run_15_tri_unmasked} shows the triangle plot corresponding to the scenario an artificial DM signal is inserted into the \textit{Fermi} data (2.4$\%$ DM injection), but with the 3FGL unmasked. The DM flux is not recovered.

\begin{figure*}[p!]
	\leavevmode
	\begin{center}
	\includegraphics[width=\textwidth]{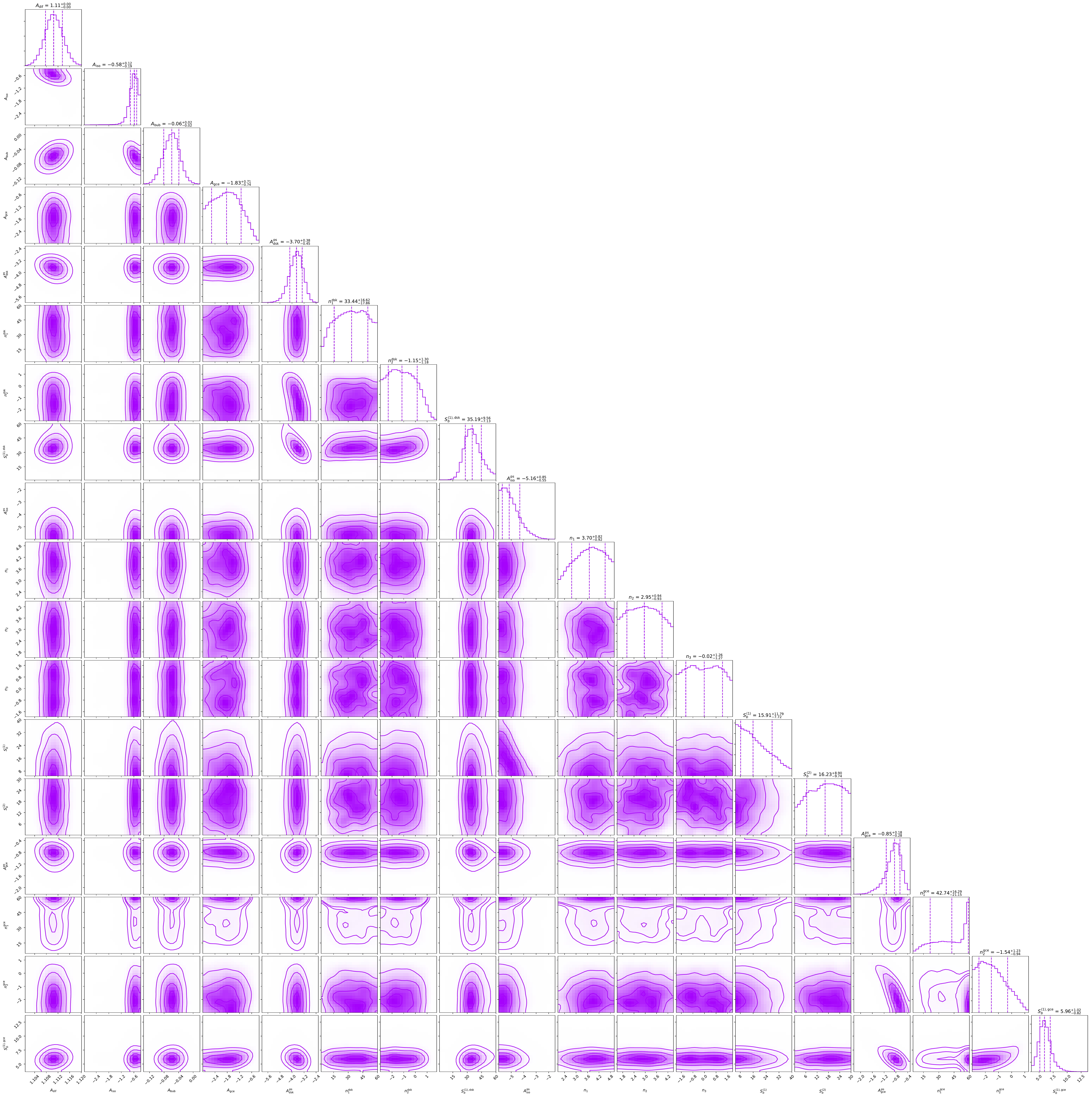} 
	\end{center}
	\vspace{-.50cm}
	\caption{Triangle plot for analysis of a larger ($6.7\%$ flux post-injection) artificial DM signal is inserted into the \textit{Fermi} data. The DM flux is again not recovered.}     
	\vspace{3in}
	\label{fig:Run_18_tri}
\end{figure*}

Figure~\ref{fig:Run_18_tri} shows the triangle plot corresponding to the scenario a larger (6.7$\%$) artificial DM signal inserted into the \textit{Fermi} data. The DM flux is again not recovered.

\begin{figure*}[p!]
	\leavevmode
	\begin{center}
	\includegraphics[width=\textwidth]{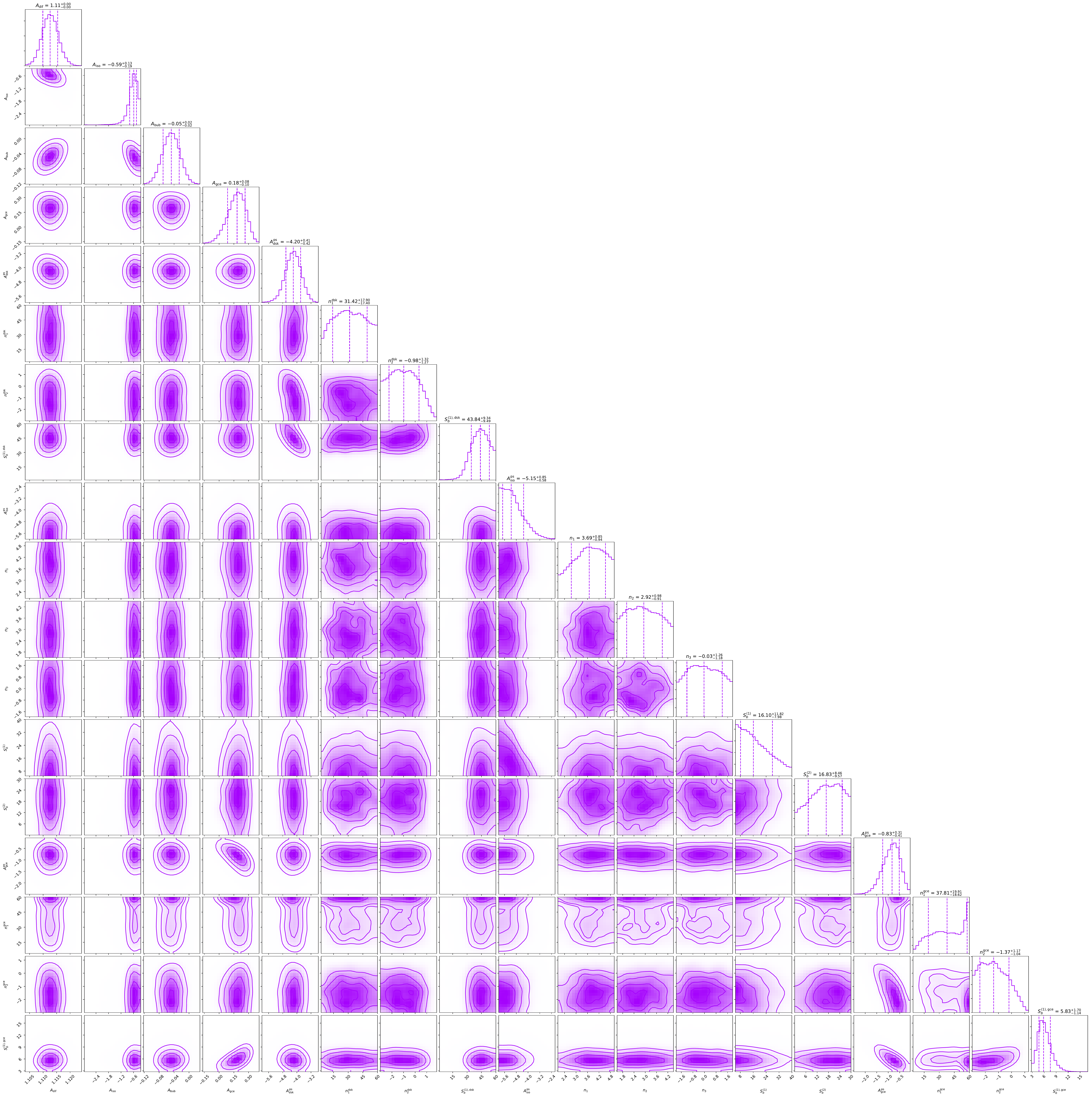} 
	\end{center}
	\vspace{-.50cm}
	\caption{Triangle plot for analysis of an even larger ($15.2\%$ flux post-injection) artificial DM signal is inserted into the \textit{Fermi} data. Some DM flux is now recovered, but some of it is still misattributed to PS.}     
	\vspace{3in}
	\label{fig:Run_20_tri}
\end{figure*}

Figure~\ref{fig:Run_20_tri} shows the triangle plot corresponding to the scenario an even larger (15.2$\%$) artificial DM signal inserted into the \textit{Fermi} data. Some DM flux is now recovered, but some of it is still misattributed to PS.

We note that in some cases, the index describing the slope of the source count function above the highest break ($n_1$) is not well converged within the prior range. We have tried increasing the prior range and do eventually find convergence; however, the physical difference between slopes at the maximum of the prior range and higher values are negligible, both corresponding to an extremely sharp cutoff in the source count function above the highest break.

\end{document}